\documentclass[reprint,showpacs,superscriptaddress,amsmath,amssymb,aps,prc]{revtex4-1}

\usepackage{graphicx}
\usepackage{dcolumn}
\usepackage{bm}
\usepackage{xcolor} 

\begin{document}


\title{Constraints for stellar electron-capture rates on $^{86}$Kr via the $^{86}$Kr($t$,$^{3}$He$+\gamma$)$^{86}$Br reaction and the implications for core-collapse supernovae}

\author{R.~Titus}
\affiliation{National Superconducting Cyclotron Laboratory, Michigan State University, East Lansing, MI 48824,
  USA}
\affiliation{Joint Institute for Nuclear Astrophysics - Center for the Evolution of the Elements,  Michigan State University, East Lansing, MI 48824,
USA}
\affiliation{Department of Physics and Astronomy, Michigan State University, East Lansing, MI 48824, USA}

\author{E.M.~Ney}
\affiliation{Department of Physics and Astronomy, The University of North Carolina at Chapel Hill, Chapel Hill, NC 27599, USA}

\author{R.G.T.~Zegers}
\email{zegers@nscl.msu.edu}
\affiliation{National Superconducting Cyclotron Laboratory, Michigan State University, East Lansing, MI 48824,
  USA}
\affiliation{Joint Institute for Nuclear Astrophysics - Center for the Evolution of the Elements,  Michigan State University, East Lansing, MI 48824,
USA}
\affiliation{Department of Physics and Astronomy, Michigan State University, East Lansing, MI 48824, USA}

\author{D.~Bazin}
\affiliation{National Superconducting Cyclotron Laboratory, Michigan State University, East Lansing, MI 48824,
  USA}
\affiliation{Department of Physics and Astronomy, Michigan State University, East Lansing, MI 48824, USA}

\author{J.~Belarge}
\affiliation{National Superconducting Cyclotron Laboratory, Michigan State University, East Lansing, MI 48824,
  USA}

\author{P.C.~Bender}
\affiliation{Department of Physics, University of Massachusetss Lowell, Lowell, MA 01854, USA}

\author{B.A.~Brown}
\affiliation{National Superconducting Cyclotron Laboratory, Michigan State University, East Lansing, MI 48824,
  USA}
\affiliation{Joint Institute for Nuclear Astrophysics - Center for the Evolution of the Elements,  Michigan State University, East Lansing, MI 48824,
USA}
\affiliation{Department of Physics and Astronomy, Michigan State University, East Lansing, MI 48824, USA}

\author{C.M.~Campbell}
\affiliation{Lawrence Berkeley National Laboratory, Berkeley, CA 94720, USA}

\author{B.~Elman}
\affiliation{National Superconducting Cyclotron Laboratory, Michigan State University, East Lansing, MI 48824,
  USA}
\affiliation{Department of Physics and Astronomy, Michigan State University, East Lansing, MI 48824, USA}

\author{J.~Engel}
\affiliation{Department of Physics and Astronomy, The University of North Carolina at Chapel Hill, Chapel Hill, NC 27599, USA}

\author{A.~Gade}
\affiliation{National Superconducting Cyclotron Laboratory, Michigan State University, East Lansing, MI 48824,
  USA}
\affiliation{Joint Institute for Nuclear Astrophysics - Center for the Evolution of the Elements,  Michigan State University, East Lansing, MI 48824,
USA}
\affiliation{Department of Physics and Astronomy, Michigan State University, East Lansing, MI 48824, USA}

\author{B.~Gao}
\affiliation{Institute of Modern Physics, Chinese Academy of Sciences, Lanzhou 730000, China}

\author{E.~Kwan}
\affiliation{National Superconducting Cyclotron Laboratory, Michigan State University, East Lansing, MI 48824,
  USA}

\author{S.~Lipschutz}
\affiliation{National Superconducting Cyclotron Laboratory, Michigan State University, East Lansing, MI 48824,
  USA}
\affiliation{Joint Institute for Nuclear Astrophysics - Center for the Evolution of the Elements,  Michigan State University, East Lansing, MI 48824,
USA}
\affiliation{Department of Physics and Astronomy, Michigan State University, East Lansing, MI 48824, USA}

\author{B.~Longfellow}
\affiliation{National Superconducting Cyclotron Laboratory, Michigan State University, East Lansing, MI 48824,
  USA}
\affiliation{Department of Physics and Astronomy, Michigan State University, East Lansing, MI 48824, USA}

\author{E.~Lunderberg}
\affiliation{National Superconducting Cyclotron Laboratory, Michigan State University, East Lansing, MI 48824,
  USA}
\affiliation{Department of Physics and Astronomy, Michigan State University, East Lansing, MI 48824, USA}

\author{T.~Mijatovi\'{c}}
\affiliation{National Superconducting Cyclotron Laboratory, Michigan State University, East Lansing, MI 48824,
  USA}

\author{S.~Noji}
\affiliation{National Superconducting Cyclotron Laboratory, Michigan State University, East Lansing, MI 48824,
  USA}
\affiliation{Joint Institute for Nuclear Astrophysics - Center for the Evolution of the Elements,  Michigan State University, East Lansing, MI 48824,
USA}

\author{J.~Pereira}
\affiliation{National Superconducting Cyclotron Laboratory, Michigan State University, East Lansing, MI 48824,
  USA}
\affiliation{Joint Institute for Nuclear Astrophysics - Center for the Evolution of the Elements,  Michigan State University, East Lansing, MI 48824,
USA}

\author{J.~Schmitt}
\affiliation{National Superconducting Cyclotron Laboratory, Michigan State University, East Lansing, MI 48824,
  USA}
\affiliation{Joint Institute for Nuclear Astrophysics - Center for the Evolution of the Elements,  Michigan State University, East Lansing, MI 48824,
USA}
\affiliation{Department of Physics and Astronomy, Michigan State University, East Lansing, MI 48824, USA}

\author{C.~Sullivan}
\affiliation{National Superconducting Cyclotron Laboratory, Michigan State University, East Lansing, MI 48824,
  USA}
\affiliation{Joint Institute for Nuclear Astrophysics - Center for the Evolution of the Elements,  Michigan State University, East Lansing, MI 48824,
USA}
\affiliation{Department of Physics and Astronomy, Michigan State University, East Lansing, MI 48824, USA}

\author{D.~Weisshaar}
\affiliation{National Superconducting Cyclotron Laboratory, Michigan State University, East Lansing, MI 48824,
  USA}

\author{J.C.~Zamora}
\affiliation{Instituto de F\'isica, Universidade de S$\tilde{a}$o Paulo, 05508-090 S$\tilde{a}$o Paulo, Brazil}


\date{\today}

\begin{abstract}
In the late stages of stellar core-collapse, prior to core bounce, electron captures on medium-heavy nuclei drive deleptonization and simulations require the use of accurate reaction rates.  Nuclei with neutron number near $N=50$, just above atomic number $Z=28$, play an important role, but rates used in astrophysical simulations rely primarily on a relatively simple single-state approximation. In order to improve the accuracy of astrophysical simulations, experimental data are needed to test the electron-capture rates and to guide the development of better theoretical models. This work presents the results of the $^{86}$Kr($t$,$^{3}$He+$\gamma$) experiment at the NSCL, from which an upper limit for the Gamow-Teller strength up to an excitation energy in $^{86}$Br of 5 MeV is extracted. The derived upper limit for the electron-capture rate on $^{86}$Kr indicates that the rate estimated through the single-state approximation is too high and that rates based on Gamow-Teller strengths estimated in shell-model and QRPA calculations are more accurate. The QRPA calculations tested in this manner were used for estimating the electron capture rates for 78 isotopes near $N=50$ and above $Z=28$. The impact of using these new electron-capture rates in simulations of supernovae instead of the rates based on the single-state approximation is investigated, indicating a significant reduction in the deleptonization that affects multi-messenger signals, such as the emission of neutrinos and gravitational waves.
\end{abstract}



\maketitle

\section{\label{sec:introduction}Introduction}
Reactions mediated by the weak nuclear force, such as electron captures and $\beta$ decays, are known to play important roles in many stellar phenomena. In particular, the rates at which nuclei capture electrons at high stellar densities and temperatures affects the evolution of core-collapse supernovae \cite{Fryer1999,Heger2003,janka07, 10.1093/ptep/pts067,BUR13,RevModPhys.75.819}.  The late stages of core-collapse supernovae, immediately before the explosion of the star, are heavily dependent on electron-capture rates on medium-heavy, neutron-rich nuclei \cite{PhysRevD.95.063019,janka07,hix03,A_bbal79,PhysRevLett.90.241102,Sull2016}. In recent sensitivity studies \cite{Sull2016,TITUS2018,furusawa2017,1906.05114}, the electron captures on nuclei surrounding the $N=50$ shell closure above $^{78}$Ni (here referred to as the ``high-sensitivity region'') were shown to have a significant effect on the change in electron fraction of the star during the period of deleptonization until core bounce.

Electron-capture rates are sensitive to Gamow-Teller transitions ($\Delta L=0$, $\Delta S=1$, $\Delta J=1$) in the $\beta ^+$ direction. Such transitions can be measured directly via $\beta^+$-decay experiments but are limited to probing transitions within a finite $Q$-value window. For neutron-rich systems, which are of greatest importance for core-collapse supernovae, the $Q$ values of such reactions are negative. Hence, no direct information can be obtained from $\beta^{+}$ decay, although $\beta^{-}$ decay data from the electron-capture daughter to the mother can be used to estimate the ground-state to ground-state transition strength. Gamow-Teller strengths may also be measured indirectly via charge-exchange reactions, which are not limited by a $Q$-value window, yielding information about transitions at higher excitation energies. Additionally, because there is a well-known proportionality between the charge-exchange cross section and the Gamow-Teller strength \cite{Taddeucci1987125,Zegers:2006,PhysRevLett.99.202501}, the strength distribution, and associated electron-capture rates, can be extracted model-independently.

While it would be preferable to derive electron-capture rates for astrophysical simulations based on measured Gamow-Teller strengths, this is not feasible for two reasons. First, there are thousands of nuclei that participate in these astrophysical processes, making it difficult to perform charge-exchange experiments on all of them in a timely manner. Second, transitions from thermally populated low-lying excited states \cite{LANGANKE2000481} and high-temperature unblocking effects occur in stellar environments \cite{PhysRevC.63.032801}, but cannot be explored in the laboratory. For these reasons, the majority of electron-capture rates must come from theoretical calculations. Experimental data are needed to validate and benchmark current theoretical models and to guide the improvement of these models.

The neutrino interaction library NuLib \cite{OCO15,Sull2016a}, which is used in a variety of astrophysical simulations, including the spherically-symmetric, general-relativistic hydrodynamics code GR1D, \cite{OCO15,OCO10} used in this work, contains electron-capture rates on approximately 4000 nuclei. The electron-capture rates within the tables are derived from a variety of theoretical calculations and experimental data \cite{fuller82,ODA1994231,RevModPhys.75.819,LAN01a,hix03,PhysRevC.63.032801,pruet1,Suzuki2016}. For light nuclei (up to the $pf$ shell), shell-model calculations have been used for computing electron-capture rates, which are well-tested against data (see Ref. \cite{cole_ecrate_pf} and references therein). However, for medium-heavy and heavy nuclei, and nuclei near the drip lines, the calculations rely on a variety of other approaches. For a large number of nuclei for which no rates based on microscopic calculations are available for the density and temperature ranges of relevance for astrophysical simulations, a ``single-state approximation'' is presently used. It is based on the following parametrization \cite{PhysRevLett.90.241102, PhysRevC.95.025805}:
\begin{equation}
\label{eqn:approx}
\lambda_{EC}=\frac{\textrm{ln}2 \cdot B}{K} \Bigg(\frac{T}{m_e c^2} \Bigg)^5 \big[F_4(\eta) - 2\chi F_3(\eta) + \chi^2 F_2(\eta)\big],
\end{equation}
where $m_e$ is the electron mass, $K=6146$ s, $F_k$ are Fermi integrals of rank $k$ and degeneracy $\eta$, $\chi=(Q- \Delta E)/T$, $\eta=\chi+\mu_e/T$, and $T$ and $\mu_e$ are the temperature and electron chemical potential, respectively. $B$, the effective Gamow-Teller transition strength, is fixed for all isotopes to 4.6. $\Delta E$, the effective excitation energy was originally fixed \cite{PhysRevLett.90.241102} to a single value for all nuclei, but following Ref. \cite{PhysRevC.95.025805}, is adjusted based on the neutron and proton numbers of the parent nucleus.

For the nuclei in the high-sensitivity region, one presently relies primarily on this single-state approximation to calculate the electron-capture rates, potentially leading to a significant overestimation of the rates \cite{TITUS2018}, because it does not account fully for Pauli-blocking effects, which become increasingly prominent for progressively more neutron-rich nuclei. In order to achieve more accurate astrophysical simulations, it is necessary to obtain more accurate electron-capture rates in the high-sensitivity region. Therefore, an effort was started to better constrain and guide the theoretical development by performing charge-exchange experiments on nuclei at and near the $N=50$ shell closure, starting with $^{86}$Kr, $^{88}$Sr \cite{juan2019}, and $^{93}$Nb. In parallel, new theoretical calculations were pursued that can be compared with the single-state approximation, and  benchmarked by the data. These efforts are closely integrated with astrophysical simulations, in order to have immediate feedback on sensitivities of astrophysical phenomena to variations in electron-capture rates derived from experimental and theoretical strength distributions.

This work describes the experimental results of a ($t$,$^{3}$He$+\gamma$) charge-exchange reaction experiment on $^{86}$Kr and the comparison with shell-model and quasi-particle random-phase approximation (QRPA) calculations. The latter model is also used to create a new addition to the electron-capture rate tables, which is inserted into NuLib to estimate the impact on the late-stage evolution of core-collapse supernovae.

\section{\label{sec:experiment} Experimental Setup}
The $^{86}$Kr($t$,$^3$He+$\gamma$) experiment was performed at the National Superconducting Cyclotron Laboratory (NSCL). A primary beam of $^{16}$O, generated by the Coupled Cyclotron Facility (CCF) with an energy of 150 MeV/$u$, was impinged on a beryllium production target with a thickness of 3525 mg/cm$^2$. The A1900 fragment separator \cite{MORRISSEY200390} was used to select tritons from the reaction products by using an aluminum wedge \cite{HITT2006264}, yielding a secondary beam with an energy of approximately 115 MeV/$u$ and a purity in excess of 99\%. \textcolor{black}{A negligible amount of $^{6}$He particles was also present in the beam. The energy spread of the triton beam was about 3.3 MeV}.
The tritons were transported to a $^{86}$Kr gas target cell, which was 7 cm in diameter and 3 cm in thickness. The gas target was controlled and monitored by the Ursinus College Liquid Hydrogen Target gas handling system \cite{2010APS.DNP.EA083P}. When filled with $^{86}$Kr gas with a purity of 99.952\% to a pressure of 1210 Torr at a temperature of 295 K, the target thickness was approximately constant with a value of 20 mg/cm$^2$. The target thickness was measured by comparing ($t$,$^{3}$He) spectra in the S800 focal plane with the target filled and empty. The target cell windows were made of kapton foil (C$_{22}$H$_{10}$N$_2$O$_5$) with a thickness of 125 $\mu$m. Reactions on the $^{12}$C nuclei in the windows were used for data calibration. Production runs with the target cell both filled and empty were performed in order to determine the contribution of events from the target windows to the measured cross section. In addition, data were taken with only the upstream target-cell foil and only the downstream target-cell foil to model the background from reactions on the windows, as discussed below.

$^3$He reaction products were momentum-analyzed by the S800 magnetic spectrograph \cite{BAZIN2003629}. The triton beam was transported to the target using dispersion-matched optics \cite{FUJITA200217} \textcolor{black}{to improve the energy resolution of the reconstructed excitation-energy spectra to about 400 keV (see below) even though the beam energy spread was 3.3 MeV}.  \textcolor{black}{The magnetic rigidity of the transport beam line was 4.8 Tm (close to the present limit for operating in dispersion-matched optics). The magnetic rigidity of the S800 was set to 2.32 Tm, well below its maximum rigidity of 4 Tm}.
A timing signal from a 5-mm thick plastic scintillator, placed behind the tracking detectors at the end of the S800 focal plane, was used in conjunction with the cyclotron RF signal to obtain the time-of-flight of particles through the spectrograph.  \textcolor{black}{Together with the energy-loss signal in the plastic scintillator (the $^{3}$He particles were not stopped), $^{3}$He ejectiles were cleanly separated from background caused by unreacted tritons scattering off the S800 beam chamber in its first dipole magnet.} The scattering angles and momenta of the $^{3}$He ejectile at the target were reconstructed by using the position and angle measurements from two cathode-readout drift chambers (CRDCs) \cite{YURKON1999291}, \textcolor{black}{with a detection efficiency of about 95\%, measured relative to $^{3}$He particles detected in the scintillator}. An inverse raytrace matrix was used to determine the scattering angle and momentum at the target from the position and angle measurements in the focal plane detectors \cite{MAKINO1999338}. A missing-mass calculation yielded excitation energies of the reaction products, $^{86}$Br, along with $^{16}$F, $^{14}$C and $^{12}$B from the target windows, with an energy resolution of $\sim400$ keV full width at half maximum (FWHM). \textcolor{black}{This resolution was determined from the excitation of the $^{12}$B ground state, which is the dominant peak in the spectrum. Given the limited statistics obtained for reactions from the $^{86}$Kr gas target, the analysis was performed in bins of 500 keV wide.} Scattering angles at the target position were measured from 0$^{\textrm{o}}$ $\leq$ $\theta_{\textrm{lab}}$ $\leq$ 4$^{\textrm{o}}$ with an angular resolution of 12 mrad (FWHM). \textcolor{black}{The angular acceptance of the S800 is 100\% for scattering angles up to 50 mrad. For larger scattering angles, the reduced solid angle coverage is corrected for in the determination of the differential cross sections.}

The high resolution $\gamma$-ray detection array, GRETINA \cite{PASCHALIS201344,WEISSHAAR2017187}, was installed around the target position. To make room for the gas handling system of the krypton target cell, all thirty-two high-purity germanium crystals (36 fold segmented) were placed in the northern hemisphere of the frame, yielding about 1$\pi$ solid-angle coverage. The detectors were used to measure $\gamma$ rays from the de-excitation of $^{86}$Br, in an attempt to extract weak Gamow-Teller transitions from among the other transitions that could occur. Owing to the excellent photo-peak energy resolution and efficiency (\textcolor{black}{$\sim$6\% efficiency for $E_{\gamma}=1$ MeV as measured with $^{152}$Eu source}, and $\sim$4\% for $E_{\gamma}=2$ MeV), these measurements can be used to identify low-lying weak transitions which are not easily identified in the ($t$,$^{3}$He) singles data. This technique has been used successfully in past charge-exchange experiments to extract transitions with Gamow-Teller strengths of as low as $\sim$ 0.01 \cite{PhysRevC.92.024312,PhysRevLett.112.252501}. \textcolor{black}{The live-time of the combined data acquisition systems of S800 and GRETINA was about 90\%. The S800 singles $^{3}$He event rate was about 30 events per second. The true-to-random coincidence ratio for S800-GRETINA coincidences was about 125.}

\section{\label{sec:results}Experimental Results}
\subsection{\label{sec:calibration}Calibration of the Absolute Cross Section}
The windows of the gas target cell introduced reactions on $^{12}$C, $^{16}$O and $^{14}$N contaminants into the experimental data. Although this complicated the analysis as discussed below, it also provided an advantage. Because these reactions were present in all of the data, it simplified the data calibration and overall normalization. Calibrations of the experimental data and determination of the absolute normalization factors were performed by using the $^{12}$C($t$,$^3$He)$^{12}$B(1$^+$; g.s.) reaction, which has been studied in detail previously \cite{PhysRevC.90.025801} and the absolute cross sections are known. \textcolor{black}{That previous measurement used nearly the identical setup as for the experiment described here, except that in the previous measurement the beam intensity was carefully measured and monitored by in-beam scintillators, to reduce the systematic uncertainties in the absolute normalization}. Hence, the cross section for $^{86}$Kr($t$,$^{3}$He) reactions were determined relative to the known cross section for the  $^{12}$C($t$,$^3$He)$^{12}$B(1$^+$; g.s.) reaction (after accounting for the difference in the number of $^{12}$C and $^{86}$Kr particles in the target and its foils) since this eliminated the need to carefully monitor the beam intensity or to correct for detection efficiencies and other factors that can affect the absolute cross section measurement, as these are identical for reactions on $^{12}$ and $^{86}$.  Data for the $^{14}$N($t$,$^3$He)$^{14}$C reaction (previously measured by using the ($d$,$^{2}$He) reaction \cite{negret_n14}) \textcolor{black}{were used as additional checks on the energy and angular calibration.}

\begin{figure}[htp]
\begin{center}
\includegraphics[width=\linewidth]{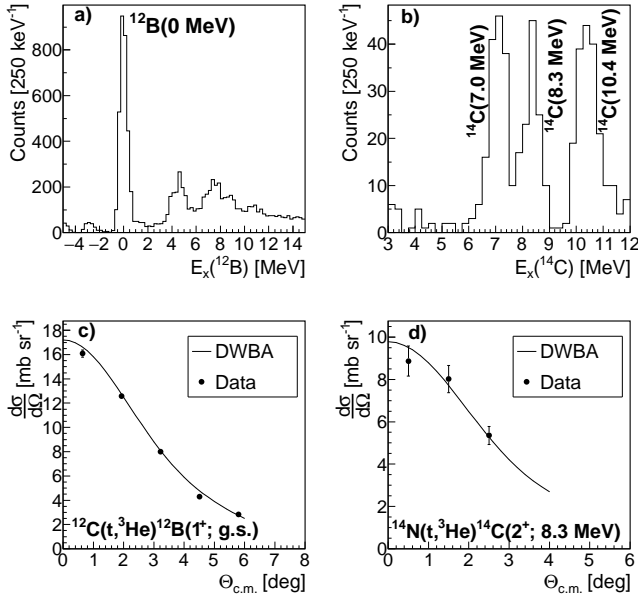}
\caption{($t$,$^{3}$He) excitation-energy spectra for $^{12}$C (a) and $^{14}$N (b) for laboratory scattering angles below $4^{\circ}$. The $^{12}$B(1$^+$; g.s.) and ($t$,$^3$He) transitions to $^{14}$C(2$^+$; 7.0 MeV), $^{14}$C(2$^+$; 8.3 MeV) $^{14}$C(2$^+$; 10.4 MeV) states are identified. Panels (c) and (d) contain the differential cross sections for the $^{12}$C($t$,$^3$He)$^{12}$B(1$^+$; g.s.) and $^{14}$N($t$,$^3$He)$^{14}$C(2$^+$; 8.3 MeV) excitations, respectively.}
\label{fig:carbon_nitrogen}
\end{center}
\end{figure}

The corresponding excitation-energy spectra extracted for these reactions are shown in Figs. \ref{fig:carbon_nitrogen}(a) and \ref{fig:carbon_nitrogen}(b). In Fig.~\ref{fig:carbon_nitrogen}(a) the $^{12}$C($t$,$^3$He)$^{12}$B(1$^+$; g.s.) transition can clearly be identified. In Fig.~\ref{fig:carbon_nitrogen}(b), several excitations belonging to ($t$,$^3$He) reactions on $^{14}$N are identified at their appropriate excitation energies. In Figs.~\ref{fig:carbon_nitrogen}(c) and \ref{fig:carbon_nitrogen}(d), the measured differential cross sections for the $^{12}$C($t$,$^3$He)$^{12}$B(1$^+$; g.s.) and $^{14}$N($t$,$^3$He)$^{14}$C(2$^+$; 8.3 MeV) excitations are shown. Both of these are well-known Gamow-Teller transitions and are compared with differential cross sections calculated in the distorted-wave Born approximation (DWBA); see Ref. \cite{PhysRevC.90.025801} and  Sec.~\ref{sec:singles} for further details. Note that for the $^{14}$N($t$,$^3$He)$^{14}$C(2$^+$; 8.3 MeV) reaction at scattering angles beyond $3^{\circ}$, charge-exchange reactions on hydrogen contaminated the signal and were excluded from Fig.~\ref{fig:carbon_nitrogen}(d).

\subsection{\label{sec:singles}Singles Analysis}
\begin{figure}[htp]
\begin{center}
\includegraphics[width=\linewidth]{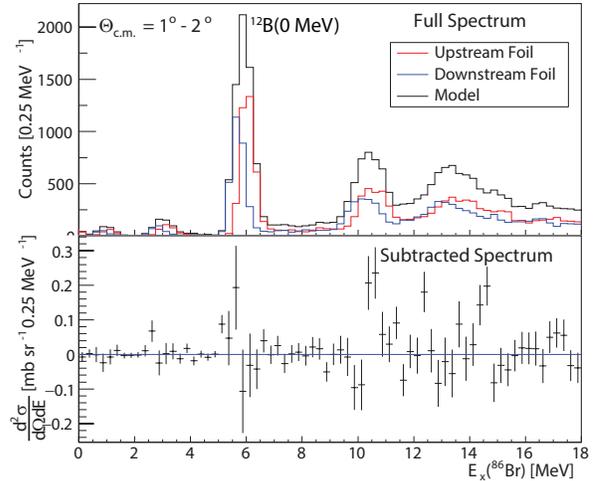}
\caption{(Top) Excitation-energy spectra from the separate measurements with the upstream (red) and downstream (blue) windows of the $^{86}$Kr gas cell. The spectra, calculated in the excitation-energy scale of the $^{86}$Kr($t$,$^{3}$He) reaction, are modified to account for energy loss and energy- and angular-straggling effects in the foils as discussed in the text for the purpose of creating a background model for events originating from the cell windows when used together, indicated by the black line. (Bottom) Difference between the ($t$,$^{3}$He) spectrum taken with the empty target cell and the background model.}
\label{fig:emptycell}
\end{center}
\end{figure}

Compared to the events from reactions on the target windows, the $^{86}$Kr($t$,$^3$He) signal was small. Therefore, it was necessary to subtract the target-window events from the krypton-cell data, in order to extract the reactions on $^{86}$Kr. To perform this subtraction accurately, data were taken on the upstream and downstream foils of the target cell separately, from which the background to the data could be modeled. To test the subtraction procedure, the data from the individual upstream and downstream windows were first used to reconstruct the empty-cell data (with both upstream and downstream windows in place). In order to accomplish this, corrections accounting for the energy losses and energy and angular straggling of the $^3$H and $^3$He particles in the windows were made. As an example, the results for one angular bin (1$^{\circ}$--2$^{\circ}$) are shown in Fig.~\ref{fig:emptycell}. In the top panel, there are several peaks, corresponding to events from $^{12}$C, $^{16}$O and $^{14}$N in the windows. Spectra from the individual upstream target window and the downstream target window are shown as well. For these, the energy-loss and energy- and angular-straggling corrections have already been applied, which is why the two spectra are slightly offset. The background model is the sum of these two contributions. When the model was subtracted from the empty-cell data, the bottom plot of Fig.~\ref{fig:emptycell} resulted. From this, it is clear that the background model matched the data from the empty cell quite well, although for $E_x >$ 5 MeV, a combination of statistical and systematic uncertainties caused some deviations from zero. These could, for example, be due to minor changes in the beam properties during the experiment.

After using the background-subtraction method on the empty cell, the process was modified to subtract the target-window events from the cell filled with krypton gas. The additional step necessary for this analysis was to include the energy loss due to the $^3$H and $^3$He particles passing through the $^{86}$Kr-filled cell. Due to the bulging of the target foils (up to 5 mm at the center of the foils) when the cell was filled with gas, in combination with the 5-cm tall beam-spot size in the dispersive plane associated with operating in dispersion-matched optics \cite{HITT2006264}, the $^{86}$Kr target thickness was not uniform. This effect was included in the background model. \textcolor{black}{By varying the energy-losses within reasonable experimental uncertainties ($\sim 50$ keV; the average energy loss of the beam and ejectile through the target is about 1.25 MeV), the subtraction was optimized, based on the reproduction of the strong $^{12}$C peak in the excitation-energy spectra.}

\begin{figure}[htp]
\begin{center}
\includegraphics[width=\linewidth]{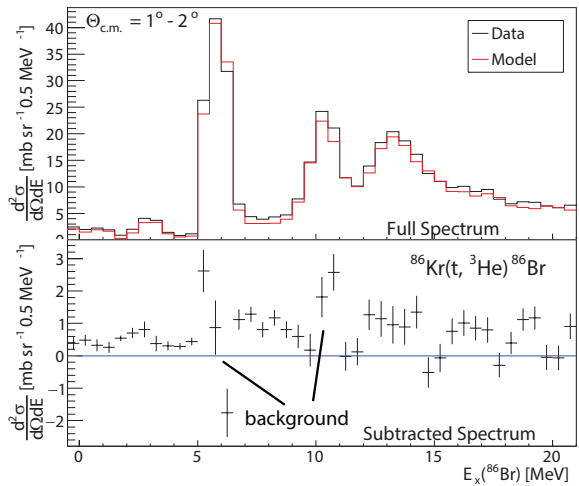}
\caption{Differential cross sections for the $^{86}$Kr($t$,$^{3}$He) reaction for $1^{\circ}<\Theta_{\textrm{c.m.}}<2^{\circ}$ before (top) and after (bottom) the background subtraction procedure. The two peaks at $\sim$5.5 MeV and $\sim$10.5 MeV in the spectrum before background subtraction are from target-window contaminants.}
\label{fig:cross_section}
\end{center}
\end{figure}

An example of the cross sections before and after the background subtraction, for the 1$^{\circ}$--2$^{\circ}$ angular bin, is shown in Fig.~\ref{fig:cross_section}. The top panel shows the full spectrum (target windows plus krypton) and the background model using the optimized parameters for energy loss and energy smearing within the target. The bottom panel shows the $^{86}$Kr($t$,$^3$He) spectrum after the subtraction of simulated background. Some systematic uncertainties remain, as evidenced, for example, by the remaining structure near the location of the $^{12}$B(1$^+$; g.s.) peak. Based on the optimization process of the background model parameters described above, the systematic uncertainties in the extracted cross sections for the $^{86}$Kr($t$,$^{3}$He)$^{86}$Br reaction were estimated. For $E_x$($^{86}$Br) $<5$ MeV the systematic uncertainties in the cross sections were $\sim$25\% determine for energy-loss shifts of up to 50 keV. For $E_x$($^{86}$Br) $\geq$ 5 MeV, where the cross sections for the reactions on $^{12}$C and $^{16}$O are very high compared to the reactions on $^{86}$Kr, the systematic uncertainties were too large to obtain sufficiently reliable cross sections for the $^{86}$Kr($t$,$^3$He) reaction. Hence, for the remainder of the analysis, only the data for $E_x$($^{86}$Br) $<5$ MeV were analyzed.

By using the newly subtracted cross sections, angular distributions were extracted from the data and a multipole decomposition analysis (MDA) was performed \cite{BONIN1984349,Ichimura:2006,PhysRevC.90.025801}. Through this method, the Gamow-Teller component ($\Delta L=0$) was extracted from the cross section. The angular distribution for each 0.5-MeV wide excitation-energy bin from 0 MeV up to 5 MeV in $^{86}$Br was fitted with a linear combination of angular distributions associated with monopole ($\Delta L=0$), dipole ($\Delta L=1$) and quadrupole ($\Delta L=2$) transitions. The components were calculated using the distorted-wave Born-approximation (DWBA) code {\footnotesize{FOLD}} \cite{FOLD}, in which the Love-Franey $NN$-interaction at 140 MeV/$u$ \cite{PhysRevC.31.488} was double-folded over the transition densities of the $^{86}$Kr-$^{86}$Br and $t$-$^3$He systems. One-body transitions densities for the $\Delta L=0,2$ transitions were obtained from a shell-model calculation using the {\footnotesize{NUSHELLX}} code \cite{nushellx_msu} in the SNE model space \cite{PhysRevC.65.021302} using the {\footnotesize{jj44pna}} interaction \cite{PhysRevC.70.044314} (for details, see below). One-body transition densities for the $\Delta L=1$ component were obtained from a normal-modes calculation \cite{HOFSTEE1995729}, using the {\footnotesize{NORMOD}} code \cite{Normod_code}. Transition densities for $t$ and $^3$He particles were taken from variational Monte Carlo calculations \cite{pieper_wiringa_2001}. The optical potential used for the DWBA calculation was from elastic $^3$He scattering on $^{90}$Zr \cite{PhysRevC.67.064612} for the outgoing $^{3}$He channel. For the incoming triton channel, the real and imaginary potential depths were scaled to 85\% of the values for $^{3}$He, following Ref.~\cite{VANDERWERF1989305}. Although transitions associated with angular-momentum transfers $\Delta L>2$ can contribute to the spectra, they are expected to be small for the small momentum transfers considered here and have angular distributions at forward scattering angles that are similar to the angular distribution for $\Delta L=2$ transitions and thus not included as separate components.

\begin{figure}[htp]
\begin{center}
\includegraphics[width=\linewidth]{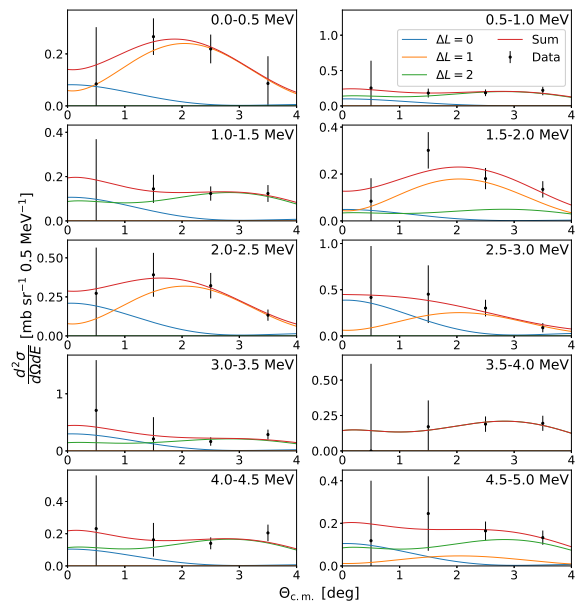}
\caption{The multipole decomposition analysis (MDA) results for each of the 0.5-MeV wide excitation-energy bins for $E_x$($^{86}$Br)$<5$ MeV. The experimental data is given by the black points. The fitting function is a linear combination of three multipole components: $\Delta L=0$, $\Delta L=1$ and $\Delta L=2$. The error bars on the experimental data points include both statistical and systematic uncertainties.}
\label{fig:mda}
\end{center}
\end{figure}

The results of the MDA are shown in Fig.~\ref{fig:mda} for each of the 0.5-MeV wide excitation-energy bins below $E_x$($^{86}$Br) $=$ 5 MeV. Some of the distributions, such as the 0 to 0.5 MeV bin, show an angular distribution that is dominated by dipole strength, whereas other bins, such as the 2.0 to 2.5 MeV bin, illustrate the case in which there could be a larger $\Delta L=0$ component. Due to the large error bars on the experimental data, which include both statistical and systematic contributions, the uncertainties in the MDA fitting parameters for each component were also large. As such, the extraction of the Gamow-Teller strength from the MDA results also carried relatively large uncertainties.

\begin{figure}[htp]
\begin{center}
\includegraphics[width=\linewidth]{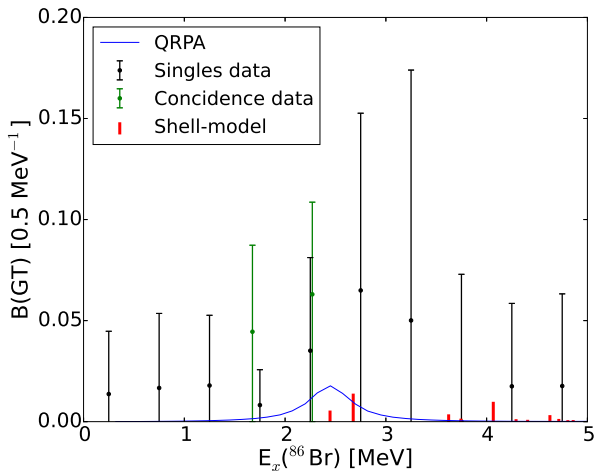}
\caption{Gamow-Teller strength distribution extracted from the $^{86}$Kr($t$,$^{3}$He) data and comparison with shell-model and QRPA calculations, as described in the text.}
\label{fig:bgt}
\end{center}
\end{figure}

As mentioned above, there is a well-known proportionality between the cross section at zero momentum transfer ($q=0$) and the Gamow-Teller strength, given by the following expression \cite{Taddeucci1987125,Zegers:2006,PhysRevLett.99.202501,PhysRevC.83.054614}:
\begin{equation} \label{proportionality}
\Bigg[\frac{d\sigma}{d\Omega} (q=0) \Bigg ] _{\mathrm{GT}} = \widehat{\sigma} \mathrm{B(GT)}.
\end{equation}
In this equation, $\frac{d\sigma}{d\Omega} (q=0 \textrm{ MeV/c})$ refers to the $\Delta L=0$ component of the cross section, extrapolated to zero momentum transfer, and $\widehat{\sigma}$ is the unit cross section, calculated using the empirical expression $\widehat{\sigma} = 109 A^{-0.65}$ mb/sr, where $A$ is the mass number of the target nucleus \cite{PhysRevLett.99.202501,Taddeucci1987125,PhysRevC.90.025801,Sasano:2009,PhysRevC.77.064303}. Following these references, to obtain the cross sections at $q=0$, the extracted cross sections at $0^{\circ}$ from the MDA at finite reaction $Q$ values were extrapolated to $Q=0$ MeV by using the DWBA calculations. The extracted Gamow-Teller strength distribution is shown in Fig.~\ref{fig:bgt}. Because of the uncertainties associated with the analysis, it is only possible to provide an upper limit on the extracted Gamow-Teller strength, as the error bars on the values are consistent with zero strength. More detailed constraints will be obtained by analysis of the $\gamma$-ray data and will be discussed in Sec.~\ref{sec:gammas}.

\subsection{\label{sec:gammas}Coincidence Analysis}
In an effort to improve the constraints on the Gamow-Teller strength from $^{86}$Kr, the S800-GRETINA coincidence data were used. The decay spectrum of $^{86}$Br, the residual nucleus of the $^{86}$Kr($t$,$^3$He) reaction, is partially known, though spin and parity assignments for some levels are not available or tentatively assigned \cite{NEGRET20151,PhysRevC.94.044328}. Excited states at 2447 keV, 2665 keV, 3226 keV, and 3365 keV were (tentatively) identified as $1^{+}$ states \textcolor{black}{(the threshold for decay by neutron emission is at 5.128 MeV)}. However, no $\gamma$-ray peaks associated with de-excitations from those states were observed in the present data, \textcolor{black}{even though the detection efficiency is still relatively high for such energies and photons with energies of up to 8 MeV have clearly been identified in a previous similar experiment (see Fig. 3 of Ref. \cite{PhysRevLett.112.252501})}. It can be concluded that although these states are populated in the $\beta^{-}$ decay of $^{86}$Se, they are not populated significantly in the $\beta^{+}$ direction. Observed $\gamma$ lines from nuclei in the target windows (in the region of interest from $^{14}$C and $^{16}$N), were eliminated from further analysis. Some known $\gamma$ lines from $^{86}$Br were observed, as well as a number of $\gamma$ lines previously not observed for $^{86}$Br. Table \ref{gamma_br} lists the $\gamma$ rays that originated from $^{86}$Br, along with their spin-parity assignments where available from previous experiments. For $\gamma$ energies below $\sim$500 keV, it was not always possible to distinguish a small peak from background due to Compton scattering of more energetic $\gamma$ rays that deposit less than their full energy in the detector. For example, a small peak at 191 keV was observed that could correspond to a known $\gamma$ decay \cite{PhysRevC.94.044328} and included in Table \ref{gamma_br}. However, as discussed below, no corresponding clear peak in the $^{86}$Kr($t$,$^{3}$He) spectrum was found and this peak could be due to a statistical fluctuation. Other similarly weak peaks that were difficult to separate from background, and for which no known $\gamma$ lines exist, were excluded from further analysis and not included in Table \ref{gamma_br} if no clear peak  in the $^{86}$Kr($t$,$^{3}$He) spectrum was found.

\begin{table}[!ht]
\caption{Overview of $\gamma$ rays (first column) identified as being emitted by excited states in $^{86}$Br produced in the $^{86}$Kr($t$,$^3$He) reaction. The second column contains spin-parity information, if available. The third column lists particular excited states in $^{86}$Br that are associated with the observed $\gamma$ rays, if identified. The fourth column denotes the evaluation of the angular-momentum transfer associated with these excitations, as determined from the MDA and detailed in the text.}
\label{gamma_br}
\centering
\begin{ruledtabular}
\begin{tabular}{c c c c}
$E_\gamma$  & $J^{\pi}$  & $E_x$ \footnote{The uncertainty is approximately 0.3 MeV} & Transition \\
 (keV)      & from \cite{NEGRET20151,PhysRevC.94.044328} & (MeV) & (tentative) \\
\hline
77  & 4$^-$ & $\sim$0.2 & $\Delta L \geq 1$ \\
191\footnote{The signal for this peak was weak compared to background in the present data and the observation is uncertain (see text)} & 4$^-$ & - & -  \\
207  & 2$^-$ & peak 1: $\sim$3.1 & $\Delta L \geq 1$ \\
     &             & peak 2: $\sim$3.6 & $\Delta L \geq 1$ \\
382  & 2$^-$ & $\sim$2.6 & $\Delta L = 1$ \\
932  & unknown & peak 1: $\sim$0.9 & $\Delta L \geq 2$ \\
     &         & peak 2: $\sim$2.3 & $\Delta L = 1$ \\
942  & unknown &  $\sim$1.7 & $\Delta L \geq 1$ \\
1427 & unknown & $\sim$2.3 & $\Delta L = 1$ \\
1753  & unknown & $\sim$1.7 & $\Delta L = 0,2$ \\
2361  & unknown\footnote{\textcolor{black}{In Ref. \cite{PhysRevC.94.044328}, a $\gamma$ line was reported at 2362 keV.  However, it was associated with the decay from a state at 2797 keV, which is unlikely to be consistent with the observation of a peak at 2.4 MeV in the $^{86}$Kr($t$,$^{3}$He) spectrum. On the basis of this information only, one cannot rule out that both states contribute to the excitation-energy  and $\gamma$ spectra. However, several other $\gamma$ lines associated with the decay from the state at 2797 keV observed in Ref. \cite{PhysRevC.94.044328} are not seen here, but should have been if the 2361 keV line belonged to the decay of the state at 2797 keV. We conclude that the 2361-keV line observed here is distinct from the 2362-keV line of Ref.~\cite{PhysRevC.94.044328}.}}   & $\sim$2.4 & $\Delta L = 0,2$ \\
\end{tabular}
\end{ruledtabular}
\end{table}

\begin{figure*}[htp]
\begin{center}
\includegraphics[width=0.49\linewidth]{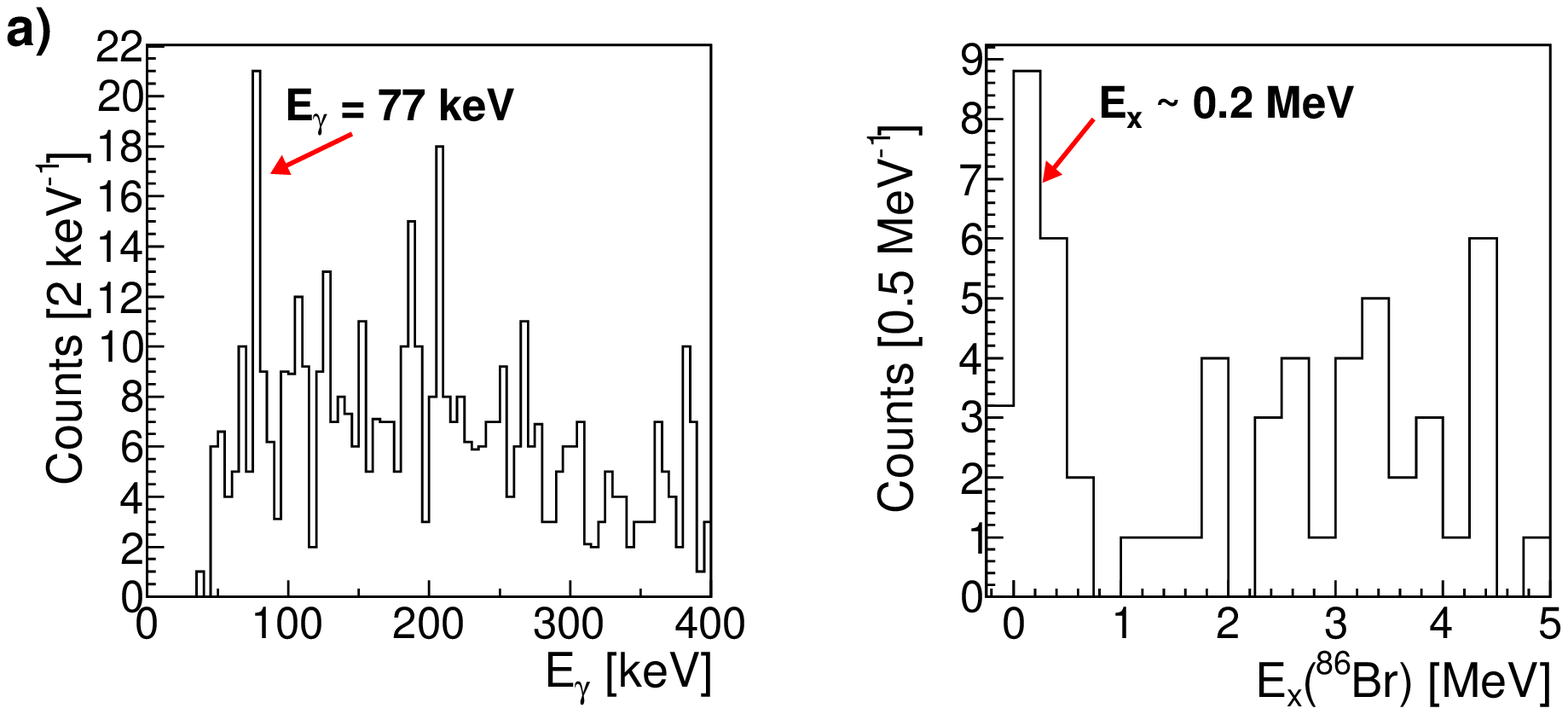}
\includegraphics[width=0.49\linewidth]{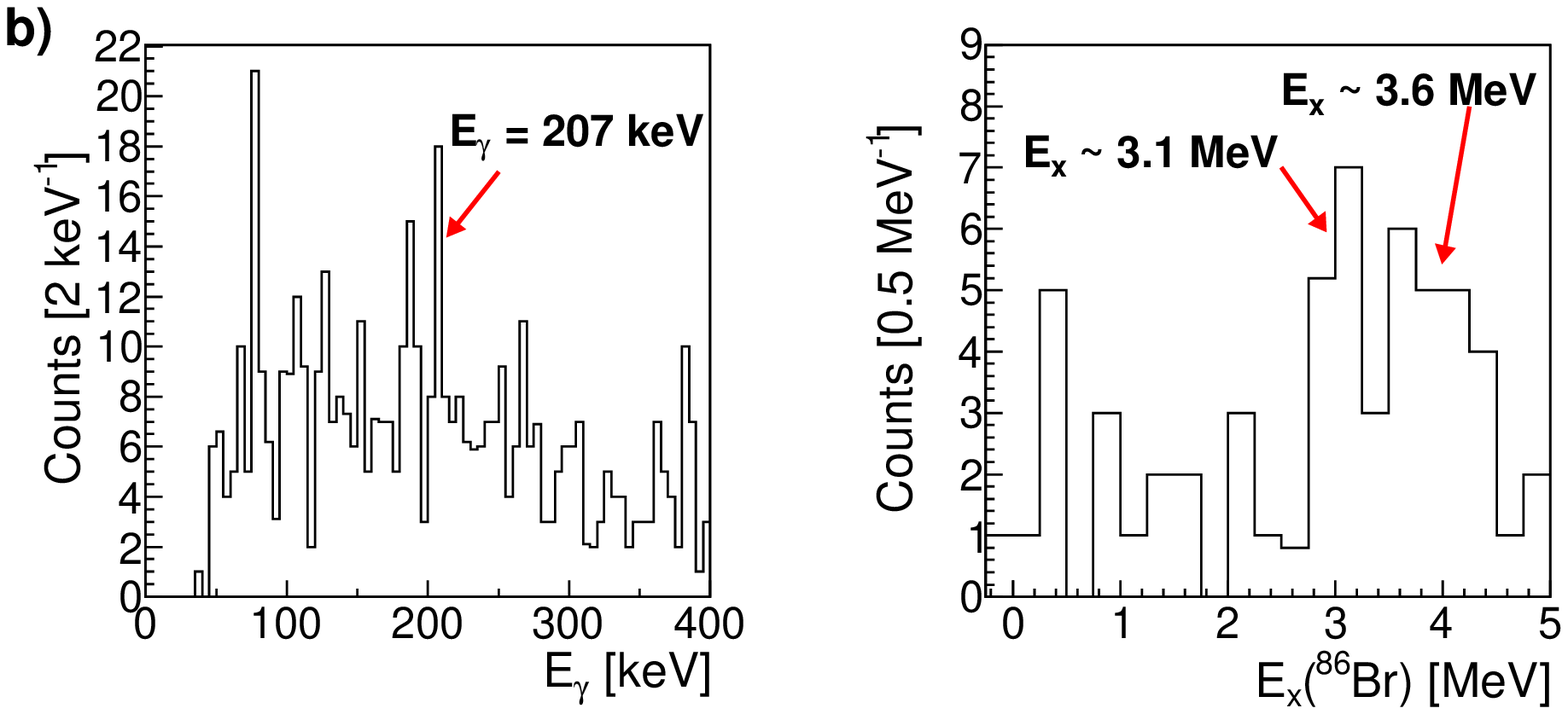}
\includegraphics[width=0.49\linewidth]{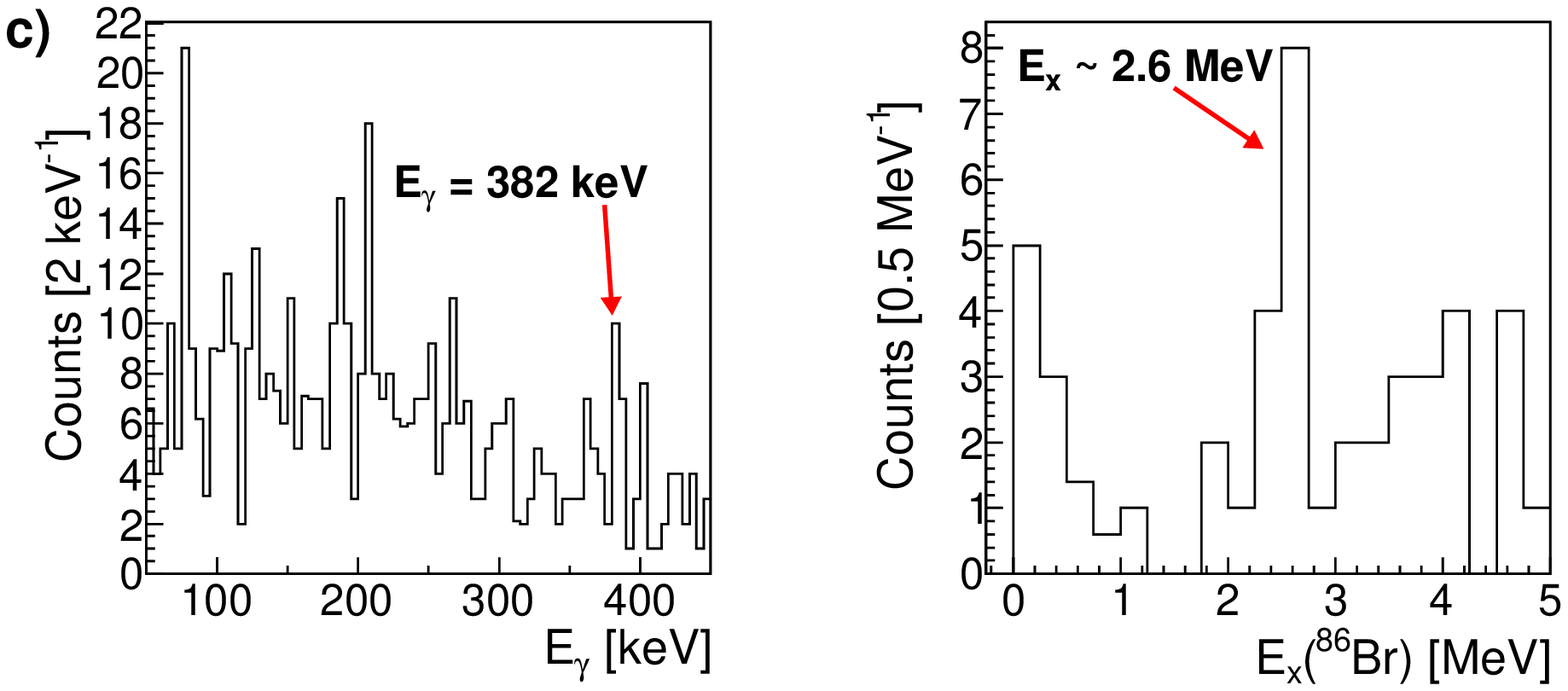}
\includegraphics[width=0.49\linewidth]{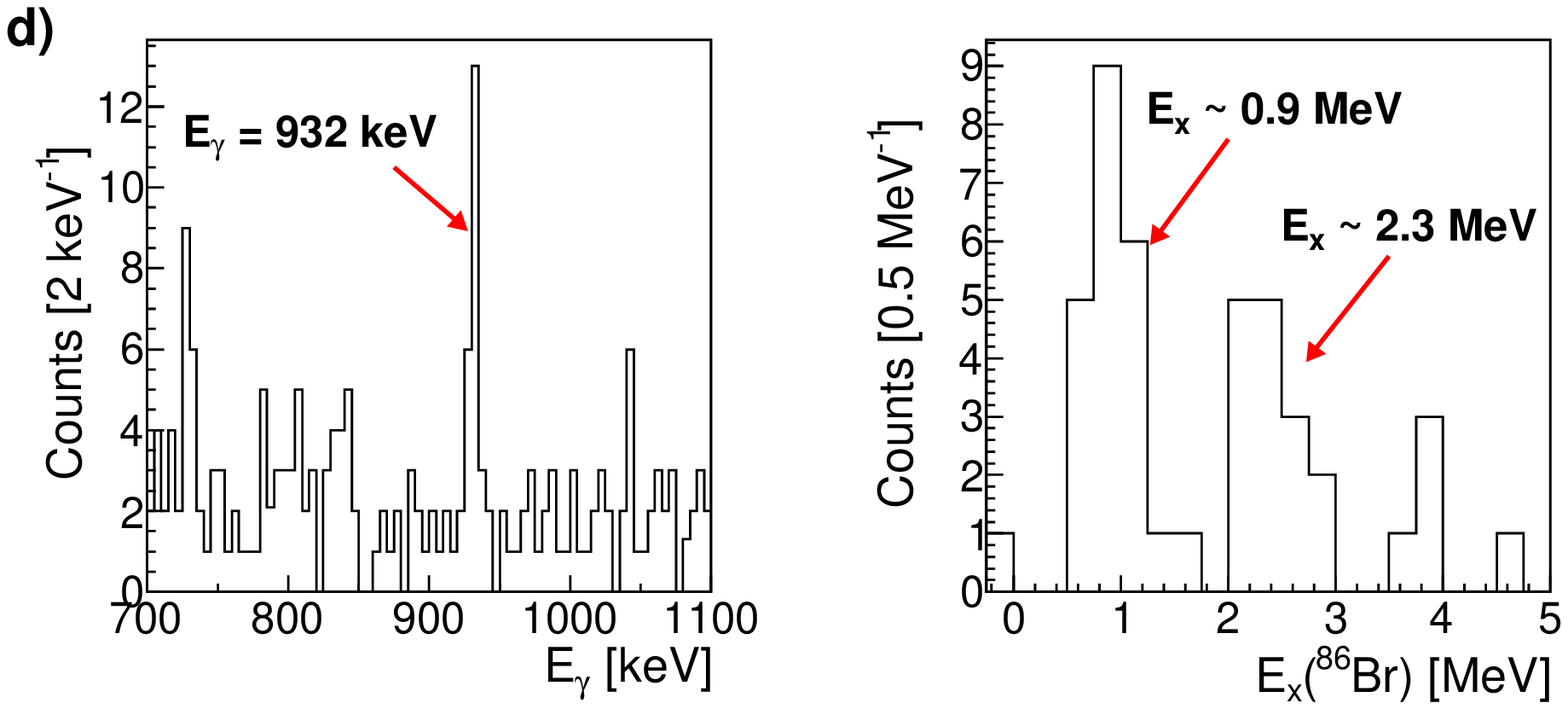}
\includegraphics[width=0.49\linewidth]{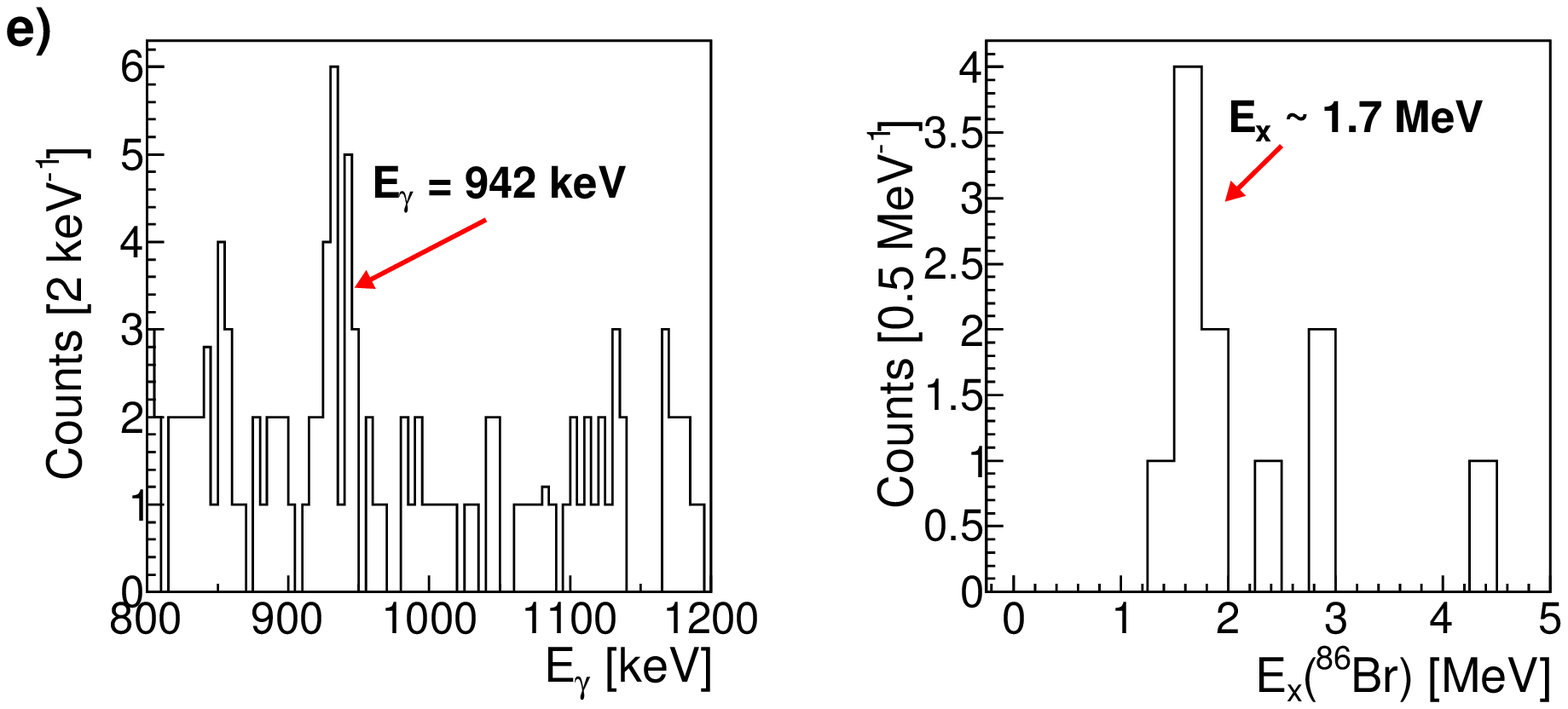}
\includegraphics[width=0.49\linewidth]{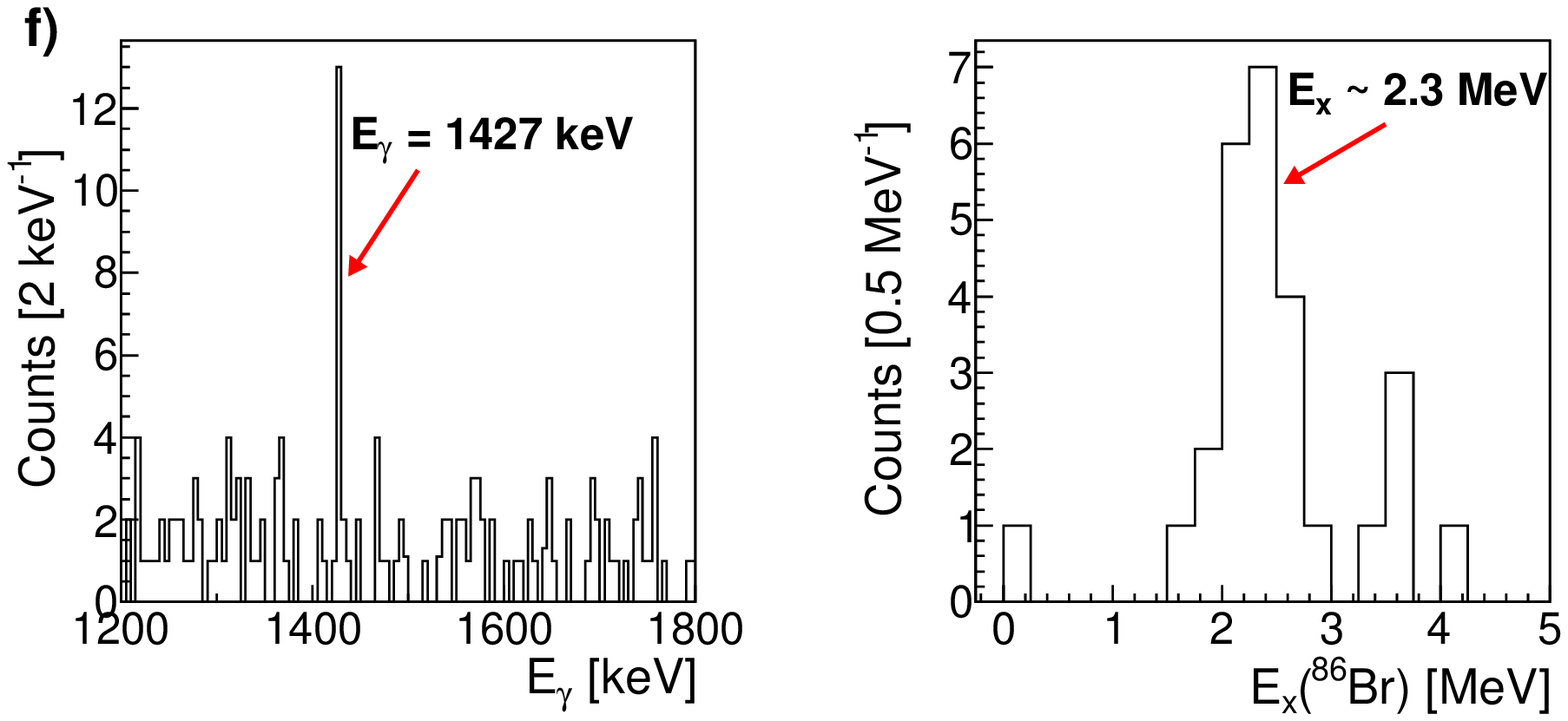}
\includegraphics[width=0.49\linewidth]{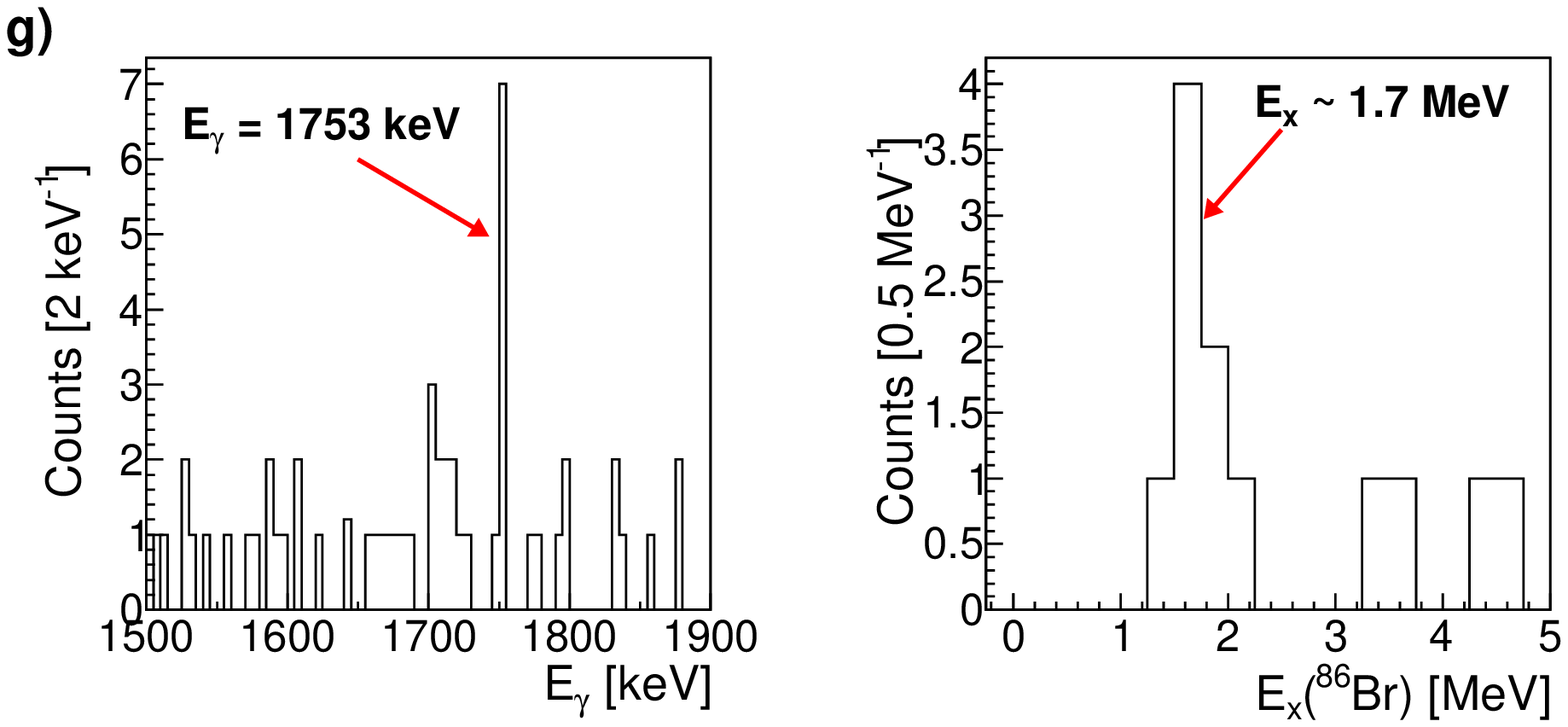}
\includegraphics[width=0.49\linewidth]{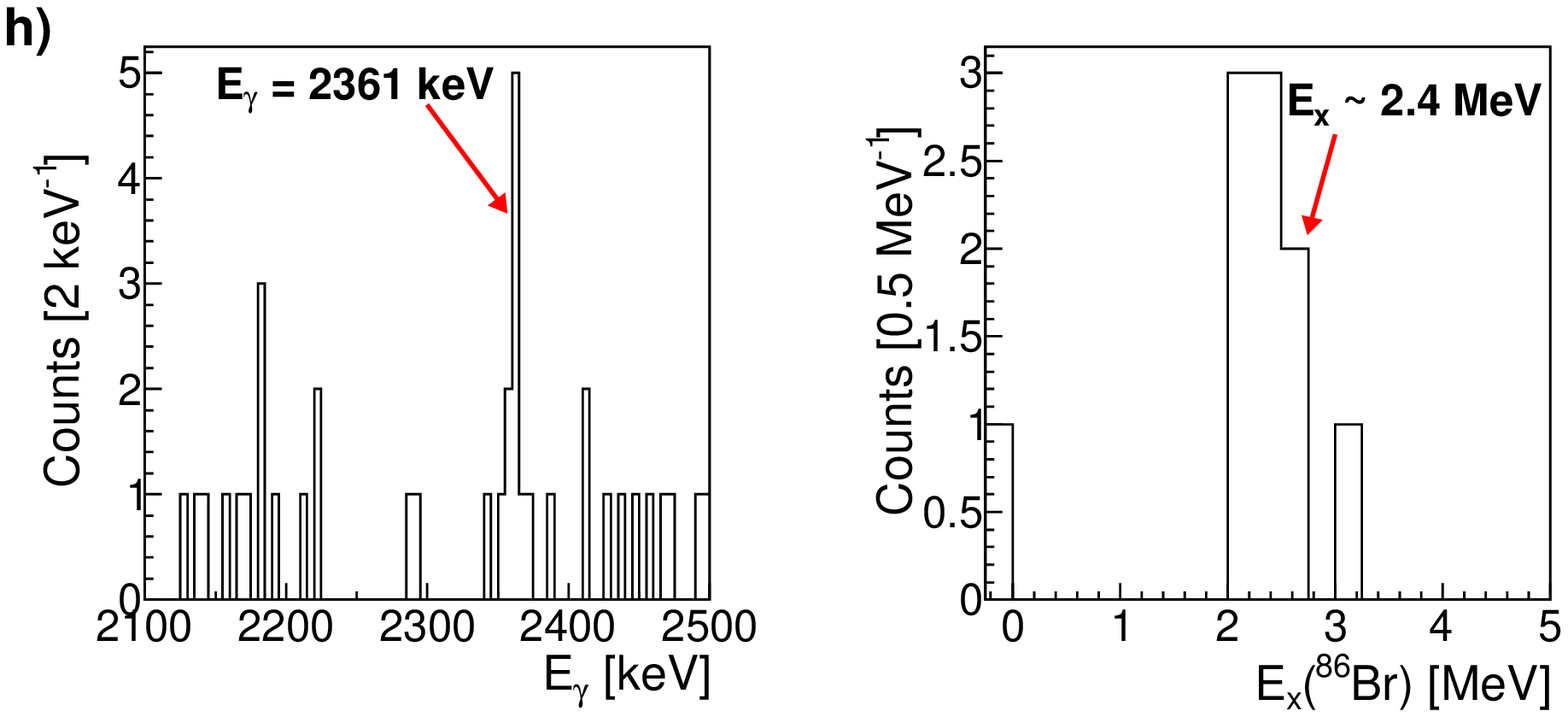}
\caption{(a-h) Left panels: $\gamma$-rays identified as being emitted from $^{86}$Br, rather than from reactions in the target-window foils, are indicated by red arrow. A gate with a width of $\sim$5 keV was placed around each of the $\gamma$-ray peaks. Right panels: $^{86}$Kr($t$,$^3$He) excitation energy spectra gated on $\gamma$ lines emitted by $^{86}$Br as identified in the panel on the left. The arrows in the excitation energy spectra are identified as excitations of specific states in $^{86}$Br, as listed in Table \ref{gamma_br}. A gate with a width of $\sim$1.5 MeV was placed on each of the excitation-energy peaks. The uncertainty in the excitation energy is approximately 0.3 MeV.}
\label{fig:gamma_ex}
\end{center}
\end{figure*}

By placing a 5 keV gate around each of the $^{86}$Br $\gamma$ lines, and examining the $^{86}$Br excitation-energy spectrum, it was possible to identify specific states in $^{86}$Br that produced the $\gamma$ ray. The $E_{\gamma}$ spectra for each $\gamma$ line identified to come from $^{86}$Br, and the associated $^{86}$Kr($t$,$^3$He) excitation-energy spectra gated on that $\gamma$ line, are shown in Fig.~\ref{fig:gamma_ex}.  The peaks in the $^{86}$Br excitation-energy spectra thus identified for each $\gamma$ ray are listed in the third column of Table \ref{gamma_br}. If the peak in the excitation-energy spectrum appears at (approximately) the same energy as the $\gamma$ ray, it is indicative of a decay directly to the ground state, or to a low-lying state in $^{86}$Br. Conversely, for the cases in which the peak in the excitation-energy spectrum is greater than the $\gamma$ energy, the decay must have proceeded through a decay chain.

Aside from the possible 191-keV $\gamma$ line discussed above, three other known lines from the decay of $^{86}$Br were identified in the $\gamma$-energy spectrum below 500 keV, at 77 keV, 207 keV, and 382 keV. By gating on these lines, the excitation-energy spectra in $^{86}$Br, determined from the ($t$,$^3$He) data were created, as shown on the right-hand side in Figs. \ref{fig:gamma_ex}(a), \ref{fig:gamma_ex}(b) and \ref{fig:gamma_ex}(c). Although the statistics are limited, peaks in these spectra can be observed that belong to excitations whose decay is associated with the corresponding $\gamma$ lines. For example, in the excitation-energy spectrum gated on the 77-keV $\gamma$ line, a low-lying peak at $E_x$($^{86}$Br) $\approx0.2$ MeV is found (indicated by the arrow), likely due to the de-excitation of a low-lying state. In addition, some distributed events at higher excitation energy are found, either due to higher-lying excitations that decay through the 77-keV $\gamma$ line, or that are associated with background from Compton scattering of more energetic $\gamma$ lines in GRETINA that appear under the 77-keV $\gamma$ line, as discussed above.

The same procedure was followed for all $\gamma$ lines that were determined to come from $^{86}$Br, as shown in Figs. \ref{fig:gamma_ex}(a-h). For $\gamma$ lines above $\sim500$ keV, the excitation-energy spectra obtained are relatively clean, as the Compton background from more energetic $\gamma$ lines is small. The extracted peaks are listed, per $\gamma$ line, in the third column of Table \ref{gamma_br}, and indicated with arrows in the figures. The uncertainty in these excitation energies is approximately 0.3 MeV. Subsequently, a gate with a width of 1.5 MeV was made around each of the peaks marked by arrows in Fig.~\ref{fig:gamma_ex} and listed in Table \ref{gamma_br}, and the differential cross sections were extracted and an MDA was performed. In this analysis, each peak was assumed to have a particular spin-assignment, i.e. to be due to the excitation of a single state. Hence, each peak was associated with a combination of $\Delta L=0$ and $\Delta L=2$ components (for a 1$^+$ state), a $\Delta L=1$ component (for 0$^-$, 1$^-$ or 2$^-$ excitations), or $\Delta L=2$ or higher, for quadrupole and higher multipole excitations. However, due to the limited statistics, it was only possible to make tentative assignments.

\begin{figure}[htp]
\begin{center}
\includegraphics[width=\linewidth]{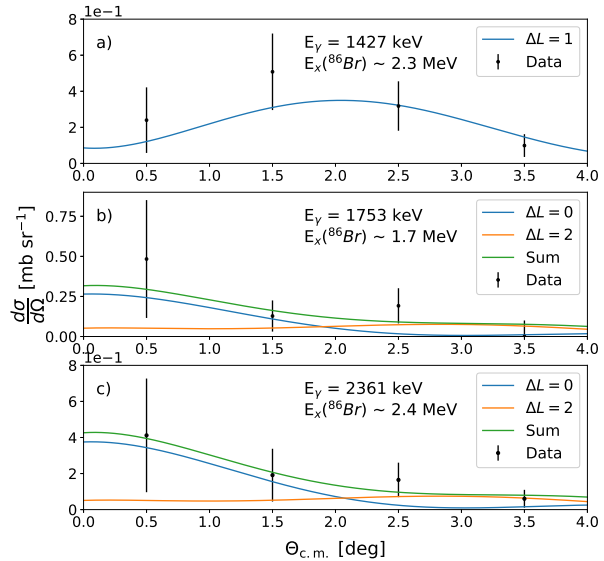}
\caption{MDA for excited states at $\sim$2.3 MeV (associated with the decay by a 1427-keV $\gamma$ ray), at $\sim$1.7 MeV (associated with the decay by a 1753-keV $\gamma$ ray), and at $\sim$2.4 MeV (associated with the decay by a 2361-keV $\gamma$ ray) in panels (a), (b) and (c), respectively. The angular distribution in panel (a) peaks at finite angles and is well described with a dipole angular distribution. The states at $\sim$1.7 MeV (panel (b)) and $\sim$2.4 MeV (panel (c)) are potentially associated with Gamow-Teller transitions as they peak at forward scattering angles. For both of these, a fit with a linear combination of $\Delta L=0$ and $\Delta L=2$ angular distributions is shown. The $\Delta L=0$ component was used to extract a tentative Gamow-Teller strength associated with each transition.}
\label{fig:gamma_mda}
\end{center}
\end{figure}

Three examples of this MDA are shown in Fig.~\ref{fig:gamma_mda}, and the extracted angular momentum transfers determined for each peak are listed in the fourth column of Table \ref{gamma_br}. In Fig.~\ref{fig:gamma_mda}a, the angular distribution associated with the $\gamma$ line of 1427 keV and $E_x$($^{86}$Br) $\approx$ 2.3 MeV is shown. This angular distribution is best described by a $\Delta L=1$ transition. For two peaks, at $E_x$($^{86}$Br) $\approx$ 1.7 MeV (associated with a 1753 keV $\gamma$ line) and at $E_x$($^{86}$Br) $\approx$ 2.4 MeV (associated with a 2361 keV $\gamma$ line), the differential cross section peaked at forward scattering angles, as shown in Figs. \ref{fig:gamma_mda}(b) and \ref{fig:gamma_mda}(c), respectively. Although the statistical uncertainties are large, the associated transitions could be due to Gamow-Teller excitations. Following the procedure from Sec.~\ref{sec:singles}, and accounting for the $\gamma$ detection efficiency, the Gamow-Teller strengths associated with the 1.7 MeV and 2.4 MeV states were extracted: 0.045 $^{+0.043}_{-0.045}$ for the 1.7-MeV state and 0.063 $^{+0.046}_{-0.063}$ for the 2.4-MeV state. Since it cannot be ruled out that these transitions are not due to Gamow-Teller excitations, the lower error bars extend to $B(\textrm{GT})=0$. However, the results from this $\gamma$-ray analysis (also shown in Fig.~\ref{fig:bgt}) provide more detailed constraints on the Gamow-Teller strength from $^{86}$Kr to $^{86}$Br for excitation energies below 5 MeV. Finally, since the summed energy of the 932 keV and 1427 keV lines (2359 keV) is close to 2361 keV (given the energy resolution of GRETINA of 2.2 keV for $E_{\gamma}=1332$ keV \cite{PASCHALIS201344}), and all three $\gamma$ lines are associated with an excited state at about 2.3 MeV, we cannot rule out that these come from the same excited state. If true, it is unlikely that this state can be assigned to a spin-parity of $1^{+}$, as the center-of-mass $^{3}$He angular distribution associated with the 1427 keV line is quite clearly not of monopole character. Given the statistical limitations, it was not possible to perform a $\gamma$-$\gamma$ correlation analysis to study this possibility in more detail.

\subsection{\label{theory_comparison}Comparison to Theory}
It is informative to compare the extracted experimental results with theoretical models that could be used to calculate Gamow-Teller strengths for the purpose of estimating electron-capture rates for astrophysical simulations. Therefore, as part of this work, two calculations were performed. The first calculation is a shell-model calculation, which was performed with the {\footnotesize{NUSHELLX}} code \cite{nushellx_msu}, using the {\footnotesize{jj44pna}} effective interaction \cite{PhysRevC.70.044314} in the SNE model space \cite{PhysRevC.65.021302,PhysRevC.50.R2270}, and a renormalized G-matrix using the charge-dependent (CD-Bonn) nucleon-nucleon interaction \cite{PhysRevLett.91.162503}. This model space and interaction assume an inert $^{78}$Ni core, on top of which protons can populate the $0f_{5/2}$, $1p_{3/2}$, $1p_{1/2}$, and $0g_{9/2}$ orbitals, and neutrons can populate the $0g_{7/2}$, $1d_{5/2}$, $1d_{3/2}$, $2s_{1/2}$  and $0h_{11/2}$ orbitals. Because of the restrictions of the model space, Gamow-Teller excitations must be associated with $\pi \textrm{0}g_{9/2}\rightarrow$ $\nu \textrm{0}g_{7/2}$ transitions.

For this simple model space, the total Gamow-Teller strength can be estimated \cite{1985NuPhA.444..402T} as $B(\textrm{GT}^+) = \langle \pi$0$g_{9/2} \rangle B_{sp}$, where $B_{sp}$ is the single-particle strength for the $\pi \textrm{0}g_{9/2}\rightarrow$ $\nu \textrm{0}g_{7/2}$ transition, and $\langle \pi$0$g_{9/2} \rangle$ is the occupation number for protons in the 0$g_{9/2}$ orbital. The calculated occupation number for $^{86}$Kr is 0.40, resulting in a total Gamow-Teller strength of 0.71, of which 0.16 was estimated to reside at $E_x$($^{86}$Br) $<5$ MeV, and the remainder distributed over many weak transitions up to high excitation energies. The occupation number determined from experiment for $^{86}$Kr is uncertain; Ref. \cite{PFEIFFER1986381} does not report any $\Delta L=4$ strength in the $^{86}$Kr to $^{85}$Br reaction, while Ref. \cite{PhysRevC.5.117} reports tentative $\Delta L=4$ strength in a state at $E_x=2.31$ MeV with a deduced occupation of 1.12. We note that proton 0$g_{9/2}$ occupation numbers for $^{90}$Zr and $^{88}$Sr of 1.0 and 0.7 were reported, respectively \cite{PFEIFFER1986381}, suggesting the calculated value of 0.40 in the shell model for $^{86}$Kr is not unreasonable.

It is necessary to take into account the consequences of model-space truncation; in Ref. \cite{1985NuPhA.444..402T}, this is divided into two parts that are expressed in terms of hindrance factors $h_{\textrm{high}}$ and $h_{\textrm{c.p.}}$. $h_{\textrm{high}}$ is associated with configurations beyond the ($0g$, $1d$, $2s$) model space, which corresponds to the admixtures of two-particle two-hole states with unperturbed energies of $2 \hbar \omega$ and higher in the oscillator basis. Such behavior has been extensively studied in lighter nuclei; for the ($0d$, $1s$) model space, the empirical value of $h_{\textrm{high}}$ is 1.67 \cite{BROWN1985347}, which qualitatively agrees with calculations that include the $2 \hbar \omega$ admixture. For the ($0f$, $1p$) model space, the empirical value for $h_{\textrm{high}}$ is 1.81(1) \cite{PhysRevC.53.R2602}. The latter is assumed here, because it is consistent with the value observed for heavier nuclei \cite{GAA85}. The factor $h_{\textrm{c.p.}}$ (where c.p. stands for core polarization) is introduced for the truncation from ($0g$, $1d$, $2s$) to the model space used in the calculations for this work. In particular, the $\nu$0$g_{9/2}$ orbital is assumed to be filled and the $\pi$0$g_{9/2}$ is assumed to be empty. $h_{\textrm{c.p.}}$ takes into account the mixing between the 0$g_{9/2}$ and 0$g_{7/2}$ spin-orbit partners, and in Ref. \cite{1985NuPhA.444..402T}, the calculated associated hindrance factor for the ($\pi$0$g_{9/2}$)$^n$ configurations. With these hindrance factors, $h_{\textrm{high}}$ and $h_{\textrm{c.p.}}$, the Gamow-Teller strength can be written in the following form:
\begin{equation} \label{quenched_bgt}
B'(\mathrm{GT}) = \frac{B(\mathrm{GT}^+)}{h_{\textrm{c.p.}}h_{\textrm{high}}}.
\end{equation}
The hindrances observed in the $\beta ^+$ decay of nuclei from $^{94}$Ru to $^{98}$Cd are consistent with the calculations in Ref. \cite{1985NuPhA.444..402T}, which are based on Eq.~(\ref{quenched_bgt}) \cite{PhysRevC.50.R2270}. In this work, $h_{\textrm{c.p.}} = 5.0$, from the result of the ($\pi$0$g_{9/2}$)$^2$ configuration in Table 5 of Ref. \cite{1985NuPhA.444..402T}. It assumed that this can be applied to the calculations of $^{86}$Kr. Hence, the total hindrance factor $h = h_{\textrm{c.p.}}h_{\textrm{high}} \approx 9$, which was applied to the shell-model calculations.

The second theoretical calculation is a quasiparticle random-phase approximation (QRPA) calculation. It was performed using the axially-deformed Skyrme Finite Amplitude Method (FAM) \cite{PhysRevC.84.014314,PhysRevC.90.024308}. This method has recently been extended to odd-$A$ nuclei in the equal filling approximation \cite{PhysRevC.94.055802}, making it a candidate for calculating Gamow-Teller strengths and electron-capture rates for all nuclei in the high-sensitivity region. The Skyrme functional and single-particle space are the same as those used in Ref. \cite{PhysRevC.93.014304}, in which a single set of parameters, including an effective axial-coupling constant, $g_A$, of 1.0, were fixed. The width of the states in the QRPA calculation was set to 0.25 MeV.

For both of the theoretical models, the first peak in the strength distribution was placed at the excitation energy of the first known $1^{+}$ state in $^{86}$Br, 2.446 MeV. The results for both sets of calculations are included in Fig.~\ref{fig:bgt}. Both the shell-model and QRPA calculations yield comparable amounts of Gamow-Teller strength up to $E_x$($^{86}$Br) $=5$ MeV: 0.035 for the shell-model calculation and 0.024 for the QRPA calculation. The total strength obtained from the $^{86}$Kr($t$,$^3$He+$\gamma$) coincidence data is 0.108$^{+0.063}_{-0.108}$. Although the experimental error bars are large, these data set an upper limit on the summed Gamow-Teller strength, with which the theory is consistent.

\section{\label{sec:ecrates} Electron-Capture Rates}
\subsection{\label{sec:calculations} Calculation of Electron-Capture Rates}
Using the Gamow-Teller strength distributions calculated in Secs.~\ref{sec:singles} and \ref{sec:gammas}, electron-capture rates were calculated for a wide range of stellar densities and temperatures of relevance for astrophysical phenomena. These calculations were performed using the {\footnotesize{ECRATES}} code \cite{ecrates_reyes,0067-0049-126-2-501,ValdezTHESIS}, which takes the reaction $Q$ values and the Gamow-Teller strengths as inputs.

\begin{figure}[htp]
\begin{center}
\includegraphics[width=\linewidth]{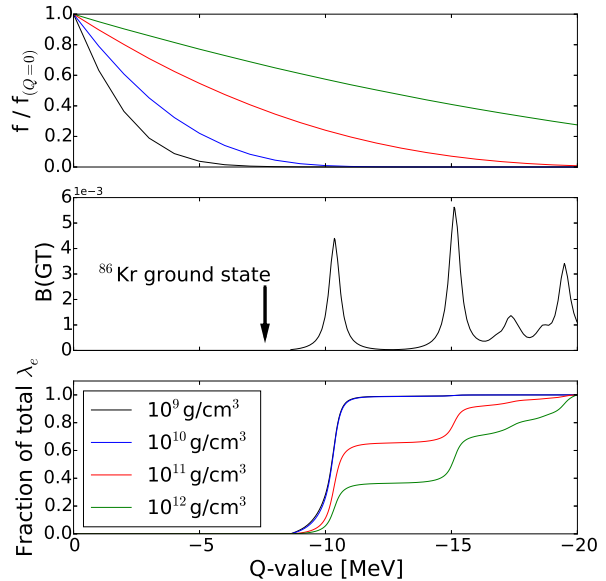}
\caption{Top: the phase-space factor used in the calculation of the electron-capture rates as a function of electron-capture $Q$ value for stellar densities ranging from $10^{9}-10^{12}$ g/cm$^{3}$. Note that the factors are normalized to unity at $Q=0$. Center: Gamow-Teller strength distribution for $^{86}$Kr as calculated in QRPA. Bottom: Summed fractions of the total electron-capture rate as a function of $Q$, obtained by multiplying the phase space factors in the top panel with the strength distribution of the center panel.}
\label{fig:kr_ecrate}
\end{center}
\end{figure}

Electron-capture rates are calculated using the following expression:
\begin{equation}
\lambda _{EC} = \mathrm{ln}(2) \sum \limits_{ij} f_{ij}(T,\rho, U_F) B(\mathrm{GT})_{ij}.
\end{equation}
$B(\textrm{GT})$ is the Gamow-Teller strength distribution, derived either from experimental data or theoretical calculations, and $f(T, \rho, U_F)$ is the phase-space factor, which depends on the stellar density, $\rho$, temperature, $T$, and chemical potential, $U_F$. It is informative to examine these two components of the electron-capture rates. The Gamow-Teller strength distribution for $^{86}$Kr as a function of $Q$ value, as obtained from the QRPA calculation is shown in the middle panel of Fig.~\ref{fig:kr_ecrate}. \textcolor{black}{The first peak in the Gamow-Teller distribution as a function of $Q$ value corresponds to the peak of the QRPA calculation shown in the Gamow-Teller strength distribution of  Fig.~\ref{fig:bgt} as a function of excitation energy}. The phase-space factor, normalized to unity at $Q = 0$, is shown in the top panel of Fig.~\ref{fig:kr_ecrate} as a function of $Q$ value, for a range of stellar densities from 10$^9$ to 10$^{12}$ g/cm$^3$. By multiplying the phase-space factor by the Gamow-Teller strength and then summing over the entire distribution, the total electron-capture rate for the nucleus is produced.  The bottom panel of Fig.~\ref{fig:kr_ecrate} shows the summed fraction to the total electron-capture rate as a function of $Q$ value.

At relatively low stellar densities, the phase-space factor drops quickly with decreasing $Q$, with the consequence that Gamow-Teller transitions to states at less negative $Q$ dominate the electron-capture rate. As the density increases, the phase-space factor drops off more slowly with decreasing $Q$, and Gamow-Teller transitions to states at more negative $Q$ start to contribute to the electron-capture rate, as becomes clear from the bottom panel of Fig.~\ref{fig:kr_ecrate}. Even at densities in excess of  $10^{11}$ g/cm$^{3}$, the contribution from the transition to the lowest state is still the strongest single contribution to the rate. This is because the threshold electron-capture $Q$ value for the case of $^{86}$Kr is rather negative (-7.607 MeV), and the first Gamow-Teller transition only appears at $Q\approx -10$ MeV. The situation for $^{86}$Kr described in Fig.~\ref{fig:kr_ecrate} is exemplary for the neutron-rich nuclei in the $N=50$ region. Due to the relatively large, negative $Q$-values for electron-capture on these nuclei, the details of the Gamow-Teller strength distributions, including the location of the lowest-lying 1$^+$ state, are important for estimating accurate electron-capture rates, even at relatively high densities.

\begin{figure}[htp]
\begin{center}
\includegraphics[width=\linewidth]{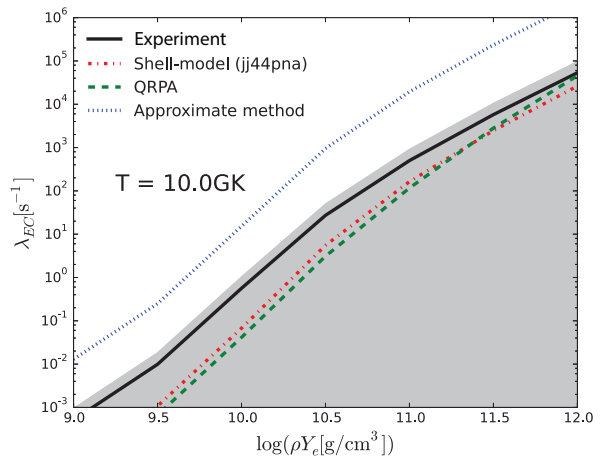}
\caption{A comparison of the experimentally determined electron-capture rates on $^{86}$Kr, at a temperatures of 10 GK over the range of stellar densities relevant during deleptonization in the collapse phase of core-collapse supernovae, as derived from experimental data. The results are compared with rates derived from the shell-model and QRPA calculations detailed in the text, as well as the single-state approximation from Eq.~(\ref{eqn:approx}).}
\label{fig:ecrates}
\end{center}
\end{figure}

The results of the electron-capture rate calculations for this work are shown in Fig.~\ref{fig:ecrates}, at a temperature of 10 GK, and for densities of relevance for the collapse phase of core-collapse supernovae. The black solid line represents the electron-capture rates that are calculated from the Gamow-Teller strength extracted from the $\gamma$-ray analysis in Sec.~\ref{sec:gammas}. Since this rate represents an upper limit for $E_x <$ 5 MeV, the uncertainty band (in grey) extends down to zero. It is important to note that, because it was only possible to extract Gamow-Teller strength up to $E_{x}=5$ MeV, transitions to states at higher excitation energies are not included in the electron-capture rate calculations derived from the data. However, as explained above, the contributions from these states to the overall rate are expected to be relatively small at the lower end of the density scale presented here and slowly increase at higher excitation energies. Also shown in Fig.~\ref{fig:ecrates} are the rates determined from the theoretical strength distributions described above, and the single-state approximation presently implemented in NuLib \cite{fuller85,PhysRevLett.90.241102,PhysRevC.95.025805}. For the latter, $\Delta E$ was 2.5 MeV for the case of $^{86}$Kr.

The electron-capture rates derived from the shell-model and QRPA calculations are consistent with the experimental result, as they both fall within the experimental uncertainties. Conversely, the rates obtained by the single-state approximation are much higher, exceeding the electron-capture rates estimated based on the data by about two orders of magnitude. At high stellar temperatures, Pauli unblocking effects will increase the electron-capture rates \cite{PhysRevC.63.032801}, but in cases such as $^{86}$Kr, where Pauli blocking is not complete at zero temperature, such increases are likely small \cite{PhysRevC.63.032801}. The placement of a single state at one fixed excitation energy of 2.5 MeV with a Gamow-Teller strength of 4.6 is inconsistent with the present data. \textcolor{black}{If a single-state approximation were to be used to represent the present experimental results, a Gamow-Teller strength of less than 0.03 or an excitation energy in excess of 20 MeV would be required}.
Microscopic models are needed to more accurately estimate electron-capture rates for astrophysical simulations. These models can be tested at zero temperature against available experimental data. We note that a similar conclusion was drawn on the basis of a recent $^{88}$Sr($t$,$^{3}$He$+\gamma$) experiment \cite{juan2019}.

\subsection{ \label{sec:rate_table} New Rate Table}
Because of the importance of the region of nuclei surrounding the $N=50$ shell closure, a new electron-capture rate table was developed for the use in astrophysical simulations that contained, for 78 nuclei in and around the high-sensitivity region \cite{TITUS2018}, rates calculated on the basis of the QRPA framework described in Sec.~\ref{theory_comparison}. QRPA calculations were chosen over shell-model calculations in this case because calculations were needed for a large number of nuclei both above and below the $N=50$ shell closure. In addition, these QRPA calculations can be extended in the future to include temperature-dependent effects.

The nuclei included were $^{75-76}$Fe, $^{75-78}$Co, $^{75-80}$Ni, $^{75-82}$Cu, $^{75-84}$Zn, $^{75-85}$Ga, $^{76-85}$Ge, $^{75-85}$As, $^{80-85}$Se, $^{82-85}$Br, $^{84-86}$Kr, $^{88}$Sr, $^{90}$Zr, and $^{93}$Nb. The ground-state $Q$ value was obtained from experimental data where available, and from the Hartree-Fock-Bogoliubov solution according to the approximation in Ref.~\cite{PhysRevC.60.014302} for nuclei lacking experimental data. Additionally, the spin and parity of the ground states of the relevant nuclei were obtained from experimental assignments, and from the Gallagher and Moszkowski rule \cite{PhysRev.111.1282}, for nuclei lacking definite assignments. Although these calculations do not yet contain temperature-dependent effects that might increase the electron-capture rates, these simulations provide important insights in the maximum effects that can be caused by the overestimation of the electron-capture rates in the single-state approximation.

\section{\label{sec:simulations} Core-Collapse Supernova Simulations}
The core-collapse supernova simulations for this work were performed using the neutrino-transport code, NuLib \cite{OCO15}, and the general-relativistic, spherically-symmetric hydrodynamics code, GR1D \cite{OCO10,OCO15}. The progenitor used in the simulations was a well-known 15 solar-mass, solar-metallicity star (s15WW95) \cite{WOO95}. The SFHo equation of state and nuclear statistical equilibrium distributions were used \cite{0004-637X-774-1-17}. The goal of this study was to determine the effect of the newly calculated rate table on the late-stage evolution of the collapsing star. As such, two simulations were performed: the first was a base simulation in which the electron-capture rates were calculated based on the single-state approximation \cite{fuller85,PhysRevLett.90.241102,PhysRevC.95.025805}. The second simulation used the new rate table based on the QRPA calculations.

\begin{figure}[htp]
\begin{center}
\includegraphics[trim=0.3cm 0.5cm 0cm 0cm,clip,width=\linewidth]{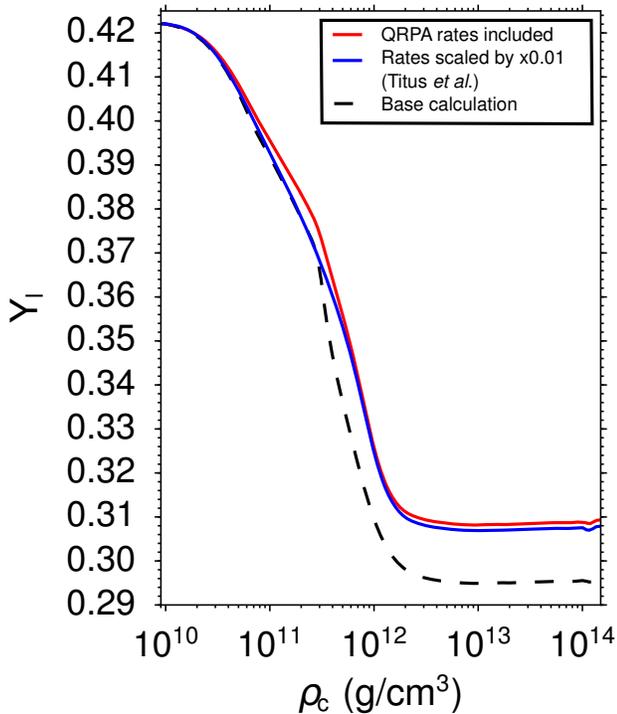}
\caption{Lepton fraction as a function of core density. The results of the original simulation, without the addition of the new rate table, are compared to the case in which the new QRPA rates are included. Also shown is a simulation from the previous sensitivity study \cite{TITUS2018}, in which the electron-capture rates for the high-sensitivity region nuclei are scaled by a factor of 0.01.}
\label{fig:ylep}
\end{center}
\end{figure}

A comparison of the lepton fraction ($Y_l$) as a function of central density ($\rho _c$) for these two cases is shown in Fig.~\ref{fig:ylep}. For densities of 10$^{10}$ to 10$^{12}$ g/cm$^3$, strong deleptonization of the matter in the core of the star occurs. The opacity of the matter is relatively low, which allows electron neutrinos to escape the star. At a density of 10$^{12}$ g/cm$^3$, the neutrinos become trapped, preventing further deleptonization and causing the lepton fraction to saturate.

Comparing the results of the base simulation with the simulation in which the new QRPA rates were included, it is clear that the new rates strongly affected the final lepton fraction. The original simulation attains a final lepton fraction of 0.294, while the simulation including the new rates reaches a final lepton fraction of 0.312. This constitutes a 14\% reduction in the decrease of the lepton fraction with the addition of the new, more accurate rates.  Also shown in Fig.~\ref{fig:ylep} are the results of a simulation from a previous sensitivity study \cite{TITUS2018}, which illustrates the case in which the electron-capture rates for the nuclei in the high-sensitivity region calculated with the single-state approximate method, were scaled by a factor of 0.01. This result is comparable to the simulation that uses the new QRPA rates. Because of this similarity, the consequences of using the QRPA-derived electron-capture rates in the high-sensitivity region are comparable to those discussed in detail in Ref. \cite{TITUS2018}, and will strongly impact physical observables, such as the peak neutrino luminosity \cite{Sull2016} and the frequency of gravitational waves emitted from the collapsing star \cite{PhysRevD.95.063019}, which are both potential multimessenger signals that can be used to better understand and model core-collapse supernovae.

\section{\label{sec:conclusions}Conclusions}
In an effort to constrain the electron-capture rates on nuclei near $N=50$ above $^{78}$Ni, which play an important role in the late-stage evolution of core-collpase supernovae, the $^{86}$Kr($t$,$^{3}$He$+\gamma$) reaction at 115 MeV$/u$ was investigated. Due to the need to subtract events from reactions on foils that maintained the $^{86}$Kr gas target, the uncertainties in the extracted strengths are larger than has been achieved in other ($t$,$^{3}$He) experiments at NSCL. Still, an upper limit for the Gamow-Teller strengths for $E_{x}<5$ MeV could be obtained, which was strengthened through the investigation of the $\gamma$-decay spectra. Theoretical Gamow-Teller strength calculations by using the shell model and QRPA were consistent with the experimental upper limit.

The stellar electron-capture rate on $^{86}$Kr derived from the upper limit of the extracted Gamow-Teller strength was used to test a single-state approximation used in astrophysical simulations, as well as electron-capture rates derived from the Gamow-Teller strengths calculated in the shell-model and QRPA. The electron-capture rates based on the single-state approximation are too high compared to the rates based on the experimental data, whereas those based on the microscopic calculations are consistent. Although at high stellar temperatures Pauli unblocking can increase the electron capture rates as compared to the rates extracted and calculated at zero temperature, the effect is likely significantly smaller than the difference between the experimental upper limit and the rate based on the single-state approximation, assuming that the protons partially fill the 0$g_{9/2}$ orbit and Gamow-Teller transitions are not completely Pauli blocked, even at zero temperature.

Based on these results, simulations of the late stages of the evolution of core-collapse supernovae were performed that utilized a new set of rates for nuclei in the high-sensitivity region near $N=50$ based on Gamow-Teller strengths from QRPA calculations tested in these experiments. The results indicate that the reduced electron-capture rates in the high-sensitivity region strongly affects the deleptonization during the collapse phase, with significant consequences for the late evolution and potentially observable neutrino and gravitation-wave multi-messenger signals, as discussed in detail in Ref. \cite{TITUS2018}. These simulations will benefit from further theoretical work to improve the electron-capture rates on nuclei of relevance for late stellar evolution, in particular by including temperature-dependent effects in the QRPA calculations. Finally, the development of techniques to measure the $\beta^{+}$ Gamow-Teller strengths on unstable nuclei will be important to test the theoretical calculations in the high-sensitivity region. Such developments are being pursued at NSCL through the use of the ($d$,$^{2}$He) and ($^{7}$Li,$^{7}$Be) reactions in inverse kinematics.

\section{\label{sec:acknowledgments}Acknowledgments}
We thank the NSCL staff for their support during the preparations for and conducting of the experiment. J.C.Z. thanks the support by Funda\c{c}$\tilde{a}$o de Amparo a Pesquisa do Estado de S$\tilde{a}$o Paulo (FAPESP) under Grant No. 2018/04965-4.

This work was supported by the US National Science Foundation (NSF) under Cooperative Agreement PHY-156554 (NSCL), PHY-1430152 (JINA Center for the Evolution of the Elements), and PHY-1811855. GRETINA was funded by the US Department of Energy, in the Office of Nuclear Physics of the Office of Science. Operation of the array at NSCL was supported by DOE under Grants No. DE-SC0014537 (NSCL) and No. DE-AC02-05CH11231 (LBNL).

\bibliographystyle{apsrev4-1}
\bibliography{kr_bib}

\begin{thebibliography}{80}%
\makeatletter
\providecommand \@ifxundefined [1]{%
 \@ifx{#1\undefined}
}%
\providecommand \@ifnum [1]{%
 \ifnum #1\expandafter \@firstoftwo
 \else \expandafter \@secondoftwo
 \fi
}%
\providecommand \@ifx [1]{%
 \ifx #1\expandafter \@firstoftwo
 \else \expandafter \@secondoftwo
 \fi
}%
\providecommand \natexlab [1]{#1}%
\providecommand \enquote  [1]{``#1''}%
\providecommand \bibnamefont  [1]{#1}%
\providecommand \bibfnamefont [1]{#1}%
\providecommand \citenamefont [1]{#1}%
\providecommand \href@noop [0]{\@secondoftwo}%
\providecommand \href [0]{\begingroup \@sanitize@url \@href}%
\providecommand \@href[1]{\@@startlink{#1}\@@href}%
\providecommand \@@href[1]{\endgroup#1\@@endlink}%
\providecommand \@sanitize@url [0]{\catcode `\\12\catcode `\$12\catcode
  `\&12\catcode `\#12\catcode `\^12\catcode `\_12\catcode `\%12\relax}%
\providecommand \@@startlink[1]{}%
\providecommand \@@endlink[0]{}%
\providecommand \url  [0]{\begingroup\@sanitize@url \@url }%
\providecommand \@url [1]{\endgroup\@href {#1}{\urlprefix }}%
\providecommand \urlprefix  [0]{URL }%
\providecommand \Eprint [0]{\href }%
\providecommand \doibase [0]{http://dx.doi.org/}%
\providecommand \selectlanguage [0]{\@gobble}%
\providecommand \bibinfo  [0]{\@secondoftwo}%
\providecommand \bibfield  [0]{\@secondoftwo}%
\providecommand \translation [1]{[#1]}%
\providecommand \BibitemOpen [0]{}%
\providecommand \bibitemStop [0]{}%
\providecommand \bibitemNoStop [0]{.\EOS\space}%
\providecommand \EOS [0]{\spacefactor3000\relax}%
\providecommand \BibitemShut  [1]{\csname bibitem#1\endcsname}%
\let\auto@bib@innerbib\@empty
\bibitem [{\citenamefont {Fryer}(1999)}]{Fryer1999}%
  \BibitemOpen
  \bibfield  {author} {\bibinfo {author} {\bibfnamefont {C.~L.}\ \bibnamefont
  {Fryer}},\ }\href {\doibase 10.1086/307647} {\bibfield  {journal} {\bibinfo
  {journal} {Astrophys. J.}\ }\textbf {\bibinfo {volume} {522}},\ \bibinfo
  {pages} {413} (\bibinfo {year} {1999})}\BibitemShut {NoStop}%
\bibitem [{\citenamefont {Heger}\ \emph {et~al.}(2003)\citenamefont {Heger},
  \citenamefont {Fryer}, \citenamefont {Woosley}, \citenamefont {Langer},\ and\
  \citenamefont {Hartmann}}]{Heger2003}%
  \BibitemOpen
  \bibfield  {author} {\bibinfo {author} {\bibfnamefont {A.}~\bibnamefont
  {Heger}}, \bibinfo {author} {\bibfnamefont {C.~L.}\ \bibnamefont {Fryer}},
  \bibinfo {author} {\bibfnamefont {S.~E.}\ \bibnamefont {Woosley}}, \bibinfo
  {author} {\bibfnamefont {N.}~\bibnamefont {Langer}}, \ and\ \bibinfo {author}
  {\bibfnamefont {D.~H.}\ \bibnamefont {Hartmann}},\ }\href {\doibase
  10.1086/375341} {\bibfield  {journal} {\bibinfo  {journal} {Astrophys. J.}\
  }\textbf {\bibinfo {volume} {591}},\ \bibinfo {pages} {288} (\bibinfo {year}
  {2003})}\BibitemShut {NoStop}%
\bibitem [{\citenamefont {Janka}\ \emph {et~al.}(2007)\citenamefont {Janka},
  \citenamefont {Langanke}, \citenamefont {Marek}, \citenamefont
  {Mart{\'i}nez-Pinedo},\ and\ \citenamefont {M{\"u}ller}}]{janka07}%
  \BibitemOpen
  \bibfield  {author} {\bibinfo {author} {\bibfnamefont {H.-T.}\ \bibnamefont
  {Janka}}, \bibinfo {author} {\bibfnamefont {K.}~\bibnamefont {Langanke}},
  \bibinfo {author} {\bibfnamefont {A.}~\bibnamefont {Marek}}, \bibinfo
  {author} {\bibfnamefont {G.}~\bibnamefont {Mart{\'i}nez-Pinedo}}, \ and\
  \bibinfo {author} {\bibfnamefont {B.}~\bibnamefont {M{\"u}ller}},\
  }\href@noop {} {\bibfield  {journal} {\bibinfo  {journal} {Phys. Rep.}\
  }\textbf {\bibinfo {volume} {442}},\ \bibinfo {pages} {38} (\bibinfo {year}
  {2007})}\BibitemShut {NoStop}%
\bibitem [{\citenamefont {Janka}\ \emph {et~al.}(2012)\citenamefont {Janka},
  \citenamefont {Hanke}, \citenamefont {Hüdepohl}, \citenamefont {Marek},
  \citenamefont {Müller},\ and\ \citenamefont
  {Obergaulinger}}]{10.1093/ptep/pts067}%
  \BibitemOpen
  \bibfield  {author} {\bibinfo {author} {\bibfnamefont {H.-T.}\ \bibnamefont
  {Janka}}, \bibinfo {author} {\bibfnamefont {F.}~\bibnamefont {Hanke}},
  \bibinfo {author} {\bibfnamefont {L.}~\bibnamefont {Hüdepohl}}, \bibinfo
  {author} {\bibfnamefont {A.}~\bibnamefont {Marek}}, \bibinfo {author}
  {\bibfnamefont {B.}~\bibnamefont {Müller}}, \ and\ \bibinfo {author}
  {\bibfnamefont {M.}~\bibnamefont {Obergaulinger}},\ }\href@noop {} {\bibfield
   {journal} {\bibinfo  {journal} {Progress of Theoretical and Experimental
  Physics}\ }\textbf {\bibinfo {volume} {2012}} (\bibinfo {year}
  {2012})}\BibitemShut {NoStop}%
\bibitem [{\citenamefont {Burrows}(2013)}]{BUR13}%
  \BibitemOpen
  \bibfield  {author} {\bibinfo {author} {\bibfnamefont {A.}~\bibnamefont
  {Burrows}},\ }\href {\doibase 10.1103/RevModPhys.85.24} {\bibfield  {journal}
  {\bibinfo  {journal} {Rev. Mod. Phys.}\ }\textbf {\bibinfo {volume}
  {\bf{85}}},\ \bibinfo {pages} {245} (\bibinfo {year} {2013})}\BibitemShut
  {NoStop}%
\bibitem [{\citenamefont {Langanke}\ and\ \citenamefont
  {Mart\'{\i}nez-Pinedo}(2003)}]{RevModPhys.75.819}%
  \BibitemOpen
  \bibfield  {author} {\bibinfo {author} {\bibfnamefont {K.}~\bibnamefont
  {Langanke}}\ and\ \bibinfo {author} {\bibfnamefont {G.}~\bibnamefont
  {Mart\'{\i}nez-Pinedo}},\ }\href@noop {} {\bibfield  {journal} {\bibinfo
  {journal} {Rev. Mod. Phys.}\ }\textbf {\bibinfo {volume} {75}},\ \bibinfo
  {pages} {819} (\bibinfo {year} {2003})}\BibitemShut {NoStop}%
\bibitem [{\citenamefont {Richers}\ \emph {et~al.}(2017)\citenamefont
  {Richers}, \citenamefont {Ott}, \citenamefont {Abdikamalov}, \citenamefont
  {O'Connor},\ and\ \citenamefont {Sullivan}}]{PhysRevD.95.063019}%
  \BibitemOpen
  \bibfield  {author} {\bibinfo {author} {\bibfnamefont {S.}~\bibnamefont
  {Richers}}, \bibinfo {author} {\bibfnamefont {C.~D.}\ \bibnamefont {Ott}},
  \bibinfo {author} {\bibfnamefont {E.}~\bibnamefont {Abdikamalov}}, \bibinfo
  {author} {\bibfnamefont {E.}~\bibnamefont {O'Connor}}, \ and\ \bibinfo
  {author} {\bibfnamefont {C.}~\bibnamefont {Sullivan}},\ }\href@noop {}
  {\bibfield  {journal} {\bibinfo  {journal} {Phys. Rev. D}\ }\textbf {\bibinfo
  {volume} {95}},\ \bibinfo {pages} {063019} (\bibinfo {year}
  {2017})}\BibitemShut {NoStop}%
\bibitem [{\citenamefont {Hix}\ \emph {et~al.}(2003)\citenamefont {Hix},
  \citenamefont {Messer}, \citenamefont {Mezzacappa}, \citenamefont
  {Liebend{\"{o}}rfer}, \citenamefont {Sampaio}, \citenamefont {Langanke},
  \citenamefont {Dean},\ and\ \citenamefont {Mart{\'{i}}nez-Pinedo}}]{hix03}%
  \BibitemOpen
  \bibfield  {author} {\bibinfo {author} {\bibfnamefont {W.~R.}\ \bibnamefont
  {Hix}}, \bibinfo {author} {\bibfnamefont {O.~E.~B.}\ \bibnamefont {Messer}},
  \bibinfo {author} {\bibfnamefont {A.}~\bibnamefont {Mezzacappa}}, \bibinfo
  {author} {\bibfnamefont {M.}~\bibnamefont {Liebend{\"{o}}rfer}}, \bibinfo
  {author} {\bibfnamefont {J.}~\bibnamefont {Sampaio}}, \bibinfo {author}
  {\bibfnamefont {K.}~\bibnamefont {Langanke}}, \bibinfo {author}
  {\bibfnamefont {D.~J.}\ \bibnamefont {Dean}}, \ and\ \bibinfo {author}
  {\bibfnamefont {G.}~\bibnamefont {Mart{\'{i}}nez-Pinedo}},\ }\href@noop {}
  {\bibfield  {journal} {\bibinfo  {journal} {Phys. Rev. Lett.}\ }\textbf
  {\bibinfo {volume} {91}},\ \bibinfo {pages} {201102} (\bibinfo {year}
  {2003})}\BibitemShut {NoStop}%
\bibitem [{\citenamefont {Bethe}\ \emph {et~al.}(1979)\citenamefont {Bethe},
  \citenamefont {Brown}, \citenamefont {Applegate},\ and\ \citenamefont
  {Lattimer}}]{A_bbal79}%
  \BibitemOpen
  \bibfield  {author} {\bibinfo {author} {\bibfnamefont {H.}~\bibnamefont
  {Bethe}}, \bibinfo {author} {\bibfnamefont {G.}~\bibnamefont {Brown}},
  \bibinfo {author} {\bibfnamefont {J.}~\bibnamefont {Applegate}}, \ and\
  \bibinfo {author} {\bibfnamefont {J.}~\bibnamefont {Lattimer}},\ }\href@noop
  {} {\bibfield  {journal} {\bibinfo  {journal} {Nucl. Phys. A}\ }\textbf
  {\bibinfo {volume} {324}},\ \bibinfo {pages} {487} (\bibinfo {year}
  {1979})}\BibitemShut {NoStop}%
\bibitem [{\citenamefont {Langanke}\ \emph {et~al.}(2003)\citenamefont
  {Langanke}, \citenamefont {Mart\'{\i}nez-Pinedo}, \citenamefont {Sampaio},
  \citenamefont {Dean}, \citenamefont {Hix}, \citenamefont {Messer},
  \citenamefont {Mezzacappa}, \citenamefont {Liebend\"orfer}, \citenamefont
  {Janka},\ and\ \citenamefont {Rampp}}]{PhysRevLett.90.241102}%
  \BibitemOpen
  \bibfield  {author} {\bibinfo {author} {\bibfnamefont {K.}~\bibnamefont
  {Langanke}}, \bibinfo {author} {\bibfnamefont {G.}~\bibnamefont
  {Mart\'{\i}nez-Pinedo}}, \bibinfo {author} {\bibfnamefont {J.~M.}\
  \bibnamefont {Sampaio}}, \bibinfo {author} {\bibfnamefont {D.~J.}\
  \bibnamefont {Dean}}, \bibinfo {author} {\bibfnamefont {W.~R.}\ \bibnamefont
  {Hix}}, \bibinfo {author} {\bibfnamefont {O.~E.~B.}\ \bibnamefont {Messer}},
  \bibinfo {author} {\bibfnamefont {A.}~\bibnamefont {Mezzacappa}}, \bibinfo
  {author} {\bibfnamefont {M.}~\bibnamefont {Liebend\"orfer}}, \bibinfo
  {author} {\bibfnamefont {H.-T.}\ \bibnamefont {Janka}}, \ and\ \bibinfo
  {author} {\bibfnamefont {M.}~\bibnamefont {Rampp}},\ }\href@noop {}
  {\bibfield  {journal} {\bibinfo  {journal} {Phys. Rev. Lett.}\ }\textbf
  {\bibinfo {volume} {90}},\ \bibinfo {pages} {241102} (\bibinfo {year}
  {2003})}\BibitemShut {NoStop}%
\bibitem [{\citenamefont {Sullivan}\ \emph {et~al.}(2016)\citenamefont
  {Sullivan}, \citenamefont {O'Connor}, \citenamefont {Zegers}, \citenamefont
  {Grubb},\ and\ \citenamefont {Austin}}]{Sull2016}%
  \BibitemOpen
  \bibfield  {author} {\bibinfo {author} {\bibfnamefont {C.}~\bibnamefont
  {Sullivan}}, \bibinfo {author} {\bibfnamefont {E.}~\bibnamefont {O'Connor}},
  \bibinfo {author} {\bibfnamefont {R.~G.~T.}\ \bibnamefont {Zegers}}, \bibinfo
  {author} {\bibfnamefont {T.}~\bibnamefont {Grubb}}, \ and\ \bibinfo {author}
  {\bibfnamefont {S.~M.}\ \bibnamefont {Austin}},\ }\href@noop {} {\bibfield
  {journal} {\bibinfo  {journal} {Astrophys. J.}\ }\textbf {\bibinfo {volume}
  {816}},\ \bibinfo {pages} {44} (\bibinfo {year} {2016})}\BibitemShut
  {NoStop}%
\bibitem [{\citenamefont {Titus}\ \emph {et~al.}(2018)\citenamefont {Titus},
  \citenamefont {Sullivan}, \citenamefont {Zegers}, \citenamefont {Brown},\
  and\ \citenamefont {Gao}}]{TITUS2018}%
  \BibitemOpen
  \bibfield  {author} {\bibinfo {author} {\bibfnamefont {R.}~\bibnamefont
  {Titus}}, \bibinfo {author} {\bibfnamefont {C.}~\bibnamefont {Sullivan}},
  \bibinfo {author} {\bibfnamefont {R.~G.~T.}\ \bibnamefont {Zegers}}, \bibinfo
  {author} {\bibfnamefont {B.~A.}\ \bibnamefont {Brown}}, \ and\ \bibinfo
  {author} {\bibfnamefont {B.}~\bibnamefont {Gao}},\ }\href
  {http://stacks.iop.org/0954-3899/45/i=1/a=014004} {\bibfield  {journal}
  {\bibinfo  {journal} {J. Phys. G}\ }\textbf {\bibinfo {volume} {45}},\
  \bibinfo {pages} {014004} (\bibinfo {year} {2018})}\BibitemShut {NoStop}%
\bibitem [{\citenamefont {Furusawa}\ \emph {et~al.}(2017)\citenamefont
  {Furusawa}, \citenamefont {Nagakura}, \citenamefont {Sumiyoshi},
  \citenamefont {Kato},\ and\ \citenamefont {Yamada}}]{furusawa2017}%
  \BibitemOpen
  \bibfield  {author} {\bibinfo {author} {\bibfnamefont {S.}~\bibnamefont
  {Furusawa}}, \bibinfo {author} {\bibfnamefont {H.}~\bibnamefont {Nagakura}},
  \bibinfo {author} {\bibfnamefont {K.}~\bibnamefont {Sumiyoshi}}, \bibinfo
  {author} {\bibfnamefont {C.}~\bibnamefont {Kato}}, \ and\ \bibinfo {author}
  {\bibfnamefont {S.}~\bibnamefont {Yamada}},\ }\href@noop {} {\bibfield
  {journal} {\bibinfo  {journal} {Phys. Rev. C}\ }\textbf {\bibinfo {volume}
  {95}},\ \bibinfo {pages} {025809} (\bibinfo {year} {2017})}\BibitemShut
  {NoStop}%
\bibitem [{\citenamefont {Pascal}\ \emph {et~al.}(2019)\citenamefont {Pascal},
  \citenamefont {Giraud}, \citenamefont {Fantina}, \citenamefont {Gulminelli},
  \citenamefont {Novak}, \citenamefont {Oertel},\ and\ \citenamefont
  {Raduta}}]{1906.05114}%
  \BibitemOpen
  \bibfield  {author} {\bibinfo {author} {\bibfnamefont {A.}~\bibnamefont
  {Pascal}}, \bibinfo {author} {\bibfnamefont {S.}~\bibnamefont {Giraud}},
  \bibinfo {author} {\bibfnamefont {A.}~\bibnamefont {Fantina}}, \bibinfo
  {author} {\bibfnamefont {F.}~\bibnamefont {Gulminelli}}, \bibinfo {author}
  {\bibfnamefont {J.}~\bibnamefont {Novak}}, \bibinfo {author} {\bibfnamefont
  {M.}~\bibnamefont {Oertel}}, \ and\ \bibinfo {author} {\bibfnamefont
  {A.}~\bibnamefont {Raduta}},\ }\href@noop {} {\enquote {\bibinfo {title}
  {Impact of electron capture rates on nuclei far from stability on
  core-collapse supernovae},}\ } (\bibinfo {year} {2019}),\ \Eprint
  {http://arxiv.org/abs/arXiv:1906.05114} {arXiv:1906.05114} \BibitemShut
  {NoStop}%
\bibitem [{\citenamefont {Taddeucci}\ \emph {et~al.}(1987)\citenamefont
  {Taddeucci}, \citenamefont {Goulding}, \citenamefont {Carey}, \citenamefont
  {Byrd}, \citenamefont {Goodman}, \citenamefont {Gaarde}, \citenamefont
  {Larsen}, \citenamefont {Horen}, \citenamefont {Rapaport},\ and\
  \citenamefont {Sugarbaker}}]{Taddeucci1987125}%
  \BibitemOpen
  \bibfield  {author} {\bibinfo {author} {\bibfnamefont {T.}~\bibnamefont
  {Taddeucci}}, \bibinfo {author} {\bibfnamefont {C.}~\bibnamefont {Goulding}},
  \bibinfo {author} {\bibfnamefont {T.}~\bibnamefont {Carey}}, \bibinfo
  {author} {\bibfnamefont {R.}~\bibnamefont {Byrd}}, \bibinfo {author}
  {\bibfnamefont {C.}~\bibnamefont {Goodman}}, \bibinfo {author} {\bibfnamefont
  {C.}~\bibnamefont {Gaarde}}, \bibinfo {author} {\bibfnamefont
  {J.}~\bibnamefont {Larsen}}, \bibinfo {author} {\bibfnamefont
  {D.}~\bibnamefont {Horen}}, \bibinfo {author} {\bibfnamefont
  {J.}~\bibnamefont {Rapaport}}, \ and\ \bibinfo {author} {\bibfnamefont
  {E.}~\bibnamefont {Sugarbaker}},\ }\href@noop {} {\bibfield  {journal}
  {\bibinfo  {journal} {Nucl. Phys. A}\ }\textbf {\bibinfo {volume} {469}},\
  \bibinfo {pages} {125} (\bibinfo {year} {1987})}\BibitemShut {NoStop}%
\bibitem [{\citenamefont {Zegers}\ \emph {et~al.}(2006)\citenamefont {Zegers},
  \citenamefont {Akimune}, \citenamefont {Austin}, \citenamefont {Bazin},
  \citenamefont {{van den Berg}}, \citenamefont {Berg}, \citenamefont {Brown},
  \citenamefont {Brown}, \citenamefont {Cole}, \citenamefont {Daito},
  \citenamefont {Fujita}, \citenamefont {Fujiwara}, \citenamefont {Gal\`es},
  \citenamefont {Harakeh}, \citenamefont {Hashimoto}, \citenamefont {Hayami},
  \citenamefont {Hitt}, \citenamefont {Howard}, \citenamefont {Itoh},
  \citenamefont {J\"anecke}, \citenamefont {Kawabata}, \citenamefont {Kawase},
  \citenamefont {Kinoshita}, \citenamefont {Nakamura}, \citenamefont
  {Nakanishi}, \citenamefont {Nakayama}, \citenamefont {Okumura}, \citenamefont
  {Richter}, \citenamefont {Roberts}, \citenamefont {Sherrill}, \citenamefont
  {Shimbara}, \citenamefont {Steiner}, \citenamefont {Uchida}, \citenamefont
  {Ueno}, \citenamefont {Yamagata},\ and\ \citenamefont {Yosoi}}]{Zegers:2006}%
  \BibitemOpen
  \bibfield  {author} {\bibinfo {author} {\bibfnamefont {R.~G.~T.}\
  \bibnamefont {Zegers}}, \bibinfo {author} {\bibfnamefont {H.}~\bibnamefont
  {Akimune}}, \bibinfo {author} {\bibfnamefont {S.~M.}\ \bibnamefont {Austin}},
  \bibinfo {author} {\bibfnamefont {D.}~\bibnamefont {Bazin}}, \bibinfo
  {author} {\bibfnamefont {A.~M.}\ \bibnamefont {{van den Berg}}}, \bibinfo
  {author} {\bibfnamefont {G.~P.~A.}\ \bibnamefont {Berg}}, \bibinfo {author}
  {\bibfnamefont {B.~A.}\ \bibnamefont {Brown}}, \bibinfo {author}
  {\bibfnamefont {J.}~\bibnamefont {Brown}}, \bibinfo {author} {\bibfnamefont
  {A.~L.}\ \bibnamefont {Cole}}, \bibinfo {author} {\bibfnamefont
  {I.}~\bibnamefont {Daito}}, \bibinfo {author} {\bibfnamefont
  {Y.}~\bibnamefont {Fujita}}, \bibinfo {author} {\bibfnamefont
  {M.}~\bibnamefont {Fujiwara}}, \bibinfo {author} {\bibfnamefont
  {S.}~\bibnamefont {Gal\`es}}, \bibinfo {author} {\bibfnamefont {M.~N.}\
  \bibnamefont {Harakeh}}, \bibinfo {author} {\bibfnamefont {H.}~\bibnamefont
  {Hashimoto}}, \bibinfo {author} {\bibfnamefont {R.}~\bibnamefont {Hayami}},
  \bibinfo {author} {\bibfnamefont {G.~W.}\ \bibnamefont {Hitt}}, \bibinfo
  {author} {\bibfnamefont {M.~E.}\ \bibnamefont {Howard}}, \bibinfo {author}
  {\bibfnamefont {M.}~\bibnamefont {Itoh}}, \bibinfo {author} {\bibfnamefont
  {J.}~\bibnamefont {J\"anecke}}, \bibinfo {author} {\bibfnamefont
  {T.}~\bibnamefont {Kawabata}}, \bibinfo {author} {\bibfnamefont
  {K.}~\bibnamefont {Kawase}}, \bibinfo {author} {\bibfnamefont
  {M.}~\bibnamefont {Kinoshita}}, \bibinfo {author} {\bibfnamefont
  {T.}~\bibnamefont {Nakamura}}, \bibinfo {author} {\bibfnamefont
  {K.}~\bibnamefont {Nakanishi}}, \bibinfo {author} {\bibfnamefont
  {S.}~\bibnamefont {Nakayama}}, \bibinfo {author} {\bibfnamefont
  {S.}~\bibnamefont {Okumura}}, \bibinfo {author} {\bibfnamefont {W.~A.}\
  \bibnamefont {Richter}}, \bibinfo {author} {\bibfnamefont {D.~A.}\
  \bibnamefont {Roberts}}, \bibinfo {author} {\bibfnamefont {B.~M.}\
  \bibnamefont {Sherrill}}, \bibinfo {author} {\bibfnamefont {Y.}~\bibnamefont
  {Shimbara}}, \bibinfo {author} {\bibfnamefont {M.}~\bibnamefont {Steiner}},
  \bibinfo {author} {\bibfnamefont {M.}~\bibnamefont {Uchida}}, \bibinfo
  {author} {\bibfnamefont {H.}~\bibnamefont {Ueno}}, \bibinfo {author}
  {\bibfnamefont {T.}~\bibnamefont {Yamagata}}, \ and\ \bibinfo {author}
  {\bibfnamefont {M.}~\bibnamefont {Yosoi}},\ }\href@noop {} {\bibfield
  {journal} {\bibinfo  {journal} {Phys. Rev. C}\ }\textbf {\bibinfo {volume}
  {74}},\ \bibinfo {pages} {024309} (\bibinfo {year} {2006})}\BibitemShut
  {NoStop}%
\bibitem [{\citenamefont {Zegers}\ \emph {et~al.}(2007)\citenamefont {Zegers},
  \citenamefont {Adachi}, \citenamefont {Akimune}, \citenamefont {Austin},
  \citenamefont {van~den Berg}, \citenamefont {Brown}, \citenamefont {Fujita},
  \citenamefont {Fujiwara}, \citenamefont {Gal\`es}, \citenamefont {Guess},
  \citenamefont {Harakeh}, \citenamefont {Hashimoto}, \citenamefont {Hatanaka},
  \citenamefont {Hayami}, \citenamefont {Hitt}, \citenamefont {Howard},
  \citenamefont {Itoh}, \citenamefont {Kawabata}, \citenamefont {Kawase},
  \citenamefont {Kinoshita}, \citenamefont {Matsubara}, \citenamefont
  {Nakanishi}, \citenamefont {Nakayama}, \citenamefont {Okumura}, \citenamefont
  {Ohta}, \citenamefont {Sakemi}, \citenamefont {Shimbara}, \citenamefont
  {Shimizu}, \citenamefont {Scholl}, \citenamefont {Simenel}, \citenamefont
  {Tameshige}, \citenamefont {Tamii}, \citenamefont {Uchida}, \citenamefont
  {Yamagata},\ and\ \citenamefont {Yosoi}}]{PhysRevLett.99.202501}%
  \BibitemOpen
  \bibfield  {author} {\bibinfo {author} {\bibfnamefont {R.~G.~T.}\
  \bibnamefont {Zegers}}, \bibinfo {author} {\bibfnamefont {T.}~\bibnamefont
  {Adachi}}, \bibinfo {author} {\bibfnamefont {H.}~\bibnamefont {Akimune}},
  \bibinfo {author} {\bibfnamefont {S.~M.}\ \bibnamefont {Austin}}, \bibinfo
  {author} {\bibfnamefont {A.~M.}\ \bibnamefont {van~den Berg}}, \bibinfo
  {author} {\bibfnamefont {B.~A.}\ \bibnamefont {Brown}}, \bibinfo {author}
  {\bibfnamefont {Y.}~\bibnamefont {Fujita}}, \bibinfo {author} {\bibfnamefont
  {M.}~\bibnamefont {Fujiwara}}, \bibinfo {author} {\bibfnamefont
  {S.}~\bibnamefont {Gal\`es}}, \bibinfo {author} {\bibfnamefont {C.~J.}\
  \bibnamefont {Guess}}, \bibinfo {author} {\bibfnamefont {M.~N.}\ \bibnamefont
  {Harakeh}}, \bibinfo {author} {\bibfnamefont {H.}~\bibnamefont {Hashimoto}},
  \bibinfo {author} {\bibfnamefont {K.}~\bibnamefont {Hatanaka}}, \bibinfo
  {author} {\bibfnamefont {R.}~\bibnamefont {Hayami}}, \bibinfo {author}
  {\bibfnamefont {G.~W.}\ \bibnamefont {Hitt}}, \bibinfo {author}
  {\bibfnamefont {M.~E.}\ \bibnamefont {Howard}}, \bibinfo {author}
  {\bibfnamefont {M.}~\bibnamefont {Itoh}}, \bibinfo {author} {\bibfnamefont
  {T.}~\bibnamefont {Kawabata}}, \bibinfo {author} {\bibfnamefont
  {K.}~\bibnamefont {Kawase}}, \bibinfo {author} {\bibfnamefont
  {M.}~\bibnamefont {Kinoshita}}, \bibinfo {author} {\bibfnamefont
  {M.}~\bibnamefont {Matsubara}}, \bibinfo {author} {\bibfnamefont
  {K.}~\bibnamefont {Nakanishi}}, \bibinfo {author} {\bibfnamefont
  {S.}~\bibnamefont {Nakayama}}, \bibinfo {author} {\bibfnamefont
  {S.}~\bibnamefont {Okumura}}, \bibinfo {author} {\bibfnamefont
  {T.}~\bibnamefont {Ohta}}, \bibinfo {author} {\bibfnamefont {Y.}~\bibnamefont
  {Sakemi}}, \bibinfo {author} {\bibfnamefont {Y.}~\bibnamefont {Shimbara}},
  \bibinfo {author} {\bibfnamefont {Y.}~\bibnamefont {Shimizu}}, \bibinfo
  {author} {\bibfnamefont {C.}~\bibnamefont {Scholl}}, \bibinfo {author}
  {\bibfnamefont {C.}~\bibnamefont {Simenel}}, \bibinfo {author} {\bibfnamefont
  {Y.}~\bibnamefont {Tameshige}}, \bibinfo {author} {\bibfnamefont
  {A.}~\bibnamefont {Tamii}}, \bibinfo {author} {\bibfnamefont
  {M.}~\bibnamefont {Uchida}}, \bibinfo {author} {\bibfnamefont
  {T.}~\bibnamefont {Yamagata}}, \ and\ \bibinfo {author} {\bibfnamefont
  {M.}~\bibnamefont {Yosoi}},\ }\href@noop {} {\bibfield  {journal} {\bibinfo
  {journal} {Phys. Rev. Lett.}\ }\textbf {\bibinfo {volume} {99}},\ \bibinfo
  {pages} {202501} (\bibinfo {year} {2007})}\BibitemShut {NoStop}%
\bibitem [{\citenamefont {Langanke}\ and\ \citenamefont
  {Mart\'{\i}nez-Pinedo}(2000)}]{LANGANKE2000481}%
  \BibitemOpen
  \bibfield  {author} {\bibinfo {author} {\bibfnamefont {K.}~\bibnamefont
  {Langanke}}\ and\ \bibinfo {author} {\bibfnamefont {G.}~\bibnamefont
  {Mart\'{\i}nez-Pinedo}},\ }\href@noop {} {\bibfield  {journal} {\bibinfo
  {journal} {Nucl. Phys. A}\ }\textbf {\bibinfo {volume} {673}},\ \bibinfo
  {pages} {481} (\bibinfo {year} {2000})}\BibitemShut {NoStop}%
\bibitem [{\citenamefont {Langanke}\ \emph {et~al.}(2001)\citenamefont
  {Langanke}, \citenamefont {Kolbe},\ and\ \citenamefont
  {Dean}}]{PhysRevC.63.032801}%
  \BibitemOpen
  \bibfield  {author} {\bibinfo {author} {\bibfnamefont {K.}~\bibnamefont
  {Langanke}}, \bibinfo {author} {\bibfnamefont {E.}~\bibnamefont {Kolbe}}, \
  and\ \bibinfo {author} {\bibfnamefont {D.~J.}\ \bibnamefont {Dean}},\
  }\href@noop {} {\bibfield  {journal} {\bibinfo  {journal} {Phys. Rev. C}\
  }\textbf {\bibinfo {volume} {63}},\ \bibinfo {pages} {032801(R)} (\bibinfo
  {year} {2001})}\BibitemShut {NoStop}%
\bibitem [{\citenamefont {O'Connor}(2015)}]{OCO15}%
  \BibitemOpen
  \bibfield  {author} {\bibinfo {author} {\bibfnamefont {E.}~\bibnamefont
  {O'Connor}},\ }\href@noop {} {\bibfield  {journal} {\bibinfo  {journal}
  {Astrophys. J. Supp.}\ }\textbf {\bibinfo {volume} {219}},\ \bibinfo {pages}
  {24} (\bibinfo {year} {2015})}\BibitemShut {NoStop}%
\bibitem [{\citenamefont {Sullivan}(2015)}]{Sull2016a}%
  \BibitemOpen
  \bibfield  {author} {\bibinfo {author} {\bibfnamefont {C.}~\bibnamefont
  {Sullivan}},\ }\href@noop {} {\bibfield  {journal} {\bibinfo  {journal}
  {Zenodo}\ } (\bibinfo {year} {2015})},\ \bibinfo {note}
  {http://doi.org/10.5281/zenodo.33788}\BibitemShut {NoStop}%
\bibitem [{\citenamefont {O'Connor}\ and\ \citenamefont {Ott}(2010)}]{OCO10}%
  \BibitemOpen
  \bibfield  {author} {\bibinfo {author} {\bibfnamefont {E.}~\bibnamefont
  {O'Connor}}\ and\ \bibinfo {author} {\bibfnamefont {C.~D.}\ \bibnamefont
  {Ott}},\ }\href@noop {} {\bibfield  {journal} {\bibinfo  {journal} {Classical
  and Quantum Gravity}\ }\textbf {\bibinfo {volume} {27}},\ \bibinfo {pages}
  {114103} (\bibinfo {year} {2010})}\BibitemShut {NoStop}%
\bibitem [{\citenamefont {Fuller}\ \emph {et~al.}(1982)\citenamefont {Fuller},
  \citenamefont {Fowler},\ and\ \citenamefont {Newman}}]{fuller82}%
  \BibitemOpen
  \bibfield  {author} {\bibinfo {author} {\bibfnamefont {G.~M.}\ \bibnamefont
  {Fuller}}, \bibinfo {author} {\bibfnamefont {W.~A.}\ \bibnamefont {Fowler}},
  \ and\ \bibinfo {author} {\bibfnamefont {M.~J.}\ \bibnamefont {Newman}},\
  }\href@noop {} {\bibfield  {journal} {\bibinfo  {journal} {Astrophys. J.}\
  }\textbf {\bibinfo {volume} {252}},\ \bibinfo {pages} {715} (\bibinfo {year}
  {1982})}\BibitemShut {NoStop}%
\bibitem [{\citenamefont {Oda}\ \emph {et~al.}(1994)\citenamefont {Oda},
  \citenamefont {Hino}, \citenamefont {Muto}, \citenamefont {Takahara},\ and\
  \citenamefont {Sato}}]{ODA1994231}%
  \BibitemOpen
  \bibfield  {author} {\bibinfo {author} {\bibfnamefont {T.}~\bibnamefont
  {Oda}}, \bibinfo {author} {\bibfnamefont {M.}~\bibnamefont {Hino}}, \bibinfo
  {author} {\bibfnamefont {K.}~\bibnamefont {Muto}}, \bibinfo {author}
  {\bibfnamefont {M.}~\bibnamefont {Takahara}}, \ and\ \bibinfo {author}
  {\bibfnamefont {K.}~\bibnamefont {Sato}},\ }\href@noop {} {\bibfield
  {journal} {\bibinfo  {journal} {At. Data Nucl. Data Tables}\ }\textbf
  {\bibinfo {volume} {56}},\ \bibinfo {pages} {231} (\bibinfo {year}
  {1994})}\BibitemShut {NoStop}%
\bibitem [{\citenamefont {Langanke}\ and\ \citenamefont
  {Mart\'{i}nez-Pinedo}(2001)}]{LAN01a}%
  \BibitemOpen
  \bibfield  {author} {\bibinfo {author} {\bibfnamefont {K.}~\bibnamefont
  {Langanke}}\ and\ \bibinfo {author} {\bibfnamefont {G.}~\bibnamefont
  {Mart\'{i}nez-Pinedo}},\ }\href@noop {} {\bibfield  {journal} {\bibinfo
  {journal} {At. Data Nucl. Data Tables}\ }\textbf {\bibinfo {volume} {{79}}},\
  \bibinfo {pages} {146} (\bibinfo {year} {2001})}\BibitemShut {NoStop}%
\bibitem [{\citenamefont {Pruet}\ and\ \citenamefont {Fuller}(2003)}]{pruet1}%
  \BibitemOpen
  \bibfield  {author} {\bibinfo {author} {\bibfnamefont {J.}~\bibnamefont
  {Pruet}}\ and\ \bibinfo {author} {\bibfnamefont {G.~M.}\ \bibnamefont
  {Fuller}},\ }\href@noop {} {\bibfield  {journal} {\bibinfo  {journal}
  {Astrophys. J. Supp.}\ }\textbf {\bibinfo {volume} {149}},\ \bibinfo {pages}
  {189} (\bibinfo {year} {2003})}\BibitemShut {NoStop}%
\bibitem [{\citenamefont {Suzuki}\ \emph {et~al.}(2016)\citenamefont {Suzuki},
  \citenamefont {Toki},\ and\ \citenamefont {Nomoto}}]{Suzuki2016}%
  \BibitemOpen
  \bibfield  {author} {\bibinfo {author} {\bibfnamefont {T.}~\bibnamefont
  {Suzuki}}, \bibinfo {author} {\bibfnamefont {H.}~\bibnamefont {Toki}}, \ and\
  \bibinfo {author} {\bibfnamefont {K.}~\bibnamefont {Nomoto}},\ }\href@noop {}
  {\bibfield  {journal} {\bibinfo  {journal} {Astrophys. J.}\ }\textbf
  {\bibinfo {volume} {817}},\ \bibinfo {pages} {163} (\bibinfo {year}
  {2016})}\BibitemShut {NoStop}%
\bibitem [{\citenamefont {Cole}\ \emph {et~al.}(2012)\citenamefont {Cole},
  \citenamefont {Anderson}, \citenamefont {Zegers}, \citenamefont {Austin},
  \citenamefont {Brown}, \citenamefont {Valdez}, \citenamefont {Gupta},
  \citenamefont {Hitt},\ and\ \citenamefont {Fawwaz}}]{cole_ecrate_pf}%
  \BibitemOpen
  \bibfield  {author} {\bibinfo {author} {\bibfnamefont {A.~L.}\ \bibnamefont
  {Cole}}, \bibinfo {author} {\bibfnamefont {T.~S.}\ \bibnamefont {Anderson}},
  \bibinfo {author} {\bibfnamefont {R.~G.~T.}\ \bibnamefont {Zegers}}, \bibinfo
  {author} {\bibfnamefont {S.~M.}\ \bibnamefont {Austin}}, \bibinfo {author}
  {\bibfnamefont {B.~A.}\ \bibnamefont {Brown}}, \bibinfo {author}
  {\bibfnamefont {L.}~\bibnamefont {Valdez}}, \bibinfo {author} {\bibfnamefont
  {S.}~\bibnamefont {Gupta}}, \bibinfo {author} {\bibfnamefont {G.~W.}\
  \bibnamefont {Hitt}}, \ and\ \bibinfo {author} {\bibfnamefont
  {O.}~\bibnamefont {Fawwaz}},\ }\href@noop {} {\bibfield  {journal} {\bibinfo
  {journal} {Phys. Rev. C}\ }\textbf {\bibinfo {volume} {86}},\ \bibinfo
  {pages} {015809} (\bibinfo {year} {2012})}\BibitemShut {NoStop}%
\bibitem [{\citenamefont {Raduta}\ \emph {et~al.}(2017)\citenamefont {Raduta},
  \citenamefont {Gulminelli},\ and\ \citenamefont
  {Oertel}}]{PhysRevC.95.025805}%
  \BibitemOpen
  \bibfield  {author} {\bibinfo {author} {\bibfnamefont {A.~R.}\ \bibnamefont
  {Raduta}}, \bibinfo {author} {\bibfnamefont {F.}~\bibnamefont {Gulminelli}},
  \ and\ \bibinfo {author} {\bibfnamefont {M.}~\bibnamefont {Oertel}},\
  }\href@noop {} {\bibfield  {journal} {\bibinfo  {journal} {Phys. Rev. C}\
  }\textbf {\bibinfo {volume} {95}},\ \bibinfo {pages} {025805} (\bibinfo
  {year} {2017})}\BibitemShut {NoStop}%
\bibitem [{jua(2019)}]{juan2019}%
  \BibitemOpen
  \href@noop {} {\enquote {\bibinfo {title} {Experimental constraint on stellar
  electron-capture rates from the
  $^{88}${S}r($t$,$^{3}${H}e+$\gamma$)$^{88}${R}b reaction at 115
  {M}e{V}/$u$},}\ } (\bibinfo {year} {2019}),\ \Eprint
  {http://arxiv.org/abs/arXiv:1906.05934} {arXiv:1906.05934} \BibitemShut
  {NoStop}%
\bibitem [{\citenamefont {Morrissey}\ \emph {et~al.}(2003)\citenamefont
  {Morrissey}, \citenamefont {Sherrill}, \citenamefont {Steiner}, \citenamefont
  {Stolz},\ and\ \citenamefont {Wiedenhoever}}]{MORRISSEY200390}%
  \BibitemOpen
  \bibfield  {author} {\bibinfo {author} {\bibfnamefont {D.}~\bibnamefont
  {Morrissey}}, \bibinfo {author} {\bibfnamefont {B.}~\bibnamefont {Sherrill}},
  \bibinfo {author} {\bibfnamefont {M.}~\bibnamefont {Steiner}}, \bibinfo
  {author} {\bibfnamefont {A.}~\bibnamefont {Stolz}}, \ and\ \bibinfo {author}
  {\bibfnamefont {I.}~\bibnamefont {Wiedenhoever}},\ }\href {\doibase
  https://doi.org/10.1016/S0168-583X(02)01895-5} {\bibfield  {journal}
  {\bibinfo  {journal} {Nucl. Instr. Meth. Phys. Res. B}\ }\textbf {\bibinfo
  {volume} {204}},\ \bibinfo {pages} {90 } (\bibinfo {year}
  {2003})}\BibitemShut {NoStop}%
\bibitem [{\citenamefont {Hitt}\ \emph {et~al.}(2006)\citenamefont {Hitt},
  \citenamefont {Austin}, \citenamefont {Bazin}, \citenamefont {Cole},
  \citenamefont {Dietrich}, \citenamefont {Gade}, \citenamefont {Howard},
  \citenamefont {Reitzner}, \citenamefont {Sherrill}, \citenamefont {Simenel},
  \citenamefont {Smith}, \citenamefont {Stetson}, \citenamefont {Stolz},\ and\
  \citenamefont {Zegers}}]{HITT2006264}%
  \BibitemOpen
  \bibfield  {author} {\bibinfo {author} {\bibfnamefont {G.~W.}\ \bibnamefont
  {Hitt}}, \bibinfo {author} {\bibfnamefont {S.~M.}\ \bibnamefont {Austin}},
  \bibinfo {author} {\bibfnamefont {D.}~\bibnamefont {Bazin}}, \bibinfo
  {author} {\bibfnamefont {A.~L.}\ \bibnamefont {Cole}}, \bibinfo {author}
  {\bibfnamefont {J.}~\bibnamefont {Dietrich}}, \bibinfo {author}
  {\bibfnamefont {A.}~\bibnamefont {Gade}}, \bibinfo {author} {\bibfnamefont
  {M.~E.}\ \bibnamefont {Howard}}, \bibinfo {author} {\bibfnamefont {S.~D.}\
  \bibnamefont {Reitzner}}, \bibinfo {author} {\bibfnamefont {B.~M.}\
  \bibnamefont {Sherrill}}, \bibinfo {author} {\bibfnamefont {C.}~\bibnamefont
  {Simenel}}, \bibinfo {author} {\bibfnamefont {E.~E.}\ \bibnamefont {Smith}},
  \bibinfo {author} {\bibfnamefont {J.}~\bibnamefont {Stetson}}, \bibinfo
  {author} {\bibfnamefont {A.}~\bibnamefont {Stolz}}, \ and\ \bibinfo {author}
  {\bibfnamefont {R.~G.~T.}\ \bibnamefont {Zegers}},\ }\href {\doibase
  https://doi.org/10.1016/j.nima.2006.07.045} {\bibfield  {journal} {\bibinfo
  {journal} {Nucl. Instr. Meth. Phys. Res. A}\ }\textbf {\bibinfo {volume}
  {566}},\ \bibinfo {pages} {264 } (\bibinfo {year} {2006})}\BibitemShut
  {NoStop}%
\bibitem [{\citenamefont {{Palardy}}\ \emph {et~al.}(2010)\citenamefont
  {{Palardy}}, \citenamefont {{Ferrante}}, \citenamefont {{Riley}},\ and\
  \citenamefont {{Zegers}}}]{2010APS.DNP.EA083P}%
  \BibitemOpen
  \bibfield  {author} {\bibinfo {author} {\bibfnamefont {J.}~\bibnamefont
  {{Palardy}}}, \bibinfo {author} {\bibfnamefont {N.}~\bibnamefont
  {{Ferrante}}}, \bibinfo {author} {\bibfnamefont {L.}~\bibnamefont {{Riley}}},
  \ and\ \bibinfo {author} {\bibfnamefont {R.~G.~T.}\ \bibnamefont
  {{Zegers}}},\ }\href@noop {} {\bibfield  {journal} {\bibinfo  {journal}
  {Bulletin of the American Physical Society}\ }\textbf {\bibinfo {volume}
  {DNP}},\ \bibinfo {pages} {EA.38} (\bibinfo {year} {2010})}\BibitemShut
  {NoStop}%
\bibitem [{\citenamefont {Bazin}\ \emph {et~al.}(2003)\citenamefont {Bazin},
  \citenamefont {Caggiano}, \citenamefont {Sherrill}, \citenamefont {Yurkon},\
  and\ \citenamefont {Zeller}}]{BAZIN2003629}%
  \BibitemOpen
  \bibfield  {author} {\bibinfo {author} {\bibfnamefont {D.}~\bibnamefont
  {Bazin}}, \bibinfo {author} {\bibfnamefont {J.}~\bibnamefont {Caggiano}},
  \bibinfo {author} {\bibfnamefont {B.}~\bibnamefont {Sherrill}}, \bibinfo
  {author} {\bibfnamefont {J.}~\bibnamefont {Yurkon}}, \ and\ \bibinfo {author}
  {\bibfnamefont {A.}~\bibnamefont {Zeller}},\ }\href {\doibase
  https://doi.org/10.1016/S0168-583X(02)02142-0} {\bibfield  {journal}
  {\bibinfo  {journal} {Nucl. Instr. Meth. Phys. Res. B}\ }\textbf {\bibinfo
  {volume} {204}},\ \bibinfo {pages} {629 } (\bibinfo {year}
  {2003})}\BibitemShut {NoStop}%
\bibitem [{\citenamefont {Fujita}\ \emph {et~al.}(2002)\citenamefont {Fujita},
  \citenamefont {Fujita}, \citenamefont {Berg}, \citenamefont {Bacher},
  \citenamefont {Foster}, \citenamefont {Hara}, \citenamefont {Hatanaka},
  \citenamefont {Kawabata}, \citenamefont {Noro}, \citenamefont {Sakaguchi},
  \citenamefont {Shimbara}, \citenamefont {Shinada}, \citenamefont
  {Stephenson}, \citenamefont {Ueno},\ and\ \citenamefont
  {Yosoi}}]{FUJITA200217}%
  \BibitemOpen
  \bibfield  {author} {\bibinfo {author} {\bibfnamefont {H.}~\bibnamefont
  {Fujita}}, \bibinfo {author} {\bibfnamefont {Y.}~\bibnamefont {Fujita}},
  \bibinfo {author} {\bibfnamefont {G.~P.~A.}\ \bibnamefont {Berg}}, \bibinfo
  {author} {\bibfnamefont {A.~D.}\ \bibnamefont {Bacher}}, \bibinfo {author}
  {\bibfnamefont {C.~C.}\ \bibnamefont {Foster}}, \bibinfo {author}
  {\bibfnamefont {K.}~\bibnamefont {Hara}}, \bibinfo {author} {\bibfnamefont
  {K.}~\bibnamefont {Hatanaka}}, \bibinfo {author} {\bibfnamefont
  {T.}~\bibnamefont {Kawabata}}, \bibinfo {author} {\bibfnamefont
  {T.}~\bibnamefont {Noro}}, \bibinfo {author} {\bibfnamefont {H.}~\bibnamefont
  {Sakaguchi}}, \bibinfo {author} {\bibfnamefont {Y.}~\bibnamefont {Shimbara}},
  \bibinfo {author} {\bibfnamefont {T.}~\bibnamefont {Shinada}}, \bibinfo
  {author} {\bibfnamefont {E.~J.}\ \bibnamefont {Stephenson}}, \bibinfo
  {author} {\bibfnamefont {H.}~\bibnamefont {Ueno}}, \ and\ \bibinfo {author}
  {\bibfnamefont {M.}~\bibnamefont {Yosoi}},\ }\href {\doibase
  https://doi.org/10.1016/S0168-9002(01)01970-2} {\bibfield  {journal}
  {\bibinfo  {journal} {Nucl. Instr. Meth. Phys. Res. A}\ }\textbf {\bibinfo
  {volume} {484}},\ \bibinfo {pages} {17 } (\bibinfo {year}
  {2002})}\BibitemShut {NoStop}%
\bibitem [{\citenamefont {Yurkon}\ \emph {et~al.}(1999)\citenamefont {Yurkon},
  \citenamefont {Bazin}, \citenamefont {Benenson}, \citenamefont {Morrissey},
  \citenamefont {Sherrill}, \citenamefont {Swan},\ and\ \citenamefont
  {Swanson}}]{YURKON1999291}%
  \BibitemOpen
  \bibfield  {author} {\bibinfo {author} {\bibfnamefont {J.}~\bibnamefont
  {Yurkon}}, \bibinfo {author} {\bibfnamefont {D.}~\bibnamefont {Bazin}},
  \bibinfo {author} {\bibfnamefont {W.}~\bibnamefont {Benenson}}, \bibinfo
  {author} {\bibfnamefont {D.~J.}\ \bibnamefont {Morrissey}}, \bibinfo {author}
  {\bibfnamefont {B.~M.}\ \bibnamefont {Sherrill}}, \bibinfo {author}
  {\bibfnamefont {D.}~\bibnamefont {Swan}}, \ and\ \bibinfo {author}
  {\bibfnamefont {R.}~\bibnamefont {Swanson}},\ }\href {\doibase
  https://doi.org/10.1016/S0168-9002(98)00960-7} {\bibfield  {journal}
  {\bibinfo  {journal} {Nucl. Instr. Meth. Phys. Res. A}\ }\textbf {\bibinfo
  {volume} {422}},\ \bibinfo {pages} {291 } (\bibinfo {year}
  {1999})}\BibitemShut {NoStop}%
\bibitem [{\citenamefont {Makino}\ and\ \citenamefont
  {Berz}(1999)}]{MAKINO1999338}%
  \BibitemOpen
  \bibfield  {author} {\bibinfo {author} {\bibfnamefont {K.}~\bibnamefont
  {Makino}}\ and\ \bibinfo {author} {\bibfnamefont {M.}~\bibnamefont {Berz}},\
  }\href {\doibase https://doi.org/10.1016/S0168-9002(98)01554-X} {\bibfield
  {journal} {\bibinfo  {journal} {Nucl. Instr. Meth. Phys. Res. A}\ }\textbf
  {\bibinfo {volume} {427}},\ \bibinfo {pages} {338 } (\bibinfo {year}
  {1999})}\BibitemShut {NoStop}%
\bibitem [{\citenamefont {Paschalis}\ \emph {et~al.}(2013)\citenamefont
  {Paschalis}, \citenamefont {Lee}, \citenamefont {Macchiavelli}, \citenamefont
  {Campbell}, \citenamefont {Cromaz}, \citenamefont {Gros}, \citenamefont
  {Pavan}, \citenamefont {Qian}, \citenamefont {Clark}, \citenamefont
  {Crawford}, \citenamefont {Doering}, \citenamefont {Fallon}, \citenamefont
  {Lionberger}, \citenamefont {Loew}, \citenamefont {Petri}, \citenamefont
  {Stezelberger}, \citenamefont {Zimmermann}, \citenamefont {Radford},
  \citenamefont {Lagergren}, \citenamefont {Weisshaar}, \citenamefont
  {Winkler}, \citenamefont {Glasmacher}, \citenamefont {Anderson},\ and\
  \citenamefont {Beausang}}]{PASCHALIS201344}%
  \BibitemOpen
  \bibfield  {author} {\bibinfo {author} {\bibfnamefont {S.}~\bibnamefont
  {Paschalis}}, \bibinfo {author} {\bibfnamefont {I.}~\bibnamefont {Lee}},
  \bibinfo {author} {\bibfnamefont {A.}~\bibnamefont {Macchiavelli}}, \bibinfo
  {author} {\bibfnamefont {C.}~\bibnamefont {Campbell}}, \bibinfo {author}
  {\bibfnamefont {M.}~\bibnamefont {Cromaz}}, \bibinfo {author} {\bibfnamefont
  {S.}~\bibnamefont {Gros}}, \bibinfo {author} {\bibfnamefont {J.}~\bibnamefont
  {Pavan}}, \bibinfo {author} {\bibfnamefont {J.}~\bibnamefont {Qian}},
  \bibinfo {author} {\bibfnamefont {R.}~\bibnamefont {Clark}}, \bibinfo
  {author} {\bibfnamefont {H.}~\bibnamefont {Crawford}}, \bibinfo {author}
  {\bibfnamefont {D.}~\bibnamefont {Doering}}, \bibinfo {author} {\bibfnamefont
  {P.}~\bibnamefont {Fallon}}, \bibinfo {author} {\bibfnamefont
  {C.}~\bibnamefont {Lionberger}}, \bibinfo {author} {\bibfnamefont
  {T.}~\bibnamefont {Loew}}, \bibinfo {author} {\bibfnamefont {M.}~\bibnamefont
  {Petri}}, \bibinfo {author} {\bibfnamefont {T.}~\bibnamefont {Stezelberger}},
  \bibinfo {author} {\bibfnamefont {S.}~\bibnamefont {Zimmermann}}, \bibinfo
  {author} {\bibfnamefont {D.}~\bibnamefont {Radford}}, \bibinfo {author}
  {\bibfnamefont {K.}~\bibnamefont {Lagergren}}, \bibinfo {author}
  {\bibfnamefont {D.}~\bibnamefont {Weisshaar}}, \bibinfo {author}
  {\bibfnamefont {R.}~\bibnamefont {Winkler}}, \bibinfo {author} {\bibfnamefont
  {T.}~\bibnamefont {Glasmacher}}, \bibinfo {author} {\bibfnamefont
  {J.}~\bibnamefont {Anderson}}, \ and\ \bibinfo {author} {\bibfnamefont
  {C.}~\bibnamefont {Beausang}},\ }\href {\doibase
  https://doi.org/10.1016/j.nima.2013.01.009} {\bibfield  {journal} {\bibinfo
  {journal} {Nucl. Instr. Meth. Phys. Res. A}\ }\textbf {\bibinfo {volume}
  {709}},\ \bibinfo {pages} {44 } (\bibinfo {year} {2013})}\BibitemShut
  {NoStop}%
\bibitem [{\citenamefont {Weisshaar}\ \emph {et~al.}(2017)\citenamefont
  {Weisshaar}, \citenamefont {Bazin}, \citenamefont {Bender}, \citenamefont
  {Campbell}, \citenamefont {Recchia}, \citenamefont {Bader}, \citenamefont
  {Baugher}, \citenamefont {Belarge}, \citenamefont {Carpenter}, \citenamefont
  {Crawford}, \citenamefont {Cromaz}, \citenamefont {Elman}, \citenamefont
  {Fallon}, \citenamefont {Forney}, \citenamefont {Gade}, \citenamefont
  {Harker}, \citenamefont {Kobayashi}, \citenamefont {Langer}, \citenamefont
  {Lauritsen}, \citenamefont {Lee}, \citenamefont {Lemasson}, \citenamefont
  {Longfellow}, \citenamefont {Lunderberg}, \citenamefont {Macchiavelli},
  \citenamefont {Miki}, \citenamefont {Momiyama}, \citenamefont {Noji},
  \citenamefont {Radford}, \citenamefont {Scott}, \citenamefont {Sethi},
  \citenamefont {Stroberg}, \citenamefont {Sullivan}, \citenamefont {Titus},
  \citenamefont {Wiens}, \citenamefont {Williams}, \citenamefont {Wimmer},\
  and\ \citenamefont {Zhu}}]{WEISSHAAR2017187}%
  \BibitemOpen
  \bibfield  {author} {\bibinfo {author} {\bibfnamefont {D.}~\bibnamefont
  {Weisshaar}}, \bibinfo {author} {\bibfnamefont {D.}~\bibnamefont {Bazin}},
  \bibinfo {author} {\bibfnamefont {P.}~\bibnamefont {Bender}}, \bibinfo
  {author} {\bibfnamefont {C.}~\bibnamefont {Campbell}}, \bibinfo {author}
  {\bibfnamefont {F.}~\bibnamefont {Recchia}}, \bibinfo {author} {\bibfnamefont
  {V.}~\bibnamefont {Bader}}, \bibinfo {author} {\bibfnamefont
  {T.}~\bibnamefont {Baugher}}, \bibinfo {author} {\bibfnamefont
  {J.}~\bibnamefont {Belarge}}, \bibinfo {author} {\bibfnamefont
  {M.}~\bibnamefont {Carpenter}}, \bibinfo {author} {\bibfnamefont
  {H.}~\bibnamefont {Crawford}}, \bibinfo {author} {\bibfnamefont
  {M.}~\bibnamefont {Cromaz}}, \bibinfo {author} {\bibfnamefont
  {B.}~\bibnamefont {Elman}}, \bibinfo {author} {\bibfnamefont
  {P.}~\bibnamefont {Fallon}}, \bibinfo {author} {\bibfnamefont
  {A.}~\bibnamefont {Forney}}, \bibinfo {author} {\bibfnamefont
  {A.}~\bibnamefont {Gade}}, \bibinfo {author} {\bibfnamefont {J.}~\bibnamefont
  {Harker}}, \bibinfo {author} {\bibfnamefont {N.}~\bibnamefont {Kobayashi}},
  \bibinfo {author} {\bibfnamefont {C.}~\bibnamefont {Langer}}, \bibinfo
  {author} {\bibfnamefont {T.}~\bibnamefont {Lauritsen}}, \bibinfo {author}
  {\bibfnamefont {I.}~\bibnamefont {Lee}}, \bibinfo {author} {\bibfnamefont
  {A.}~\bibnamefont {Lemasson}}, \bibinfo {author} {\bibfnamefont
  {B.}~\bibnamefont {Longfellow}}, \bibinfo {author} {\bibfnamefont
  {E.}~\bibnamefont {Lunderberg}}, \bibinfo {author} {\bibfnamefont
  {A.}~\bibnamefont {Macchiavelli}}, \bibinfo {author} {\bibfnamefont
  {K.}~\bibnamefont {Miki}}, \bibinfo {author} {\bibfnamefont {S.}~\bibnamefont
  {Momiyama}}, \bibinfo {author} {\bibfnamefont {S.}~\bibnamefont {Noji}},
  \bibinfo {author} {\bibfnamefont {D.}~\bibnamefont {Radford}}, \bibinfo
  {author} {\bibfnamefont {M.}~\bibnamefont {Scott}}, \bibinfo {author}
  {\bibfnamefont {J.}~\bibnamefont {Sethi}}, \bibinfo {author} {\bibfnamefont
  {S.}~\bibnamefont {Stroberg}}, \bibinfo {author} {\bibfnamefont
  {C.}~\bibnamefont {Sullivan}}, \bibinfo {author} {\bibfnamefont
  {R.}~\bibnamefont {Titus}}, \bibinfo {author} {\bibfnamefont
  {A.}~\bibnamefont {Wiens}}, \bibinfo {author} {\bibfnamefont
  {S.}~\bibnamefont {Williams}}, \bibinfo {author} {\bibfnamefont
  {K.}~\bibnamefont {Wimmer}}, \ and\ \bibinfo {author} {\bibfnamefont
  {S.}~\bibnamefont {Zhu}},\ }\href {\doibase
  https://doi.org/10.1016/j.nima.2016.12.001} {\bibfield  {journal} {\bibinfo
  {journal} {Nucl. Instr. Meth. Phys. Res. A}\ }\textbf {\bibinfo {volume}
  {847}},\ \bibinfo {pages} {187 } (\bibinfo {year} {2017})}\BibitemShut
  {NoStop}%
\bibitem [{\citenamefont {Noji}\ \emph {et~al.}(2015)\citenamefont {Noji},
  \citenamefont {Zegers}, \citenamefont {Austin}, \citenamefont {Baugher},
  \citenamefont {Bazin}, \citenamefont {Brown}, \citenamefont {Campbell},
  \citenamefont {Cole}, \citenamefont {Doster}, \citenamefont {Gade},
  \citenamefont {Guess}, \citenamefont {Gupta}, \citenamefont {Hitt},
  \citenamefont {Langer}, \citenamefont {Lipschutz}, \citenamefont
  {Lunderberg}, \citenamefont {Meharchand}, \citenamefont {Meisel},
  \citenamefont {Perdikakis}, \citenamefont {Pereira}, \citenamefont {Recchia},
  \citenamefont {Schatz}, \citenamefont {Scott}, \citenamefont {Stroberg},
  \citenamefont {Sullivan}, \citenamefont {Valdez}, \citenamefont {Walz},
  \citenamefont {Weisshaar}, \citenamefont {Williams},\ and\ \citenamefont
  {Wimmer}}]{PhysRevC.92.024312}%
  \BibitemOpen
  \bibfield  {author} {\bibinfo {author} {\bibfnamefont {S.}~\bibnamefont
  {Noji}}, \bibinfo {author} {\bibfnamefont {R.~G.~T.}\ \bibnamefont {Zegers}},
  \bibinfo {author} {\bibfnamefont {S.~M.}\ \bibnamefont {Austin}}, \bibinfo
  {author} {\bibfnamefont {T.}~\bibnamefont {Baugher}}, \bibinfo {author}
  {\bibfnamefont {D.}~\bibnamefont {Bazin}}, \bibinfo {author} {\bibfnamefont
  {B.~A.}\ \bibnamefont {Brown}}, \bibinfo {author} {\bibfnamefont {C.~M.}\
  \bibnamefont {Campbell}}, \bibinfo {author} {\bibfnamefont {A.~L.}\
  \bibnamefont {Cole}}, \bibinfo {author} {\bibfnamefont {H.~J.}\ \bibnamefont
  {Doster}}, \bibinfo {author} {\bibfnamefont {A.}~\bibnamefont {Gade}},
  \bibinfo {author} {\bibfnamefont {C.~J.}\ \bibnamefont {Guess}}, \bibinfo
  {author} {\bibfnamefont {S.}~\bibnamefont {Gupta}}, \bibinfo {author}
  {\bibfnamefont {G.~W.}\ \bibnamefont {Hitt}}, \bibinfo {author}
  {\bibfnamefont {C.}~\bibnamefont {Langer}}, \bibinfo {author} {\bibfnamefont
  {S.}~\bibnamefont {Lipschutz}}, \bibinfo {author} {\bibfnamefont
  {E.}~\bibnamefont {Lunderberg}}, \bibinfo {author} {\bibfnamefont
  {R.}~\bibnamefont {Meharchand}}, \bibinfo {author} {\bibfnamefont
  {Z.}~\bibnamefont {Meisel}}, \bibinfo {author} {\bibfnamefont
  {G.}~\bibnamefont {Perdikakis}}, \bibinfo {author} {\bibfnamefont
  {J.}~\bibnamefont {Pereira}}, \bibinfo {author} {\bibfnamefont
  {F.}~\bibnamefont {Recchia}}, \bibinfo {author} {\bibfnamefont
  {H.}~\bibnamefont {Schatz}}, \bibinfo {author} {\bibfnamefont
  {M.}~\bibnamefont {Scott}}, \bibinfo {author} {\bibfnamefont {S.~R.}\
  \bibnamefont {Stroberg}}, \bibinfo {author} {\bibfnamefont {C.}~\bibnamefont
  {Sullivan}}, \bibinfo {author} {\bibfnamefont {L.}~\bibnamefont {Valdez}},
  \bibinfo {author} {\bibfnamefont {C.}~\bibnamefont {Walz}}, \bibinfo {author}
  {\bibfnamefont {D.}~\bibnamefont {Weisshaar}}, \bibinfo {author}
  {\bibfnamefont {S.~J.}\ \bibnamefont {Williams}}, \ and\ \bibinfo {author}
  {\bibfnamefont {K.}~\bibnamefont {Wimmer}},\ }\href@noop {} {\bibfield
  {journal} {\bibinfo  {journal} {Phys. Rev. C}\ }\textbf {\bibinfo {volume}
  {92}},\ \bibinfo {pages} {024312} (\bibinfo {year} {2015})}\BibitemShut
  {NoStop}%
\bibitem [{\citenamefont {Noji}\ \emph {et~al.}(2014)\citenamefont {Noji},
  \citenamefont {Zegers}, \citenamefont {Austin}, \citenamefont {Baugher},
  \citenamefont {Bazin}, \citenamefont {Brown}, \citenamefont {Campbell},
  \citenamefont {Cole}, \citenamefont {Doster}, \citenamefont {Gade},
  \citenamefont {Guess}, \citenamefont {Gupta}, \citenamefont {Hitt},
  \citenamefont {Langer}, \citenamefont {Lipschutz}, \citenamefont
  {Lunderberg}, \citenamefont {Meharchand}, \citenamefont {Meisel},
  \citenamefont {Perdikakis}, \citenamefont {Pereira}, \citenamefont {Recchia},
  \citenamefont {Schatz}, \citenamefont {Scott}, \citenamefont {Stroberg},
  \citenamefont {Sullivan}, \citenamefont {Valdez}, \citenamefont {Walz},
  \citenamefont {Weisshaar}, \citenamefont {Williams},\ and\ \citenamefont
  {Wimmer}}]{PhysRevLett.112.252501}%
  \BibitemOpen
  \bibfield  {author} {\bibinfo {author} {\bibfnamefont {S.}~\bibnamefont
  {Noji}}, \bibinfo {author} {\bibfnamefont {R.~G.~T.}\ \bibnamefont {Zegers}},
  \bibinfo {author} {\bibfnamefont {S.~M.}\ \bibnamefont {Austin}}, \bibinfo
  {author} {\bibfnamefont {T.}~\bibnamefont {Baugher}}, \bibinfo {author}
  {\bibfnamefont {D.}~\bibnamefont {Bazin}}, \bibinfo {author} {\bibfnamefont
  {B.~A.}\ \bibnamefont {Brown}}, \bibinfo {author} {\bibfnamefont {C.~M.}\
  \bibnamefont {Campbell}}, \bibinfo {author} {\bibfnamefont {A.~L.}\
  \bibnamefont {Cole}}, \bibinfo {author} {\bibfnamefont {H.~J.}\ \bibnamefont
  {Doster}}, \bibinfo {author} {\bibfnamefont {A.}~\bibnamefont {Gade}},
  \bibinfo {author} {\bibfnamefont {C.~J.}\ \bibnamefont {Guess}}, \bibinfo
  {author} {\bibfnamefont {S.}~\bibnamefont {Gupta}}, \bibinfo {author}
  {\bibfnamefont {G.~W.}\ \bibnamefont {Hitt}}, \bibinfo {author}
  {\bibfnamefont {C.}~\bibnamefont {Langer}}, \bibinfo {author} {\bibfnamefont
  {S.}~\bibnamefont {Lipschutz}}, \bibinfo {author} {\bibfnamefont
  {E.}~\bibnamefont {Lunderberg}}, \bibinfo {author} {\bibfnamefont
  {R.}~\bibnamefont {Meharchand}}, \bibinfo {author} {\bibfnamefont
  {Z.}~\bibnamefont {Meisel}}, \bibinfo {author} {\bibfnamefont
  {G.}~\bibnamefont {Perdikakis}}, \bibinfo {author} {\bibfnamefont
  {J.}~\bibnamefont {Pereira}}, \bibinfo {author} {\bibfnamefont
  {F.}~\bibnamefont {Recchia}}, \bibinfo {author} {\bibfnamefont
  {H.}~\bibnamefont {Schatz}}, \bibinfo {author} {\bibfnamefont
  {M.}~\bibnamefont {Scott}}, \bibinfo {author} {\bibfnamefont {S.~R.}\
  \bibnamefont {Stroberg}}, \bibinfo {author} {\bibfnamefont {C.}~\bibnamefont
  {Sullivan}}, \bibinfo {author} {\bibfnamefont {L.}~\bibnamefont {Valdez}},
  \bibinfo {author} {\bibfnamefont {C.}~\bibnamefont {Walz}}, \bibinfo {author}
  {\bibfnamefont {D.}~\bibnamefont {Weisshaar}}, \bibinfo {author}
  {\bibfnamefont {S.~J.}\ \bibnamefont {Williams}}, \ and\ \bibinfo {author}
  {\bibfnamefont {K.}~\bibnamefont {Wimmer}},\ }\href {\doibase
  10.1103/PhysRevLett.112.252501} {\bibfield  {journal} {\bibinfo  {journal}
  {Phys. Rev. Lett.}\ }\textbf {\bibinfo {volume} {112}},\ \bibinfo {pages}
  {252501} (\bibinfo {year} {2014})}\BibitemShut {NoStop}%
\bibitem [{\citenamefont {Scott}\ \emph {et~al.}(2014)\citenamefont {Scott},
  \citenamefont {Shimbara}, \citenamefont {Austin}, \citenamefont {Bazin},
  \citenamefont {Brown}, \citenamefont {Deaven}, \citenamefont {Fujita},
  \citenamefont {Guess}, \citenamefont {Gupta}, \citenamefont {Hitt},
  \citenamefont {Koeppe}, \citenamefont {Meharchand}, \citenamefont
  {Nagashima}, \citenamefont {Perdikakis}, \citenamefont {Prinke},
  \citenamefont {Sasano}, \citenamefont {Sullivan}, \citenamefont {Valdez},\
  and\ \citenamefont {Zegers}}]{PhysRevC.90.025801}%
  \BibitemOpen
  \bibfield  {author} {\bibinfo {author} {\bibfnamefont {M.}~\bibnamefont
  {Scott}}, \bibinfo {author} {\bibfnamefont {Y.}~\bibnamefont {Shimbara}},
  \bibinfo {author} {\bibfnamefont {S.~M.}\ \bibnamefont {Austin}}, \bibinfo
  {author} {\bibfnamefont {D.}~\bibnamefont {Bazin}}, \bibinfo {author}
  {\bibfnamefont {B.~A.}\ \bibnamefont {Brown}}, \bibinfo {author}
  {\bibfnamefont {J.~M.}\ \bibnamefont {Deaven}}, \bibinfo {author}
  {\bibfnamefont {Y.}~\bibnamefont {Fujita}}, \bibinfo {author} {\bibfnamefont
  {C.~J.}\ \bibnamefont {Guess}}, \bibinfo {author} {\bibfnamefont
  {S.}~\bibnamefont {Gupta}}, \bibinfo {author} {\bibfnamefont {G.~W.}\
  \bibnamefont {Hitt}}, \bibinfo {author} {\bibfnamefont {D.}~\bibnamefont
  {Koeppe}}, \bibinfo {author} {\bibfnamefont {R.}~\bibnamefont {Meharchand}},
  \bibinfo {author} {\bibfnamefont {M.}~\bibnamefont {Nagashima}}, \bibinfo
  {author} {\bibfnamefont {G.}~\bibnamefont {Perdikakis}}, \bibinfo {author}
  {\bibfnamefont {A.}~\bibnamefont {Prinke}}, \bibinfo {author} {\bibfnamefont
  {M.}~\bibnamefont {Sasano}}, \bibinfo {author} {\bibfnamefont
  {C.}~\bibnamefont {Sullivan}}, \bibinfo {author} {\bibfnamefont
  {L.}~\bibnamefont {Valdez}}, \ and\ \bibinfo {author} {\bibfnamefont
  {R.~G.~T.}\ \bibnamefont {Zegers}},\ }\href@noop {} {\bibfield  {journal}
  {\bibinfo  {journal} {Phys. Rev. C}\ }\textbf {\bibinfo {volume} {90}},\
  \bibinfo {pages} {025801} (\bibinfo {year} {2014})}\BibitemShut {NoStop}%
\bibitem [{\citenamefont {Negret}\ \emph {et~al.}(2006)\citenamefont {Negret},
  \citenamefont {Adachi}, \citenamefont {R.~Barrett}, \citenamefont
  {B\"{a}umer}, \citenamefont {{van den Berg}}, \citenamefont {P.~A.~Berg},
  \citenamefont {von Brentano}, \citenamefont {Frekers}, \citenamefont
  {De~Frenne}, \citenamefont {Fujita}, \citenamefont {Fujita}, \citenamefont
  {Fujita}, \citenamefont {Grewe}, \citenamefont {Haefner}, \citenamefont
  {N.~Harakeh}, \citenamefont {Hatanaka}, \citenamefont {Heyde}, \citenamefont
  {Hunyadi}, \citenamefont {Jacobs},\ and\ \citenamefont
  {Zamick}}]{negret_n14}%
  \BibitemOpen
  \bibfield  {author} {\bibinfo {author} {\bibfnamefont {A.}~\bibnamefont
  {Negret}}, \bibinfo {author} {\bibfnamefont {T.}~\bibnamefont {Adachi}},
  \bibinfo {author} {\bibfnamefont {B.}~\bibnamefont {R.~Barrett}}, \bibinfo
  {author} {\bibfnamefont {C.}~\bibnamefont {B\"{a}umer}}, \bibinfo {author}
  {\bibfnamefont {A.~M.}\ \bibnamefont {{van den Berg}}}, \bibinfo {author}
  {\bibfnamefont {G.}~\bibnamefont {P.~A.~Berg}}, \bibinfo {author}
  {\bibfnamefont {P.}~\bibnamefont {von Brentano}}, \bibinfo {author}
  {\bibfnamefont {D.}~\bibnamefont {Frekers}}, \bibinfo {author} {\bibfnamefont
  {D.}~\bibnamefont {De~Frenne}}, \bibinfo {author} {\bibfnamefont
  {H.}~\bibnamefont {Fujita}}, \bibinfo {author} {\bibfnamefont
  {K.}~\bibnamefont {Fujita}}, \bibinfo {author} {\bibfnamefont
  {Y.}~\bibnamefont {Fujita}}, \bibinfo {author} {\bibfnamefont {E.-W.}\
  \bibnamefont {Grewe}}, \bibinfo {author} {\bibfnamefont {P.}~\bibnamefont
  {Haefner}}, \bibinfo {author} {\bibfnamefont {M.}~\bibnamefont {N.~Harakeh}},
  \bibinfo {author} {\bibfnamefont {K.}~\bibnamefont {Hatanaka}}, \bibinfo
  {author} {\bibfnamefont {K.}~\bibnamefont {Heyde}}, \bibinfo {author}
  {\bibfnamefont {M.}~\bibnamefont {Hunyadi}}, \bibinfo {author} {\bibfnamefont
  {E.}~\bibnamefont {Jacobs}}, \ and\ \bibinfo {author} {\bibfnamefont
  {L.}~\bibnamefont {Zamick}},\ }\href {\doibase 10.1103/PHYSREVLETT.97.062502}
  {\bibfield  {journal} {\bibinfo  {journal} {Phys. Rev. Lett.}\ }\textbf
  {\bibinfo {volume} {97}},\ \bibinfo {pages} {062502} (\bibinfo {year}
  {2006})}\BibitemShut {NoStop}%
\bibitem [{\citenamefont {Bonin}\ \emph {et~al.}(1984)\citenamefont {Bonin},
  \citenamefont {Alamanos}, \citenamefont {Berthier}, \citenamefont {Bruge},
  \citenamefont {Faraggi}, \citenamefont {Legrand}, \citenamefont {Lugol},
  \citenamefont {Mittig}, \citenamefont {Papineau}, \citenamefont {Yavin},
  \citenamefont {Scott}, \citenamefont {Levine}, \citenamefont {Arvieux},
  \citenamefont {Farvacque},\ and\ \citenamefont {Buenerd}}]{BONIN1984349}%
  \BibitemOpen
  \bibfield  {author} {\bibinfo {author} {\bibfnamefont {B.}~\bibnamefont
  {Bonin}}, \bibinfo {author} {\bibfnamefont {N.}~\bibnamefont {Alamanos}},
  \bibinfo {author} {\bibfnamefont {B.}~\bibnamefont {Berthier}}, \bibinfo
  {author} {\bibfnamefont {G.}~\bibnamefont {Bruge}}, \bibinfo {author}
  {\bibfnamefont {H.}~\bibnamefont {Faraggi}}, \bibinfo {author} {\bibfnamefont
  {D.}~\bibnamefont {Legrand}}, \bibinfo {author} {\bibfnamefont
  {J.}~\bibnamefont {Lugol}}, \bibinfo {author} {\bibfnamefont
  {W.}~\bibnamefont {Mittig}}, \bibinfo {author} {\bibfnamefont
  {L.}~\bibnamefont {Papineau}}, \bibinfo {author} {\bibfnamefont
  {A.}~\bibnamefont {Yavin}}, \bibinfo {author} {\bibfnamefont
  {D.}~\bibnamefont {Scott}}, \bibinfo {author} {\bibfnamefont
  {M.}~\bibnamefont {Levine}}, \bibinfo {author} {\bibfnamefont
  {J.}~\bibnamefont {Arvieux}}, \bibinfo {author} {\bibfnamefont
  {L.}~\bibnamefont {Farvacque}}, \ and\ \bibinfo {author} {\bibfnamefont
  {M.}~\bibnamefont {Buenerd}},\ }\href {\doibase
  https://doi.org/10.1016/0375-9474(84)90044-7} {\bibfield  {journal} {\bibinfo
   {journal} {Nuclear Physics A}\ }\textbf {\bibinfo {volume} {430}},\ \bibinfo
  {pages} {349 } (\bibinfo {year} {1984})}\BibitemShut {NoStop}%
\bibitem [{\citenamefont {Ichimura}\ \emph {et~al.}(2006)\citenamefont
  {Ichimura}, \citenamefont {Sakai},\ and\ \citenamefont
  {Wakasa}}]{Ichimura:2006}%
  \BibitemOpen
  \bibfield  {author} {\bibinfo {author} {\bibfnamefont {M.}~\bibnamefont
  {Ichimura}}, \bibinfo {author} {\bibfnamefont {H.}~\bibnamefont {Sakai}}, \
  and\ \bibinfo {author} {\bibfnamefont {T.}~\bibnamefont {Wakasa}},\
  }\href@noop {} {\bibfield  {journal} {\bibinfo  {journal} {Prog. Part. Nucl.
  Phys.}\ }\textbf {\bibinfo {volume} {56}},\ \bibinfo {pages} {446} (\bibinfo
  {year} {2006})}\BibitemShut {NoStop}%
\bibitem [{\citenamefont {Cook}\ and\ \citenamefont {Carr}()}]{FOLD}%
  \BibitemOpen
  \bibfield  {author} {\bibinfo {author} {\bibfnamefont {J.}~\bibnamefont
  {Cook}}\ and\ \bibinfo {author} {\bibfnamefont {J.~A.}\ \bibnamefont
  {Carr}},\ }\href@noop {} {\enquote {\bibinfo {title} {computer program
  $\mathrm{FOLD}$, florida state university (unpublished)},}\ }\bibinfo {note}
  {Based on F. Petrovich and D. Stanley, Nucl. Phys. A275, 487 (1977), modified
  as described in J. Cook et al., Phys. Rev. C 30, 1538 (1984) and R. G. T.
  Zegers, S. Fracasso and G. Colò (2006), unpublished}\BibitemShut {NoStop}%
\bibitem [{\citenamefont {Franey}\ and\ \citenamefont
  {Love}(1985)}]{PhysRevC.31.488}%
  \BibitemOpen
  \bibfield  {author} {\bibinfo {author} {\bibfnamefont {M.~A.}\ \bibnamefont
  {Franey}}\ and\ \bibinfo {author} {\bibfnamefont {W.~G.}\ \bibnamefont
  {Love}},\ }\href@noop {} {\bibfield  {journal} {\bibinfo  {journal} {Phys.
  Rev. C}\ }\textbf {\bibinfo {volume} {31}},\ \bibinfo {pages} {488} (\bibinfo
  {year} {1985})}\BibitemShut {NoStop}%
\bibitem [{\citenamefont {Brown}\ \emph {et~al.}(2004)\citenamefont {Brown},
  \citenamefont {Rae}, \citenamefont {McDonald},\ and\ \citenamefont
  {Horoi}}]{nushellx_msu}%
  \BibitemOpen
  \bibfield  {author} {\bibinfo {author} {\bibfnamefont {B.~A.}\ \bibnamefont
  {Brown}}, \bibinfo {author} {\bibfnamefont {W.~D.~M.}\ \bibnamefont {Rae}},
  \bibinfo {author} {\bibfnamefont {E.}~\bibnamefont {McDonald}}, \ and\
  \bibinfo {author} {\bibfnamefont {M.}~\bibnamefont {Horoi}},\ }\href@noop {}
  {\enquote {\bibinfo {title} {{NuShellX@MSU}},}\ }\bibinfo {howpublished}
  {NSCL Report No. MSUCL-1289-2004.} (\bibinfo {year} {2004})\BibitemShut
  {NoStop}%
\bibitem [{\citenamefont {Lipoglav\ifmmode~\check{s}\else \v{s}\fi{}ek}\ \emph
  {et~al.}(2002)\citenamefont {Lipoglav\ifmmode~\check{s}\else \v{s}\fi{}ek},
  \citenamefont {Baktash}, \citenamefont {Carpenter}, \citenamefont {Dean},
  \citenamefont {Engeland}, \citenamefont {Fahlander}, \citenamefont
  {Hjorth-Jensen}, \citenamefont {Janssens}, \citenamefont {Likar},
  \citenamefont {Nyberg}, \citenamefont {Osnes}, \citenamefont {Paul},
  \citenamefont {Piechaczek}, \citenamefont {Radford}, \citenamefont {Rudolph},
  \citenamefont {Seweryniak}, \citenamefont {Sarantites}, \citenamefont
  {Vencelj},\ and\ \citenamefont {Yu}}]{PhysRevC.65.021302}%
  \BibitemOpen
  \bibfield  {author} {\bibinfo {author} {\bibfnamefont {M.}~\bibnamefont
  {Lipoglav\ifmmode~\check{s}\else \v{s}\fi{}ek}}, \bibinfo {author}
  {\bibfnamefont {C.}~\bibnamefont {Baktash}}, \bibinfo {author} {\bibfnamefont
  {M.~P.}\ \bibnamefont {Carpenter}}, \bibinfo {author} {\bibfnamefont {D.~J.}\
  \bibnamefont {Dean}}, \bibinfo {author} {\bibfnamefont {T.}~\bibnamefont
  {Engeland}}, \bibinfo {author} {\bibfnamefont {C.}~\bibnamefont {Fahlander}},
  \bibinfo {author} {\bibfnamefont {M.}~\bibnamefont {Hjorth-Jensen}}, \bibinfo
  {author} {\bibfnamefont {R.~V.~F.}\ \bibnamefont {Janssens}}, \bibinfo
  {author} {\bibfnamefont {A.}~\bibnamefont {Likar}}, \bibinfo {author}
  {\bibfnamefont {J.}~\bibnamefont {Nyberg}}, \bibinfo {author} {\bibfnamefont
  {E.}~\bibnamefont {Osnes}}, \bibinfo {author} {\bibfnamefont {S.~D.}\
  \bibnamefont {Paul}}, \bibinfo {author} {\bibfnamefont {A.}~\bibnamefont
  {Piechaczek}}, \bibinfo {author} {\bibfnamefont {D.~C.}\ \bibnamefont
  {Radford}}, \bibinfo {author} {\bibfnamefont {D.}~\bibnamefont {Rudolph}},
  \bibinfo {author} {\bibfnamefont {D.}~\bibnamefont {Seweryniak}}, \bibinfo
  {author} {\bibfnamefont {D.~G.}\ \bibnamefont {Sarantites}}, \bibinfo
  {author} {\bibfnamefont {M.}~\bibnamefont {Vencelj}}, \ and\ \bibinfo
  {author} {\bibfnamefont {C.-H.}\ \bibnamefont {Yu}},\ }\href@noop {}
  {\bibfield  {journal} {\bibinfo  {journal} {Phys. Rev. C}\ }\textbf {\bibinfo
  {volume} {65}},\ \bibinfo {pages} {021302(R)} (\bibinfo {year}
  {2002})}\BibitemShut {NoStop}%
\bibitem [{\citenamefont {Lisetskiy}\ \emph {et~al.}(2004)\citenamefont
  {Lisetskiy}, \citenamefont {Brown}, \citenamefont {Horoi},\ and\
  \citenamefont {Grawe}}]{PhysRevC.70.044314}%
  \BibitemOpen
  \bibfield  {author} {\bibinfo {author} {\bibfnamefont {A.~F.}\ \bibnamefont
  {Lisetskiy}}, \bibinfo {author} {\bibfnamefont {B.~A.}\ \bibnamefont
  {Brown}}, \bibinfo {author} {\bibfnamefont {M.}~\bibnamefont {Horoi}}, \ and\
  \bibinfo {author} {\bibfnamefont {H.}~\bibnamefont {Grawe}},\ }\href
  {\doibase 10.1103/PhysRevC.70.044314} {\bibfield  {journal} {\bibinfo
  {journal} {Phys. Rev. C}\ }\textbf {\bibinfo {volume} {70}},\ \bibinfo
  {pages} {044314} (\bibinfo {year} {2004})}\BibitemShut {NoStop}%
\bibitem [{\citenamefont {Hofstee}\ \emph {et~al.}(1995)\citenamefont
  {Hofstee}, \citenamefont {van~der Werf}, \citenamefont {van~den Berg},
  \citenamefont {Blasi}, \citenamefont {Bordewijk}, \citenamefont {Borghols},
  \citenamefont {Leo}, \citenamefont {Emery}, \citenamefont {Fortier},
  \citenamefont {Galès}, \citenamefont {Harakeh}, \citenamefont {den Heijer},
  \citenamefont {de~Jager}, \citenamefont {Langevin-Joliot}, \citenamefont
  {Micheletti}, \citenamefont {Morlet}, \citenamefont {Pignanelli},
  \citenamefont {Schippers}, \citenamefont {de~Vries}, \citenamefont {Willis},\
  and\ \citenamefont {van~der Woude}}]{HOFSTEE1995729}%
  \BibitemOpen
  \bibfield  {author} {\bibinfo {author} {\bibfnamefont {M.~A.}\ \bibnamefont
  {Hofstee}}, \bibinfo {author} {\bibfnamefont {S.~Y.}\ \bibnamefont {van~der
  Werf}}, \bibinfo {author} {\bibfnamefont {A.~M.}\ \bibnamefont {van~den
  Berg}}, \bibinfo {author} {\bibfnamefont {N.}~\bibnamefont {Blasi}}, \bibinfo
  {author} {\bibfnamefont {J.~A.}\ \bibnamefont {Bordewijk}}, \bibinfo {author}
  {\bibfnamefont {W.~T.~A.}\ \bibnamefont {Borghols}}, \bibinfo {author}
  {\bibfnamefont {R.~D.}\ \bibnamefont {Leo}}, \bibinfo {author} {\bibfnamefont
  {G.~T.}\ \bibnamefont {Emery}}, \bibinfo {author} {\bibfnamefont
  {S.}~\bibnamefont {Fortier}}, \bibinfo {author} {\bibfnamefont
  {S.}~\bibnamefont {Galès}}, \bibinfo {author} {\bibfnamefont {M.~N.}\
  \bibnamefont {Harakeh}}, \bibinfo {author} {\bibfnamefont {P.}~\bibnamefont
  {den Heijer}}, \bibinfo {author} {\bibfnamefont {C.~W.}\ \bibnamefont
  {de~Jager}}, \bibinfo {author} {\bibfnamefont {H.}~\bibnamefont
  {Langevin-Joliot}}, \bibinfo {author} {\bibfnamefont {S.}~\bibnamefont
  {Micheletti}}, \bibinfo {author} {\bibfnamefont {M.}~\bibnamefont {Morlet}},
  \bibinfo {author} {\bibfnamefont {M.}~\bibnamefont {Pignanelli}}, \bibinfo
  {author} {\bibfnamefont {J.~M.}\ \bibnamefont {Schippers}}, \bibinfo {author}
  {\bibfnamefont {H.}~\bibnamefont {de~Vries}}, \bibinfo {author}
  {\bibfnamefont {A.}~\bibnamefont {Willis}}, \ and\ \bibinfo {author}
  {\bibfnamefont {A.}~\bibnamefont {van~der Woude}},\ }\href {\doibase
  https://doi.org/10.1016/0375-9474(95)00057-8} {\bibfield  {journal} {\bibinfo
   {journal} {Nuclear Physics A}\ }\textbf {\bibinfo {volume} {588}},\ \bibinfo
  {pages} {729 } (\bibinfo {year} {1995})}\BibitemShut {NoStop}%
\bibitem [{\citenamefont {{S. Y. van der Werf}}()}]{Normod_code}%
  \BibitemOpen
  \bibfield  {author} {\bibinfo {author} {\bibnamefont {{S. Y. van der
  Werf}}},\ }\href@noop {} {}\bibinfo {note} {Computer program \textsc{NORMOD},
  unpublished}\BibitemShut {NoStop}%
\bibitem [{\citenamefont {Pieper}\ and\ \citenamefont
  {Wiringa}(2001)}]{pieper_wiringa_2001}%
  \BibitemOpen
  \bibfield  {author} {\bibinfo {author} {\bibfnamefont {S.~C.}\ \bibnamefont
  {Pieper}}\ and\ \bibinfo {author} {\bibfnamefont {R.~B.}\ \bibnamefont
  {Wiringa}},\ }\href@noop {} {\bibfield  {journal} {\bibinfo  {journal}
  {Annual Review of Nuclear and Particle Science}\ }\textbf {\bibinfo {volume}
  {51}},\ \bibinfo {pages} {53} (\bibinfo {year} {2001})}\BibitemShut {NoStop}%
\bibitem [{\citenamefont {Kamiya}\ \emph {et~al.}(2003)\citenamefont {Kamiya},
  \citenamefont {Hatanaka}, \citenamefont {Adachi}, \citenamefont {Fujita},
  \citenamefont {Hara}, \citenamefont {Kawabata}, \citenamefont {Noro},
  \citenamefont {Sakaguchi}, \citenamefont {Sakamoto}, \citenamefont {Sakemi},
  \citenamefont {Shimbara}, \citenamefont {Shimizu}, \citenamefont {Terashima},
  \citenamefont {Uchida}, \citenamefont {Wakasa}, \citenamefont {Yasuda},
  \citenamefont {Yoshida},\ and\ \citenamefont {Yosoi}}]{PhysRevC.67.064612}%
  \BibitemOpen
  \bibfield  {author} {\bibinfo {author} {\bibfnamefont {J.}~\bibnamefont
  {Kamiya}}, \bibinfo {author} {\bibfnamefont {K.}~\bibnamefont {Hatanaka}},
  \bibinfo {author} {\bibfnamefont {T.}~\bibnamefont {Adachi}}, \bibinfo
  {author} {\bibfnamefont {K.}~\bibnamefont {Fujita}}, \bibinfo {author}
  {\bibfnamefont {K.}~\bibnamefont {Hara}}, \bibinfo {author} {\bibfnamefont
  {T.}~\bibnamefont {Kawabata}}, \bibinfo {author} {\bibfnamefont
  {T.}~\bibnamefont {Noro}}, \bibinfo {author} {\bibfnamefont {H.}~\bibnamefont
  {Sakaguchi}}, \bibinfo {author} {\bibfnamefont {N.}~\bibnamefont {Sakamoto}},
  \bibinfo {author} {\bibfnamefont {Y.}~\bibnamefont {Sakemi}}, \bibinfo
  {author} {\bibfnamefont {Y.}~\bibnamefont {Shimbara}}, \bibinfo {author}
  {\bibfnamefont {Y.}~\bibnamefont {Shimizu}}, \bibinfo {author} {\bibfnamefont
  {S.}~\bibnamefont {Terashima}}, \bibinfo {author} {\bibfnamefont
  {M.}~\bibnamefont {Uchida}}, \bibinfo {author} {\bibfnamefont
  {T.}~\bibnamefont {Wakasa}}, \bibinfo {author} {\bibfnamefont
  {Y.}~\bibnamefont {Yasuda}}, \bibinfo {author} {\bibfnamefont {H.~P.}\
  \bibnamefont {Yoshida}}, \ and\ \bibinfo {author} {\bibfnamefont
  {M.}~\bibnamefont {Yosoi}},\ }\href {\doibase 10.1103/PhysRevC.67.064612}
  {\bibfield  {journal} {\bibinfo  {journal} {Phys. Rev. C}\ }\textbf {\bibinfo
  {volume} {67}},\ \bibinfo {pages} {064612} (\bibinfo {year}
  {2003})}\BibitemShut {NoStop}%
\bibitem [{\citenamefont {Werf}\ \emph {et~al.}(1989)\citenamefont {Werf},
  \citenamefont {Brandenburg}, \citenamefont {Grasduk}, \citenamefont
  {Sterrenburg}, \citenamefont {Harakeh}, \citenamefont {Greenfield},
  \citenamefont {Brown},\ and\ \citenamefont {Fujiwara}}]{VANDERWERF1989305}%
  \BibitemOpen
  \bibfield  {author} {\bibinfo {author} {\bibfnamefont {S.~V.~D.}\
  \bibnamefont {Werf}}, \bibinfo {author} {\bibfnamefont {S.}~\bibnamefont
  {Brandenburg}}, \bibinfo {author} {\bibfnamefont {P.}~\bibnamefont
  {Grasduk}}, \bibinfo {author} {\bibfnamefont {W.}~\bibnamefont
  {Sterrenburg}}, \bibinfo {author} {\bibfnamefont {M.}~\bibnamefont
  {Harakeh}}, \bibinfo {author} {\bibfnamefont {M.}~\bibnamefont {Greenfield}},
  \bibinfo {author} {\bibfnamefont {B.}~\bibnamefont {Brown}}, \ and\ \bibinfo
  {author} {\bibfnamefont {M.}~\bibnamefont {Fujiwara}},\ }\href {\doibase
  https://doi.org/10.1016/0375-9474(89)90177-2} {\bibfield  {journal} {\bibinfo
   {journal} {Nucl. Phys. A}\ }\textbf {\bibinfo {volume} {496}},\ \bibinfo
  {pages} {305 } (\bibinfo {year} {1989})}\BibitemShut {NoStop}%
\bibitem [{\citenamefont {Perdikakis}\ \emph {et~al.}(2011)\citenamefont
  {Perdikakis}, \citenamefont {Zegers}, \citenamefont {Austin}, \citenamefont
  {Bazin}, \citenamefont {Caesar}, \citenamefont {Deaven}, \citenamefont
  {Gade}, \citenamefont {Galaviz}, \citenamefont {Grinyer}, \citenamefont
  {Guess}, \citenamefont {Herlitzius}, \citenamefont {Hitt}, \citenamefont
  {Howard}, \citenamefont {Meharchand}, \citenamefont {Noji}, \citenamefont
  {Sakai}, \citenamefont {Shimbara}, \citenamefont {Smith},\ and\ \citenamefont
  {Tur}}]{PhysRevC.83.054614}%
  \BibitemOpen
  \bibfield  {author} {\bibinfo {author} {\bibfnamefont {G.}~\bibnamefont
  {Perdikakis}}, \bibinfo {author} {\bibfnamefont {R.~G.~T.}\ \bibnamefont
  {Zegers}}, \bibinfo {author} {\bibfnamefont {S.~M.}\ \bibnamefont {Austin}},
  \bibinfo {author} {\bibfnamefont {D.}~\bibnamefont {Bazin}}, \bibinfo
  {author} {\bibfnamefont {C.}~\bibnamefont {Caesar}}, \bibinfo {author}
  {\bibfnamefont {J.~M.}\ \bibnamefont {Deaven}}, \bibinfo {author}
  {\bibfnamefont {A.}~\bibnamefont {Gade}}, \bibinfo {author} {\bibfnamefont
  {D.}~\bibnamefont {Galaviz}}, \bibinfo {author} {\bibfnamefont {G.~F.}\
  \bibnamefont {Grinyer}}, \bibinfo {author} {\bibfnamefont {C.~J.}\
  \bibnamefont {Guess}}, \bibinfo {author} {\bibfnamefont {C.}~\bibnamefont
  {Herlitzius}}, \bibinfo {author} {\bibfnamefont {G.~W.}\ \bibnamefont
  {Hitt}}, \bibinfo {author} {\bibfnamefont {M.~E.}\ \bibnamefont {Howard}},
  \bibinfo {author} {\bibfnamefont {R.}~\bibnamefont {Meharchand}}, \bibinfo
  {author} {\bibfnamefont {S.}~\bibnamefont {Noji}}, \bibinfo {author}
  {\bibfnamefont {H.}~\bibnamefont {Sakai}}, \bibinfo {author} {\bibfnamefont
  {Y.}~\bibnamefont {Shimbara}}, \bibinfo {author} {\bibfnamefont {E.~E.}\
  \bibnamefont {Smith}}, \ and\ \bibinfo {author} {\bibfnamefont
  {C.}~\bibnamefont {Tur}},\ }\href {\doibase 10.1103/PhysRevC.83.054614}
  {\bibfield  {journal} {\bibinfo  {journal} {Phys. Rev. C}\ }\textbf {\bibinfo
  {volume} {83}},\ \bibinfo {pages} {054614} (\bibinfo {year}
  {2011})}\BibitemShut {NoStop}%
\bibitem [{\citenamefont {Sasano}\ \emph {et~al.}(2009)\citenamefont {Sasano},
  \citenamefont {Sakai}, \citenamefont {Yako}, \citenamefont {Wakasa},
  \citenamefont {Asaji}, \citenamefont {Fujita}, \citenamefont {Fujita},
  \citenamefont {Greenfield}, \citenamefont {Hagihara}, \citenamefont
  {Hatanaka}, \citenamefont {Kawabata}, \citenamefont {Kuboki}, \citenamefont
  {Maeda}, \citenamefont {Okamura}, \citenamefont {Saito}, \citenamefont
  {Sakemi}, \citenamefont {Sekiguchi}, \citenamefont {Shimizu}, \citenamefont
  {Takahashi}, \citenamefont {Tameshige},\ and\ \citenamefont
  {Tamii}}]{Sasano:2009}%
  \BibitemOpen
  \bibfield  {author} {\bibinfo {author} {\bibfnamefont {M.}~\bibnamefont
  {Sasano}}, \bibinfo {author} {\bibfnamefont {H.}~\bibnamefont {Sakai}},
  \bibinfo {author} {\bibfnamefont {K.}~\bibnamefont {Yako}}, \bibinfo {author}
  {\bibfnamefont {T.}~\bibnamefont {Wakasa}}, \bibinfo {author} {\bibfnamefont
  {S.}~\bibnamefont {Asaji}}, \bibinfo {author} {\bibfnamefont
  {K.}~\bibnamefont {Fujita}}, \bibinfo {author} {\bibfnamefont
  {Y.}~\bibnamefont {Fujita}}, \bibinfo {author} {\bibfnamefont {M.~B.}\
  \bibnamefont {Greenfield}}, \bibinfo {author} {\bibfnamefont
  {Y.}~\bibnamefont {Hagihara}}, \bibinfo {author} {\bibfnamefont
  {K.}~\bibnamefont {Hatanaka}}, \bibinfo {author} {\bibfnamefont
  {T.}~\bibnamefont {Kawabata}}, \bibinfo {author} {\bibfnamefont
  {H.}~\bibnamefont {Kuboki}}, \bibinfo {author} {\bibfnamefont
  {Y.}~\bibnamefont {Maeda}}, \bibinfo {author} {\bibfnamefont
  {H.}~\bibnamefont {Okamura}}, \bibinfo {author} {\bibfnamefont
  {T.}~\bibnamefont {Saito}}, \bibinfo {author} {\bibfnamefont
  {Y.}~\bibnamefont {Sakemi}}, \bibinfo {author} {\bibfnamefont
  {K.}~\bibnamefont {Sekiguchi}}, \bibinfo {author} {\bibfnamefont
  {Y.}~\bibnamefont {Shimizu}}, \bibinfo {author} {\bibfnamefont
  {Y.}~\bibnamefont {Takahashi}}, \bibinfo {author} {\bibfnamefont
  {Y.}~\bibnamefont {Tameshige}}, \ and\ \bibinfo {author} {\bibfnamefont
  {A.}~\bibnamefont {Tamii}},\ }\href@noop {} {\bibfield  {journal} {\bibinfo
  {journal} {Phys. Rev. C}\ }\textbf {\bibinfo {volume} {79}},\ \bibinfo
  {pages} {024602} (\bibinfo {year} {2009})}\BibitemShut {NoStop}%
\bibitem [{\citenamefont {Grewe}\ \emph {et~al.}(2008)\citenamefont {Grewe},
  \citenamefont {B\"aumer}, \citenamefont {Dohmann}, \citenamefont {Frekers},
  \citenamefont {Harakeh}, \citenamefont {Hollstein}, \citenamefont
  {Johansson}, \citenamefont {Langanke}, \citenamefont {Mart\'{\i}nez-Pinedo},
  \citenamefont {Nowacki}, \citenamefont {Petermann}, \citenamefont {Popescu},
  \citenamefont {Rakers}, \citenamefont {Savran}, \citenamefont {Sieja},
  \citenamefont {Simon}, \citenamefont {Thies}, \citenamefont {{van den Berg}},
  \citenamefont {W\"ortche},\ and\ \citenamefont
  {Zilges}}]{PhysRevC.77.064303}%
  \BibitemOpen
  \bibfield  {author} {\bibinfo {author} {\bibfnamefont {E.-W.}\ \bibnamefont
  {Grewe}}, \bibinfo {author} {\bibfnamefont {C.}~\bibnamefont {B\"aumer}},
  \bibinfo {author} {\bibfnamefont {H.}~\bibnamefont {Dohmann}}, \bibinfo
  {author} {\bibfnamefont {D.}~\bibnamefont {Frekers}}, \bibinfo {author}
  {\bibfnamefont {M.~N.}\ \bibnamefont {Harakeh}}, \bibinfo {author}
  {\bibfnamefont {S.}~\bibnamefont {Hollstein}}, \bibinfo {author}
  {\bibfnamefont {H.}~\bibnamefont {Johansson}}, \bibinfo {author}
  {\bibfnamefont {K.}~\bibnamefont {Langanke}}, \bibinfo {author}
  {\bibfnamefont {G.}~\bibnamefont {Mart\'{\i}nez-Pinedo}}, \bibinfo {author}
  {\bibfnamefont {F.}~\bibnamefont {Nowacki}}, \bibinfo {author} {\bibfnamefont
  {I.}~\bibnamefont {Petermann}}, \bibinfo {author} {\bibfnamefont
  {L.}~\bibnamefont {Popescu}}, \bibinfo {author} {\bibfnamefont
  {S.}~\bibnamefont {Rakers}}, \bibinfo {author} {\bibfnamefont
  {D.}~\bibnamefont {Savran}}, \bibinfo {author} {\bibfnamefont
  {K.}~\bibnamefont {Sieja}}, \bibinfo {author} {\bibfnamefont
  {H.}~\bibnamefont {Simon}}, \bibinfo {author} {\bibfnamefont {J.~H.}\
  \bibnamefont {Thies}}, \bibinfo {author} {\bibfnamefont {A.~M.}\ \bibnamefont
  {{van den Berg}}}, \bibinfo {author} {\bibfnamefont {H.~J.}\ \bibnamefont
  {W\"ortche}}, \ and\ \bibinfo {author} {\bibfnamefont {A.}~\bibnamefont
  {Zilges}},\ }\href {\doibase 10.1103/PhysRevC.77.064303} {\bibfield
  {journal} {\bibinfo  {journal} {Phys. Rev. C}\ }\textbf {\bibinfo {volume}
  {77}},\ \bibinfo {pages} {064303} (\bibinfo {year} {2008})}\BibitemShut
  {NoStop}%
\bibitem [{\citenamefont {Negret}\ and\ \citenamefont
  {Singh}(2015)}]{NEGRET20151}%
  \BibitemOpen
  \bibfield  {author} {\bibinfo {author} {\bibfnamefont {A.}~\bibnamefont
  {Negret}}\ and\ \bibinfo {author} {\bibfnamefont {B.}~\bibnamefont {Singh}},\
  }\href {\doibase https://doi.org/10.1016/j.nds.2014.12.045} {\bibfield
  {journal} {\bibinfo  {journal} {Nuclear Data Sheets}\ }\textbf {\bibinfo
  {volume} {124}},\ \bibinfo {pages} {1 } (\bibinfo {year} {2015})}\BibitemShut
  {NoStop}%
\bibitem [{\citenamefont {Urban}\ \emph {et~al.}(2016)\citenamefont {Urban},
  \citenamefont {Sieja}, \citenamefont {Materna}, \citenamefont
  {Czerwi\ifmmode~\acute{n}\else \'{n}\fi{}ski}, \citenamefont
  {Rz\c{a}ca-Urban}, \citenamefont {Blanc}, \citenamefont {Jentschel},
  \citenamefont {Mutti}, \citenamefont {K\"oster}, \citenamefont {Soldner},
  \citenamefont {de~France}, \citenamefont {Simpson}, \citenamefont {Ur},
  \citenamefont {Bernards}, \citenamefont {Fransen}, \citenamefont {Jolie},
  \citenamefont {Regis}, \citenamefont {Thomas},\ and\ \citenamefont
  {Warr}}]{PhysRevC.94.044328}%
  \BibitemOpen
  \bibfield  {author} {\bibinfo {author} {\bibfnamefont {W.}~\bibnamefont
  {Urban}}, \bibinfo {author} {\bibfnamefont {K.}~\bibnamefont {Sieja}},
  \bibinfo {author} {\bibfnamefont {T.}~\bibnamefont {Materna}}, \bibinfo
  {author} {\bibfnamefont {M.}~\bibnamefont {Czerwi\ifmmode~\acute{n}\else
  \'{n}\fi{}ski}}, \bibinfo {author} {\bibfnamefont {T.}~\bibnamefont
  {Rz\c{a}ca-Urban}}, \bibinfo {author} {\bibfnamefont {A.}~\bibnamefont
  {Blanc}}, \bibinfo {author} {\bibfnamefont {M.}~\bibnamefont {Jentschel}},
  \bibinfo {author} {\bibfnamefont {P.}~\bibnamefont {Mutti}}, \bibinfo
  {author} {\bibfnamefont {U.}~\bibnamefont {K\"oster}}, \bibinfo {author}
  {\bibfnamefont {T.}~\bibnamefont {Soldner}}, \bibinfo {author} {\bibfnamefont
  {G.}~\bibnamefont {de~France}}, \bibinfo {author} {\bibfnamefont {G.~S.}\
  \bibnamefont {Simpson}}, \bibinfo {author} {\bibfnamefont {C.~A.}\
  \bibnamefont {Ur}}, \bibinfo {author} {\bibfnamefont {C.}~\bibnamefont
  {Bernards}}, \bibinfo {author} {\bibfnamefont {C.}~\bibnamefont {Fransen}},
  \bibinfo {author} {\bibfnamefont {J.}~\bibnamefont {Jolie}}, \bibinfo
  {author} {\bibfnamefont {J.-M.}\ \bibnamefont {Regis}}, \bibinfo {author}
  {\bibfnamefont {T.}~\bibnamefont {Thomas}}, \ and\ \bibinfo {author}
  {\bibfnamefont {N.}~\bibnamefont {Warr}},\ }\href {\doibase
  10.1103/PhysRevC.94.044328} {\bibfield  {journal} {\bibinfo  {journal} {Phys.
  Rev. C}\ }\textbf {\bibinfo {volume} {94}},\ \bibinfo {pages} {044328}
  (\bibinfo {year} {2016})}\BibitemShut {NoStop}%
\bibitem [{\citenamefont {Brown}\ and\ \citenamefont
  {Rykaczewski}(1994)}]{PhysRevC.50.R2270}%
  \BibitemOpen
  \bibfield  {author} {\bibinfo {author} {\bibfnamefont {B.~A.}\ \bibnamefont
  {Brown}}\ and\ \bibinfo {author} {\bibfnamefont {K.}~\bibnamefont
  {Rykaczewski}},\ }\href@noop {} {\bibfield  {journal} {\bibinfo  {journal}
  {Phys. Rev. C}\ }\textbf {\bibinfo {volume} {50}},\ \bibinfo {pages} {R2270}
  (\bibinfo {year} {1994})}\BibitemShut {NoStop}%
\bibitem [{\citenamefont {Dillmann}\ \emph {et~al.}(2003)\citenamefont
  {Dillmann}, \citenamefont {Kratz}, \citenamefont {W\"ohr}, \citenamefont
  {Arndt}, \citenamefont {Brown}, \citenamefont {Hoff}, \citenamefont
  {Hjorth-Jensen}, \citenamefont {K\"oster}, \citenamefont {Ostrowski},
  \citenamefont {Pfeiffer}, \citenamefont {Seweryniak}, \citenamefont
  {Shergur},\ and\ \citenamefont {Walters}}]{PhysRevLett.91.162503}%
  \BibitemOpen
  \bibfield  {author} {\bibinfo {author} {\bibfnamefont {I.}~\bibnamefont
  {Dillmann}}, \bibinfo {author} {\bibfnamefont {K.-L.}\ \bibnamefont {Kratz}},
  \bibinfo {author} {\bibfnamefont {A.}~\bibnamefont {W\"ohr}}, \bibinfo
  {author} {\bibfnamefont {O.}~\bibnamefont {Arndt}}, \bibinfo {author}
  {\bibfnamefont {B.~A.}\ \bibnamefont {Brown}}, \bibinfo {author}
  {\bibfnamefont {P.}~\bibnamefont {Hoff}}, \bibinfo {author} {\bibfnamefont
  {M.}~\bibnamefont {Hjorth-Jensen}}, \bibinfo {author} {\bibfnamefont
  {U.}~\bibnamefont {K\"oster}}, \bibinfo {author} {\bibfnamefont {A.~N.}\
  \bibnamefont {Ostrowski}}, \bibinfo {author} {\bibfnamefont {B.}~\bibnamefont
  {Pfeiffer}}, \bibinfo {author} {\bibfnamefont {D.}~\bibnamefont
  {Seweryniak}}, \bibinfo {author} {\bibfnamefont {J.}~\bibnamefont {Shergur}},
  \ and\ \bibinfo {author} {\bibfnamefont {W.~B.}\ \bibnamefont {Walters}}
  (\bibinfo {collaboration} {the ISOLDE Collaboration}),\ }\href {\doibase
  10.1103/PhysRevLett.91.162503} {\bibfield  {journal} {\bibinfo  {journal}
  {Phys. Rev. Lett.}\ }\textbf {\bibinfo {volume} {91}},\ \bibinfo {pages}
  {162503} (\bibinfo {year} {2003})}\BibitemShut {NoStop}%
\bibitem [{\citenamefont {{Towner}}(1985)}]{1985NuPhA.444..402T}%
  \BibitemOpen
  \bibfield  {author} {\bibinfo {author} {\bibfnamefont {I.~S.}\ \bibnamefont
  {{Towner}}},\ }\href {\doibase 10.1016/0375-9474(85)90459-2} {\bibfield
  {journal} {\bibinfo  {journal} {Nuc. Phys. A}\ }\textbf {\bibinfo {volume}
  {444}},\ \bibinfo {pages} {402} (\bibinfo {year} {1985})}\BibitemShut
  {NoStop}%
\bibitem [{\citenamefont {Pfeiffer}\ \emph {et~al.}(1986)\citenamefont
  {Pfeiffer}, \citenamefont {Mairle}, \citenamefont {Kn{\"o}pfle},
  \citenamefont {Kihm}, \citenamefont {Seegert}, \citenamefont {Grabmayr},
  \citenamefont {Wagner}, \citenamefont {Bechtold},\ and\ \citenamefont
  {Friedrich}}]{PFEIFFER1986381}%
  \BibitemOpen
  \bibfield  {author} {\bibinfo {author} {\bibfnamefont {A.}~\bibnamefont
  {Pfeiffer}}, \bibinfo {author} {\bibfnamefont {G.}~\bibnamefont {Mairle}},
  \bibinfo {author} {\bibfnamefont {K.~T.}\ \bibnamefont {Kn{\"o}pfle}},
  \bibinfo {author} {\bibfnamefont {T.}~\bibnamefont {Kihm}}, \bibinfo {author}
  {\bibfnamefont {G.}~\bibnamefont {Seegert}}, \bibinfo {author} {\bibfnamefont
  {P.}~\bibnamefont {Grabmayr}}, \bibinfo {author} {\bibfnamefont {G.~J.}\
  \bibnamefont {Wagner}}, \bibinfo {author} {\bibfnamefont {V.}~\bibnamefont
  {Bechtold}}, \ and\ \bibinfo {author} {\bibfnamefont {L.}~\bibnamefont
  {Friedrich}},\ }\href {\doibase https://doi.org/10.1016/0375-9474(86)90313-1}
  {\bibfield  {journal} {\bibinfo  {journal} {Nuc. Phys. A}\ }\textbf {\bibinfo
  {volume} {455}},\ \bibinfo {pages} {381 } (\bibinfo {year}
  {1986})}\BibitemShut {NoStop}%
\bibitem [{\citenamefont {May}\ and\ \citenamefont
  {Lewis}(1972)}]{PhysRevC.5.117}%
  \BibitemOpen
  \bibfield  {author} {\bibinfo {author} {\bibfnamefont {E.~C.}\ \bibnamefont
  {May}}\ and\ \bibinfo {author} {\bibfnamefont {S.~A.}\ \bibnamefont
  {Lewis}},\ }\href {\doibase 10.1103/PhysRevC.5.117} {\bibfield  {journal}
  {\bibinfo  {journal} {Phys. Rev. C}\ }\textbf {\bibinfo {volume} {5}},\
  \bibinfo {pages} {117} (\bibinfo {year} {1972})}\BibitemShut {NoStop}%
\bibitem [{\citenamefont {Brown}\ and\ \citenamefont
  {Wildenthal}(1985)}]{BROWN1985347}%
  \BibitemOpen
  \bibfield  {author} {\bibinfo {author} {\bibfnamefont {B.~A.}\ \bibnamefont
  {Brown}}\ and\ \bibinfo {author} {\bibfnamefont {B.~H.}\ \bibnamefont
  {Wildenthal}},\ }\href {\doibase
  https://doi.org/10.1016/0092-640X(85)90009-9} {\bibfield  {journal} {\bibinfo
   {journal} {Atomic Data and Nucl. Data Tables}\ }\textbf {\bibinfo {volume}
  {33}},\ \bibinfo {pages} {347 } (\bibinfo {year} {1985})}\BibitemShut
  {NoStop}%
\bibitem [{\citenamefont {Mart\'{\i}nez-Pinedo}\ \emph
  {et~al.}(1996)\citenamefont {Mart\'{\i}nez-Pinedo}, \citenamefont {Poves},
  \citenamefont {Caurier},\ and\ \citenamefont {Zuker}}]{PhysRevC.53.R2602}%
  \BibitemOpen
  \bibfield  {author} {\bibinfo {author} {\bibfnamefont {G.}~\bibnamefont
  {Mart\'{\i}nez-Pinedo}}, \bibinfo {author} {\bibfnamefont {A.}~\bibnamefont
  {Poves}}, \bibinfo {author} {\bibfnamefont {E.}~\bibnamefont {Caurier}}, \
  and\ \bibinfo {author} {\bibfnamefont {A.~P.}\ \bibnamefont {Zuker}},\
  }\href@noop {} {\bibfield  {journal} {\bibinfo  {journal} {Phys. Rev. C}\
  }\textbf {\bibinfo {volume} {53}},\ \bibinfo {pages} {R2602} (\bibinfo {year}
  {1996})}\BibitemShut {NoStop}%
\bibitem [{\citenamefont {Gaarde}(1985)}]{GAA85}%
  \BibitemOpen
  \bibfield  {author} {\bibinfo {author} {\bibfnamefont {C.}~\bibnamefont
  {Gaarde}},\ }in\ \href@noop {} {\emph {\bibinfo {booktitle} {Proc. Niels Bohr
  Centennial Conference on Nuclear Structure, Copenhagen}}},\ \bibinfo {editor}
  {edited by\ \bibinfo {editor} {\bibfnamefont {R.~A.}\ \bibnamefont
  {Broglia}}, \bibinfo {editor} {\bibfnamefont {G.~B.}\ \bibnamefont
  {Hagemann}}, \ and\ \bibinfo {editor} {\bibfnamefont {B.}~\bibnamefont
  {Herskind}}}\ (\bibinfo  {publisher} {North-Holland, Amsterdam},\ \bibinfo
  {year} {1985})\ p.\ \bibinfo {pages} {449c}\BibitemShut {NoStop}%
\bibitem [{\citenamefont {Avogadro}\ and\ \citenamefont
  {Nakatsukasa}(2011)}]{PhysRevC.84.014314}%
  \BibitemOpen
  \bibfield  {author} {\bibinfo {author} {\bibfnamefont {P.}~\bibnamefont
  {Avogadro}}\ and\ \bibinfo {author} {\bibfnamefont {T.}~\bibnamefont
  {Nakatsukasa}},\ }\href {\doibase 10.1103/PhysRevC.84.014314} {\bibfield
  {journal} {\bibinfo  {journal} {Phys. Rev. C}\ }\textbf {\bibinfo {volume}
  {84}},\ \bibinfo {pages} {014314} (\bibinfo {year} {2011})}\BibitemShut
  {NoStop}%
\bibitem [{\citenamefont {Mustonen}\ \emph {et~al.}(2014)\citenamefont
  {Mustonen}, \citenamefont {Shafer}, \citenamefont {Zenginerler},\ and\
  \citenamefont {Engel}}]{PhysRevC.90.024308}%
  \BibitemOpen
  \bibfield  {author} {\bibinfo {author} {\bibfnamefont {M.~T.}\ \bibnamefont
  {Mustonen}}, \bibinfo {author} {\bibfnamefont {T.}~\bibnamefont {Shafer}},
  \bibinfo {author} {\bibfnamefont {Z.}~\bibnamefont {Zenginerler}}, \ and\
  \bibinfo {author} {\bibfnamefont {J.}~\bibnamefont {Engel}},\ }\href
  {\doibase 10.1103/PhysRevC.90.024308} {\bibfield  {journal} {\bibinfo
  {journal} {Phys. Rev. C}\ }\textbf {\bibinfo {volume} {90}},\ \bibinfo
  {pages} {024308} (\bibinfo {year} {2014})}\BibitemShut {NoStop}%
\bibitem [{\citenamefont {Shafer}\ \emph {et~al.}(2016)\citenamefont {Shafer},
  \citenamefont {Engel}, \citenamefont {Fr{\"o}hlich}, \citenamefont
  {McLaughlin}, \citenamefont {Mumpower},\ and\ \citenamefont
  {Surman}}]{PhysRevC.94.055802}%
  \BibitemOpen
  \bibfield  {author} {\bibinfo {author} {\bibfnamefont {T.}~\bibnamefont
  {Shafer}}, \bibinfo {author} {\bibfnamefont {J.}~\bibnamefont {Engel}},
  \bibinfo {author} {\bibfnamefont {C.}~\bibnamefont {Fr{\"o}hlich}}, \bibinfo
  {author} {\bibfnamefont {G.~C.}\ \bibnamefont {McLaughlin}}, \bibinfo
  {author} {\bibfnamefont {M.}~\bibnamefont {Mumpower}}, \ and\ \bibinfo
  {author} {\bibfnamefont {R.}~\bibnamefont {Surman}},\ }\href {\doibase
  10.1103/PhysRevC.94.055802} {\bibfield  {journal} {\bibinfo  {journal} {Phys.
  Rev. C}\ }\textbf {\bibinfo {volume} {94}},\ \bibinfo {pages} {055802}
  (\bibinfo {year} {2016})}\BibitemShut {NoStop}%
\bibitem [{\citenamefont {Mustonen}\ and\ \citenamefont
  {Engel}(2016)}]{PhysRevC.93.014304}%
  \BibitemOpen
  \bibfield  {author} {\bibinfo {author} {\bibfnamefont {M.~T.}\ \bibnamefont
  {Mustonen}}\ and\ \bibinfo {author} {\bibfnamefont {J.}~\bibnamefont
  {Engel}},\ }\href {\doibase 10.1103/PhysRevC.93.014304} {\bibfield  {journal}
  {\bibinfo  {journal} {Phys. Rev. C}\ }\textbf {\bibinfo {volume} {93}},\
  \bibinfo {pages} {014304} (\bibinfo {year} {2016})}\BibitemShut {NoStop}%
\bibitem [{\citenamefont {Reyes}\ \emph {et~al.}(2010)\citenamefont {Reyes},
  \citenamefont {Gupta}, \citenamefont {Schatz}, \citenamefont {Kratz},\ and\
  \citenamefont {M{\"o}ller}}]{ecrates_reyes}%
  \BibitemOpen
  \bibfield  {author} {\bibinfo {author} {\bibfnamefont {A.~D.~B.}\
  \bibnamefont {Reyes}}, \bibinfo {author} {\bibfnamefont {S.}~\bibnamefont
  {Gupta}}, \bibinfo {author} {\bibfnamefont {H.}~\bibnamefont {Schatz}},
  \bibinfo {author} {\bibfnamefont {K.}~\bibnamefont {Kratz}}, \ and\ \bibinfo
  {author} {\bibfnamefont {P.}~\bibnamefont {M{\"o}ller}},\ }\href@noop {}
  {\bibfield  {journal} {\bibinfo  {journal} {Proceedings of Science}\ }\textbf
  {\bibinfo {volume} {28}},\ \bibinfo {pages} {75} (\bibinfo {year}
  {2010})}\BibitemShut {NoStop}%
\bibitem [{\citenamefont {Timmes}\ and\ \citenamefont
  {Swesty}(2000)}]{0067-0049-126-2-501}%
  \BibitemOpen
  \bibfield  {author} {\bibinfo {author} {\bibfnamefont {F.~X.}\ \bibnamefont
  {Timmes}}\ and\ \bibinfo {author} {\bibfnamefont {F.~D.}\ \bibnamefont
  {Swesty}},\ }\href {http://stacks.iop.org/0067-0049/126/i=2/a=501} {\bibfield
   {journal} {\bibinfo  {journal} {Astrophys. J. Supp.}\ }\textbf {\bibinfo
  {volume} {126}},\ \bibinfo {pages} {501} (\bibinfo {year}
  {2000})}\BibitemShut {NoStop}%
\bibitem [{\citenamefont {Valdez}(2012)}]{ValdezTHESIS}%
  \BibitemOpen
  \bibfield  {author} {\bibinfo {author} {\bibfnamefont {L.}~\bibnamefont
  {Valdez}},\ }\emph {\bibinfo {title} {Electron-captures in Supernovae}},\
  \href@noop {} {Master's thesis},\ \bibinfo  {school} {Michigan State
  University} (\bibinfo {year} {2012})\BibitemShut {NoStop}%
\bibitem [{\citenamefont {Fuller}\ \emph {et~al.}(1985)\citenamefont {Fuller},
  \citenamefont {Fowler},\ and\ \citenamefont {Newman}}]{fuller85}%
  \BibitemOpen
  \bibfield  {author} {\bibinfo {author} {\bibfnamefont {G.~M.}\ \bibnamefont
  {Fuller}}, \bibinfo {author} {\bibfnamefont {W.~A.}\ \bibnamefont {Fowler}},
  \ and\ \bibinfo {author} {\bibfnamefont {M.~J.}\ \bibnamefont {Newman}},\
  }\href@noop {} {\bibfield  {journal} {\bibinfo  {journal} {Astrophys. J.}\
  }\textbf {\bibinfo {volume} {293}},\ \bibinfo {pages} {1} (\bibinfo {year}
  {1985})}\BibitemShut {NoStop}%
\bibitem [{\citenamefont {Engel}\ \emph {et~al.}(1999)\citenamefont {Engel},
  \citenamefont {Bender}, \citenamefont {Dobaczewski}, \citenamefont
  {Nazarewicz},\ and\ \citenamefont {Surman}}]{PhysRevC.60.014302}%
  \BibitemOpen
  \bibfield  {author} {\bibinfo {author} {\bibfnamefont {J.}~\bibnamefont
  {Engel}}, \bibinfo {author} {\bibfnamefont {M.}~\bibnamefont {Bender}},
  \bibinfo {author} {\bibfnamefont {J.}~\bibnamefont {Dobaczewski}}, \bibinfo
  {author} {\bibfnamefont {W.}~\bibnamefont {Nazarewicz}}, \ and\ \bibinfo
  {author} {\bibfnamefont {R.}~\bibnamefont {Surman}},\ }\href {\doibase
  10.1103/PhysRevC.60.014302} {\bibfield  {journal} {\bibinfo  {journal} {Phys.
  Rev. C}\ }\textbf {\bibinfo {volume} {60}},\ \bibinfo {pages} {014302}
  (\bibinfo {year} {1999})}\BibitemShut {NoStop}%
\bibitem [{\citenamefont {Gallagher}\ and\ \citenamefont
  {Moszkowski}(1958)}]{PhysRev.111.1282}%
  \BibitemOpen
  \bibfield  {author} {\bibinfo {author} {\bibfnamefont {C.~J.}\ \bibnamefont
  {Gallagher}}\ and\ \bibinfo {author} {\bibfnamefont {S.~A.}\ \bibnamefont
  {Moszkowski}},\ }\href {\doibase 10.1103/PhysRev.111.1282} {\bibfield
  {journal} {\bibinfo  {journal} {Phys. Rev.}\ }\textbf {\bibinfo {volume}
  {111}},\ \bibinfo {pages} {1282} (\bibinfo {year} {1958})}\BibitemShut
  {NoStop}%
\bibitem [{\citenamefont {Woosley}\ and\ \citenamefont {Weaver}(1995)}]{WOO95}%
  \BibitemOpen
  \bibfield  {author} {\bibinfo {author} {\bibfnamefont {S.~E.}\ \bibnamefont
  {Woosley}}\ and\ \bibinfo {author} {\bibfnamefont {T.~A.}\ \bibnamefont
  {Weaver}},\ }\href@noop {} {\bibfield  {journal} {\bibinfo  {journal} {Ap. J.
  S}\ }\textbf {\bibinfo {volume} {101}},\ \bibinfo {pages} {181} (\bibinfo
  {year} {1995})}\BibitemShut {NoStop}%
\bibitem [{\citenamefont {Steiner}\ \emph {et~al.}(2013)\citenamefont
  {Steiner}, \citenamefont {Hempel},\ and\ \citenamefont
  {Fischer}}]{0004-637X-774-1-17}%
  \BibitemOpen
  \bibfield  {author} {\bibinfo {author} {\bibfnamefont {A.~W.}\ \bibnamefont
  {Steiner}}, \bibinfo {author} {\bibfnamefont {M.}~\bibnamefont {Hempel}}, \
  and\ \bibinfo {author} {\bibfnamefont {T.}~\bibnamefont {Fischer}},\
  }\href@noop {} {\bibfield  {journal} {\bibinfo  {journal} {Astrophys. J.}\
  }\textbf {\bibinfo {volume} {774}},\ \bibinfo {pages} {17} (\bibinfo {year}
  {2013})}\BibitemShut {NoStop}%
\end{thebibliography}%

\end{document}